\documentclass[onecolumn,draftcls,11pt, twoside]{IEEEtran}

\usepackage{tabu}
\usepackage{amsmath,amsthm,graphicx,cite}
\usepackage{srcltx}
\usepackage{epsfig,amsfonts,subfigure}
\usepackage{graphicx,cite,amssymb,amsmath}
\usepackage{color}
\usepackage{comment}

\usepackage{amsthm}
\usepackage[nolist]{acronym}
\usepackage{psfrag}
\usepackage{perso}
\usepackage{enumitem}
\usepackage{bbold}
\usepackage{pgfplots}
 \pgfplotsset{compat=newest}
    \pgfplotsset{plot coordinates/math parser=false}
    \pgfplotsset{
    label style={anchor=near ticklabel},
    xlabel style={yshift=0.0em},
    ylabel style={yshift=-0.3em},
    tick label style={font=\footnotesize },
    label style={font=\footnotesize},
    legend style={font=\footnotesize},
    title style={font=\fontsize{7}}}
\usepackage{xcolor}
\definecolor{iso}{rgb}{0.7,0.7,0.7}
\usepackage{blindtext}
\usepackage{relsize}
\usepackage{mathtools}
\usepackage{bm}
\mathtoolsset{showonlyrefs}

\usepackage[section]{placeins}

\usepgflibrary{arrows}
\usetikzlibrary{patterns}
\usepgflibrary{decorations.pathmorphing}
\usepackage{soul}

\definecolor{forestgreen}{rgb}{0.13, 0.55, 0.13}

\ifthenelse{\isundefined{\Bmatrix}}{
  \newcommand{\Bmatrix}{\mathbf{B}}
}{
  \renewcommand{\Bmatrix}{\mathbf{B}}
}
\renewcommand{\Bmatrix}{\mathbf{B}}

\ifthenelse{\isundefined{\C}}{
    \newcommand{\C}[1]{\mathtt{C}_{#1}}
    }{
    \renewcommand{\C}[1]{\mathtt{C}_{#1}}
    }

\ifthenelse{\isundefined{\N}}{
    \newcommand{\N}[1]{\mathsf{N}_{#1}}
    }{
    \renewcommand{\N}[1]{\mathsf{N}_{#1}}
    }

\ifthenelse{\isundefined{\m}}{
    \newcommand{\m}{m}
    }{
    \renewcommand{\m}{m}
    }

\newcommand{\we}{A}

\ifthenelse{\isundefined{\g}}{
    \newcommand{\g}{\mathsf g}
    }{
    \renewcommand{\g}{\mathsf g}
    }

\ifthenelse{\isundefined{\G}}{
    \newcommand{\G}{\mathbf{G}}
    }{
    \renewcommand{\G}{\mathbf{G}}
    }

\ifthenelse{\isundefined{\A}}{
    \newcommand{\A}{\we}
    }{
    \renewcommand{\A}{\we}
    }

\ifthenelse{\isundefined{\argmax}}{
\newcommand{\argmax}{{\arg\max}}
}{
\renewcommand{\argmax}{{\arg\max}}
}

\makeatletter
\newcommand{\vast}{\bBigg@{4}}
\newcommand{\Vast}{\bBigg@{5}}
\makeatother

\usepackage{booktabs,setspace}
\usepackage{threeparttable}
\usepackage{tikz}
\usetikzlibrary{shapes}
\usetikzlibrary{arrows}
\usetikzlibrary{positioning}
\usetikzlibrary{arrows.meta}
\usepackage{amsmath,bm}
\usepackage{mathrsfs}
\usepackage{amsmath}

\usepackage{graphicx}
\usepackage{float}
\usepackage{subfigure} \usepackage[ruled,vlined,linesnumbered]{algorithm2e}
\usepackage{amsmath, amssymb}
\usepackage{makecell}
\usepackage{multirow}

\newcommand{\numsens}{N_{\mathrm{R}}}
\newcommand{\pulseintegfactor}{N_{\mathrm{s}}}
\newcommand{\samplesperscan}{N_{\mathrm{c}}}
\newcommand{\bGamma}{\boldsymbol{\Gamma}}

\definecolor{sunflower}{HTML}{FFBB00}
\definecolor{crismon}{HTML}{F62A00}

\begin{document}
\begin{acronym}
\acro{RCS}{radar cross section}
\acro{FC}{fusion center}
\acro{UWB}{ultra-wideband}
\acro{TBD}{Track-before-detect}
\acro{PF}{particle filter}
\acro{RSN}{radar sensor network}
\acro{WEF}{weight enumerator function}
\acro{SNR}{signal-to-noise ratio}
\acro{ER}{empty room}
\acro{IR}{impulse radio}
\acro{ToA}{time-of-arrival}
\acro{OSPA}{optimal subpattern assignment}
\acro{3D}{three-dimensional}
\acro{2D}{two-dimensional}
\acro{MHT}{multiple hypothesis tracking}
\acro{CFAR}{constant false alarm}
\acro{MTI}{moving target indication}
\acro{PHD}{probability hypothesis density}
\end{acronym}

%%%%%%%%%%%%%%%%%%%%%%%%%%%%%%%%%%%%%%%%%%%%%%%%%%%%%%%%%%%%%%%%%%%%%%%%%%%%%%%%%%%%%%%%%%%%%%%%%%%%
%%%%%%%%%%%%%%%%%%%%%%%%
\title{A Track-Before-Detect Algorithm\\ for UWB Radar Sensor Networks}

\author{Bo~Yan, Andrea~Giorgetti, and Enrico~Paolini
\thanks{This work was supported in part by MIUR under program ``Departments of Excellence (2018-2022) -- Precise-CPS,'' in part by the POR FESR 2014-2020 program under CoACh project, and in part by the Fundamental Research Funds for the Central Universities and China Postdoctoral Science Foundation (No. 2019M663633).
Part of this work has been presented at the 2020 IEEE Radar Conference (RadarConf20) \cite{yan2020:Radarconf}.

B. Yan is with the School of Aerospace Science and Technology, XIDIAN University, China, and with the Department of Electrical, Electronic, and Information Engineering (DEI), University of Bologna. A. Giorgetti and E. Paolini are with CNIT, the Department of Electrical, Electronic, and Information Engineering (DEI), University of Bologna.
Italy (e-mail: \{bo.yan, andrea.giorgetti, e.paolini\}@unibo.it).}%
}
\markboth{}{Yan \MakeLowercase{\textit{et al.}}: A Track-Before-Detect Approach for UWB Radar Sensor Networks}

\maketitle

\begin{abstract}
Precise localization and tracking of moving non-collaborative persons and objects using a network of ultra-wideband (UWB) radar nodes has been shown to represent a practical and effective approach. In UWB radar sensor networks (RSNs), existence of strong clutter, weak target echoes, and closely spaced targets are obstacles to achieving a satisfactory tracking performance. Using a track-before-detect (TBD) approach, the waveform obtained by each node during a time period are jointly processed. Both spatial information and temporal relationship between measurements are exploited in generating all possible candidate trajectories and only the best trajectories are selected as the outcome. 
The effectiveness of the developed TBD technique for UWB RSNs is confirmed by numerical simulations and by two experimental results, both carried out with actual UWB signals. In the first experiment, a human target is tracked by a monostatic radar network with an average localization error of 41.9 cm with no false alarm trajectory in a cluttered outdoor environment. In the second experiment, two targets are detected by multistatic radar network with localization errors of 25.4 cm and 19.7 cm, and detection rate of the two targets is 88.75\%, and no false alarm trajectory. 
\end{abstract}
%%%%%%%%%%%%%%%%%%%%%%%%%%%%%%%%%%%%%%%%%%%%%%%%%%%%%%%%%%%%%%%%%%%%%%%%%%%%%%%%%%%%%%%%%%%%%%%%%%%%%%%%%%
%%%%%%%%%%%%%%%%%%%%%%%%%%%%%%%%%%%%%%%%%%%%%%%%%%%%%%%%%%%%%%%%%%%%%%%%%%%%%%%%%%%%%%%%%%%%%%%%%%%%%%%%%% 
\section{Introduction}\label{sec:Intro}

\IEEEPARstart{I}{n recent} years, the \ac{UWB} wireless technology has gained an increasing importance in a number of civilian and military radar applications \cite{win2011network,withington2003enhancing,bartoletti2014sensor,chiani2018sensor,SobPaoGioMazChi:J14}. 
In an \ac{IR}\footnote{Although the scope of \ac{UWB} wireless technology has been extended beyond the \ac{IR} technique, consisting of the transmission of sequences of very short duration pulses, in this paper the term \ac{UWB} always refers to \ac{IR} \ac{UWB}.} \ac{UWB} \ac{RSN}, sometimes also referred to as an \ac{IR} \ac{UWB} wireless sensor radar, radio nodes (or simply sensors), are deployed in the surveillance area to transmit and/or receive nanosecond-duration pulses; human targets or moving objects are then detected, located, and tracked by extracting information from all gathered waveforms and fusing it at a central node. 
Owing to ultra-wide signal bandwidth, the \ac{UWB} technology allows achieving very high ranging accuracy and range resolution even in harsh environments affected by dense multipath propagation, such as indoor ones.
However, it also poses several challenges. 

As a main issue, the transmit power of \ac{UWB} radio nodes is severely limited by worldwide regulations. 
Especially when working with targets characterized by a limited \ac{RCS}, such as human beings, sensors in the network may experience poor \ac{SNR} conditions even if relatively large pulse integration factors (e.g., in the order of $10^4$) are used. In this regime, a target may not generate measurements in all scan periods, leading to a weak target problem.
In the framework of a traditional approach, consisting of a detection step in which measurements acquired in one scan are processed to perform detection followed by a tracking step in which target points at different scans are associated to form the trajectories, the weak target issue leads to miss-detection events, inaccurate localization, and even loss of the tracks. 
In \ac{UWB} \acp{RSN}, miss-detection events arise even when a weak target echo is received only by a subset of the sensors, since at least three sensors are necessary to locate the target and generate one point. 
It is worth noting that merely lowering the detection threshold to increase the detection rate is an ineffective solution, due to the considerable amount of false alarms arising as a consequence of the low \ac{SNR} and the harsh propagation environment. 

As an additional issue, the ultra-wide signal bandwidth makes the range resolution of \ac{UWB} sensors in the order of $30\,\mathrm{cm}$, smaller than the linear dimension of a target such as a human being. Therefore, a target may generate multiple measurements as it can in principle be detected in several resolution cells, leading to an extended target problem \cite{chiani2018sensor,he14:keystone}. Although a number of tracking methods have been developed to track extended targets, including methods based on the \ac{PHD} filter \cite{granstrom2012extended,granstrom2017extended,granstrom2015gamma,lundquist2013extended,leonard2019multi} and methods based on the Bernoulli filter \cite{CHAI2020107501}, it is known that, without a sufficient amount of a priori information about the target, the presence of multiple measurements deteriorates tracking performance. 
We also point out that targets in \ac{UWB} \acp{RSN} are often human beings, whose motion may be characterized by some degree of maneuverability, with course and speed changing within one or two scans. 

\ac{TBD} is a well-known approach to cope with the weak target issue. Echos of weak targets are accumulated through multiple scans to reach the detection threshold. As such, rather than declaring the presence of targets relying on the measurements collected in a single scan, measurements received in multiple scans are jointly processed, keeping record of a number of candidate trajectories, and confirming only a subset of them (hence detecting the associated targets). 
Several \ac{TBD} approaches are available in the literature (e.g., 
\cite{Mohanty81:computer,arnold93:efficient,tonissen98:ML,hyoungjun99:optimization,buzzi05:TBD,blackman1999:modern,YAN2021107821,CHAI2020107501,leonard2019multi,ristic2021bernoulli,ubeda2017adaptive,YI2019303,errasti2014track,yan2017track,wei2010tbd,yi13:efficient,carlson94:search,moyer2011multi,yan2019detection,yan2019grey,grossi2013heuristic,wang2018greedy,kwon2019particle,salmond01:particle,yi2020multi,LI2020107648,aprile2016track,wang2020complex,jiang2016track,cao2020sequential}), each tailored to specific tracking problems.
For example, in the \ac{TBD} framework, the issue of extended target has been addressed through point voting in Hough transform \cite{yan2019detection,YAN2021107821} or particle filter based probability density estimation of extended target state \cite{errasti2014track}. The existing \ac{TBD} implementations can be categorized into four classes, summarized hereafter.

\textit{Particle filter (PF) based \ac{TBD}} \cite{salmond01:particle,kwon2019particle,ubeda2017adaptive,YI2019303,errasti2014track}: Particle filters can sequentially approximate the a posteriori probability density function with any dynamic and measurement models, making PF-\ac{TBD} superior to other methods in the presence of nonlinear or non-Gaussian noise. Some \ac{PHD} filters \cite{lundquist2013extended,leonard2019multi} and Bernoulli filters \cite{CHAI2020107501} within \ac{TBD} strategies are also implemented as particle filters.

\textit{Dynamic programming based \ac{TBD}}\cite{yan2017track,wei2010tbd,yi13:efficient,aprile2016track}: DP-\ac{TBD} is a grid-based method that estimates target trajectories by searching all physically admissible paths in a discrete state space. Some grid-based \ac{TBD} techniques \cite{wang2020complex,jiang2016track,cao2020sequential} perform target detection via sliding time window and multi-frame tests.

\textit{Hough transform based \ac{TBD}} \cite{carlson94:search,moyer2011multi,yan2019detection}: The points on a straight line collapse into a single point in the transformed domain. So Hough transform-based \ac{TBD} is effective in detecting the points of the trajectory on a straight-line.

\textit{Optimization based \ac{TBD}} \cite{yan2019grey,grossi2013heuristic,wang2018greedy}: Using optimization algorithms such as the Grey Wolf optimizer or greedy algorithms, it is possible to search the optimal association of points in the solution space efficiently.

In the above-mentioned methods (e.g., \cite{granstrom2012extended,granstrom2017extended,granstrom2015gamma,lundquist2013extended,Mohanty81:computer,arnold93:efficient,tonissen98:ML,hyoungjun99:optimization,buzzi05:TBD,blackman1999:modern,YAN2021107821,CHAI2020107501,leonard2019multi,ristic2021bernoulli,ubeda2017adaptive,YI2019303,errasti2014track,yan2017track,wei2010tbd,yi13:efficient,carlson94:search,moyer2011multi,yan2019detection,yan2019grey,grossi2013heuristic,wang2018greedy,kwon2019particle,salmond01:particle}) the input is typically represented by points featuring position information and a timestamp. Since, as previously remarked, in \ac{UWB} \acp{RSN} target echoes can be very weak (sometimes lower than waveform detection threshold) and the propagation environment can be harsh (especially in indoor), the tracking system must cope with frequent miss-detection events and clutter points even when state-of-the-art point generation techniques are employed. For this reason, direct application of the above-mentioned \ac{TBD} methods to the \ac{UWB} \acp{RSN} framework, even in conjunction with \ac{UWB}-tailored waveform detection and point generation techniques, such as the one proposed in \cite{chiani2018sensor}, based on in-sensor detection, range estimation and spatial relationship between targets and sensors, tends to be ineffective since the \ac{TBD} performance is degraded by imperfect point generation. 
Therefore, in contrast with current \ac{TBD} approaches, in this paper we design a new \ac{TBD} method which, rather than focusing directly on the target points, organizes points into tracklets (short segments, corresponding to a number of subsequent scan periods) and seeks to detect  target tracklets by processing waveforms incoming from multiple sensors through multiple scans. 
As such, it builds a deeper fusion framework which includes point generation plus track detection. 
Although the proposed approach does not strictly belong to any of the \ac{TBD} categories mentioned above, it shares some commonalities with DP-TBD since in some steps it applies a grid cell strategy. However, rather than building a large number of track candidate represented by grid cells, grid cells of multiple scans are jointly processed by image processing methods to obtain the most promising tracks directly.

In the proposed method, soft waveform samples from the several sensors are jointly processed by a voting algorithm to obtain an image of surveillance area.\footnote{A different approach consists of performing a pre-processing of the \ac{UWB} waveform by a \ac{CFAR} detector \cite{chiani2018sensor}. 
However, according to the \ac{TBD} approach, we avoid any such ``waveform detection'' step to exploit even weak target echoes.} 
Multiple images, constructed through subsequent scan periods, are then stacked and jointly processed by applying efficient image processing techniques on the obtained \ac{3D} structure. 
Although the proposed processing of stacked images, coherently with the \ac{TBD} philosophy, is mainly motivated by the need to enhance detection and tracking of weak targets, it turns out to be useful in reducing clutter and coping with extended targets. Trajectory detection is performed by first generating tracklets and then by associating them, an approach helpful in the presence of maneuvering targets.

The main innovative contributions of the \ac{RSN} implementation are summarized as follows:
\begin{enumerate}
    \item Design of an efficient target imaging method using multi-sensor \ac{UWB} waveforms;  
    development of a target region partition method to refine the target image.
    The two strategies together are effective in making each target represented by a region where it exists, which avoids extended target issues.
    \item Detection rate enhancement by jointly processing information from multiple sensors and multiple scans, rather than confirm a detection at single sensor or single scan level. Possible target tracks can be obtained by waveform directly, this strategy addresses issue of weak target detection.
\end{enumerate}
Although the proposed technique is not tailored to a specific type of target, we present results obtained with human targets, showing the potential of the proposed approach in intrusion detection applications.

The paper is organized as follows. 
A system overview is provided in Section~\ref{sec:System Overview and Notation}. 
The proposed processing is described in Section~\ref{sec:processing}, while numerical and experimental results are presented in Section~\ref{sec:experimental}. Computer simulation results are first presented in Section~\ref{sec:case_study0}. Then, two cases studies with actual \ac{UWB} waveforms, both focused on detection and tracking of human targets, are presented in Section~\ref{sec:case_study1} and  Section~\ref{sec:case_study2}. In particular, Section~\ref{sec:case_study1} addresses the case of a \ac{RSN} composed of monostatic \ac{UWB} sensors, while Section~\ref{sec:case_study2} the case of a multistatic \ac{UWB} \ac{RSN}. Conclusions are drawn in Section~\ref{sec:Conclusions}. A subset of the results presented in the paper appeared in its conference version \cite{yan2020:Radarconf}. With respect to \cite{yan2020:Radarconf}, the proposed processing techniques are here addressed in a more thorough way, providing all of the details. Moreover, the multistatic \ac{UWB} \ac{RSN} case has been added and a number of additional results are presented.

\section{System Overview and Notation}\label{sec:System Overview and Notation}
In this work, we consider a \ac{UWB} \ac{RSN} aimed at detecting, locating, and tracking moving objects in the surveillance area. The \ac{RSN} may be either monostatic (i.e., composed of $\numsens \geq 3$ \ac{UWB} sensors each configured as a monostatic radar), or multi-static (i.e., composed of one \ac{UWB} transmitter and $\numsens \geq 3$ \ac{UWB} sensors configured in a receive-only mode\footnote{The transmitter may also be configured as a monostatic radar, in which case it shall be included in the set of sensors.}). In both network configurations, all nodes are placed in known positions and are connected to a \ac{FC}.

\begin{figure}[tb!]
\begin{center}
{\includegraphics[width=0.48\columnwidth]{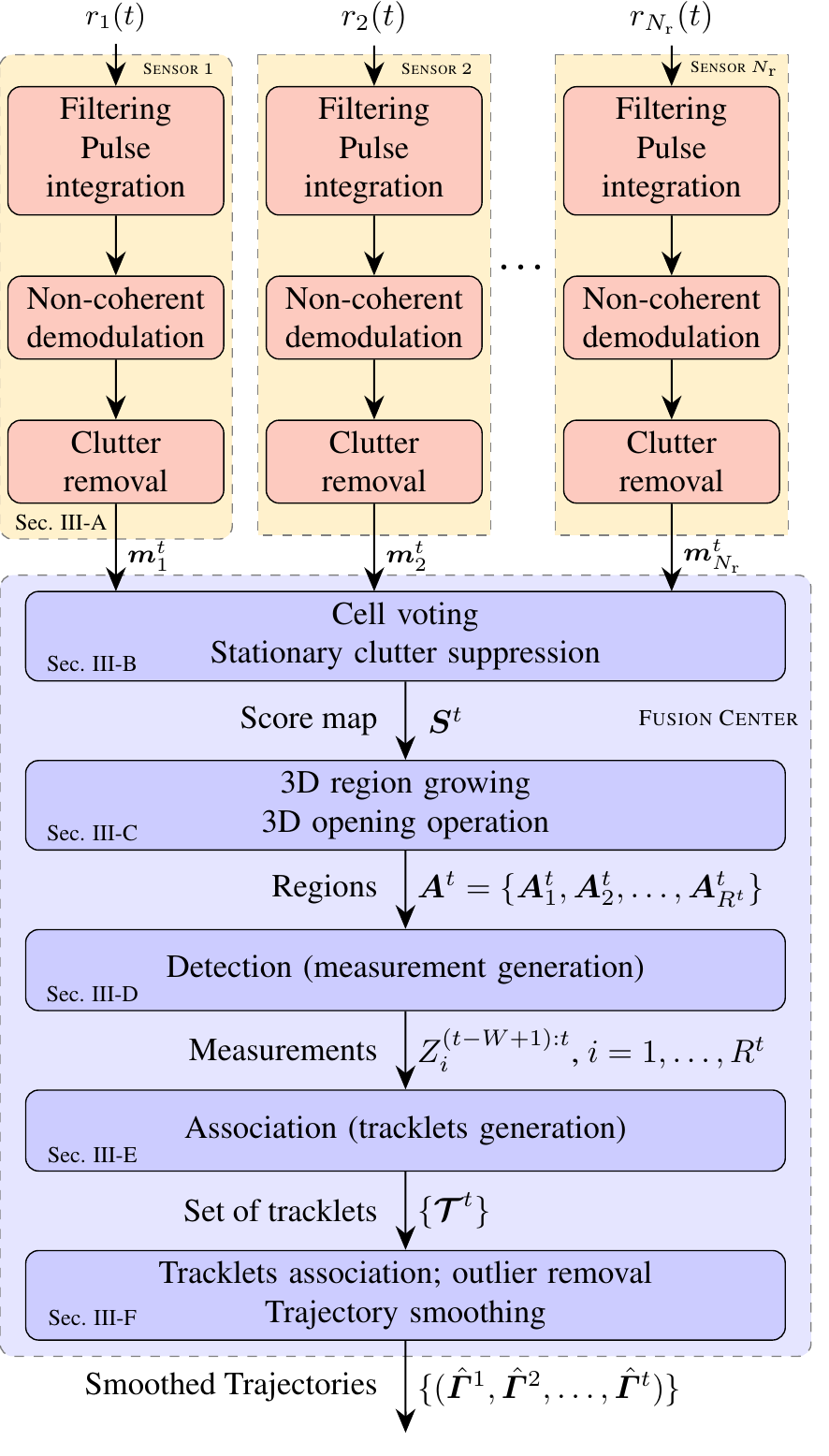}
}
\end{center}
\caption{Block diagram of the whole processing chain. }
\label{fig:blockdiag}
\end{figure}

Fig.~\ref{fig:blockdiag} shows a block diagram summarizing the whole processing performed by the \ac{RSN}. Each yellow box represents a single sensor, and the light red blocks in it are the processing steps performed by the sensor. The data preprocessed by sensors are transferred to a \ac{FC}, represented by the light blue block, performing tracking, detection, and localization according to a \ac{TBD} approach. For convenience, these blocks are labeled with the subsection in which the corresponding processing is described. In the following, we overview the main processing steps performed by the \ac{RSN}.

In every scan period $t$, each sensor performs filtering, sampling, pulse integration, non-coherent demodulation, and clutter removal; the resulting signals ${\boldsymbol{m}}_n^{t}$, $n=1,\dots,\numsens$, are delivered to the \ac{FC}.\footnote{Note that, with a slight abuse of notation, the time in the received signals $r_n(t)$ is also indicated with $t$ in Fig.~\ref{fig:blockdiag}.} 
Such signals are expected to contain the (possibly weak) target echoes plus a clutter residue due to imperfect waveform alignment, non-static clutter, and ghost artifacts \cite{chiani2018sensor}.

Within the \ac{FC}, the first stage consists of a voting process, yielding a \ac{2D} \emph{score map} of the surveillance area, $\boldsymbol{S}^t$, at scan period $t$. 
In the process, residual clutter is suppressed. %
The second stage introduces the temporal dimension in the processing. A \ac{3D} data structure $\boldsymbol{M}^t$, built up by stacking the \ac{2D} score maps, is processed in a sliding-window fashion with time window size $W$, representing the number of scans jointly handled by the \ac{TBD} algorithm. 
This processing is aimed at identifying sets of grid cells (dubbed \emph{regions}), resembling \ac{3D} trajectories that potentially correspond to targets. 
The image processing techniques named region growing and opening operation, both performed in a \ac{3D} setting, are applied to this purpose. 
The number of obtained regions is $R^t$, and the generic region is denoted by $\boldsymbol{A}_i^t$, $i=1,\dots,R^t$. 

The subsequent stage aims at extracting a set of \emph{measurements} (also referred to as \emph{points}) $Z_i^{(t-W+1):t}$ out of each region $\boldsymbol{A}_i^t$, where each measurement carries information about position, scan index, and score. %
Next, tracklet detection is performed, in which points with scan indexes belonging to a time window of size $W$ are associated to form \emph{tracklets} $\boldsymbol{\mathcal{T}}$, i.e., fragments of potential trajectories. The output of this stage is the set of detected tracklets. Tracklets are then associated with each other and combined to form trajectories. The generic trajectory point at scan period $t$ (i.e., a point considered to be a detection) is ${\bm{\varGamma}}^t=(x^t,y^t,t)$. 
More accurate trajectories can be obtained by outliers removal and application of a smoothing filter. The smoothed target trajectories represent the \ac{RSN} output. 
The point corresponding to ${\bm{\varGamma}}^t$ after outlier removal and trajectory smoothing is denoted by $\hat{\bm{\varGamma}}{}^t=(\hat{x}{}^t,\hat{y}{}^t,t)$.

The proposed processing framework deviates significantly from the one that is usually performed in \acp{RSN}. Conventional \ac{RSN} processing features a \ac{CFAR} detection step at single sensor level followed by a point detection step at \ac{FC} level. Target tracking methods (e.g., \cite{granstrom2012extended,granstrom2017extended,granstrom2015gamma,lundquist2013extended,Mohanty81:computer,arnold93:efficient,tonissen98:ML,hyoungjun99:optimization,buzzi05:TBD,blackman1999:modern,YAN2021107821,CHAI2020107501,leonard2019multi,ristic2021bernoulli,ubeda2017adaptive,YI2019303,yan2017track,wei2010tbd,yi13:efficient,carlson94:search,moyer2011multi,yan2019detection,yan2019grey,grossi2013heuristic,wang2018greedy,kwon2019particle,salmond01:particle}), fed with the detected points, can then be applied to generate the target tracks. In \ac{UWB} \acp{RSN}, however, this approach suffers from the fact that weak target echo components may be filter out throughout \ac{CFAR} detection in any sensor or in point generation at the \ac{FC}. 
Even though powerful tracking method are used to process the detected points, methods compliant with the traditional framework lead to shortage in terms of weak target detection, when no target points are generated. A comparison between the proposed approach and some more conventional methods will be described in Section~\ref{sec:case_study0}.

\section{Processing}\label{sec:processing}

%%%%%%%%%%%%%%%%%%%%%%%%%%%%%%%%%%%%%%%%%%%%%%%%%%%%%%%%%%%%%%
\subsection{In-Sensor Processing}\label{subsec:in_sensor_processing}
%%%%%%%%%%%%%%%%%%%%%%%%%%%%%%%%%%%%%%%%%%%%%%%%%%%%%%%%%%%%%%
At the $n$-th sensor, the received signal $r_n(t)$ is first processed by a bandpass filter to remove out-of-band noise spectral components and then sampled. 
Next, pulse integration is performed, where $\pulseintegfactor$ pulses received within a scan are coherently averaged, yielding a gain in terms of \ac{SNR}. 
The resulting vector in scan period $t$ is indicated by $\boldsymbol{r}_n^t$ and has length $\samplesperscan$, the number of samples per scan. 
Noncoherent demodulation, consisting of an envelope detector (implemented by a square-law device followed by a low-pass filter) is then applied to $\boldsymbol{r}_n^t$ yielding the signal ${\boldsymbol\epsilon}_n^t$ \cite{chiani2018sensor}. 

\begin{figure}[!t]
\centering
 \includegraphics[width=0.5\columnwidth]{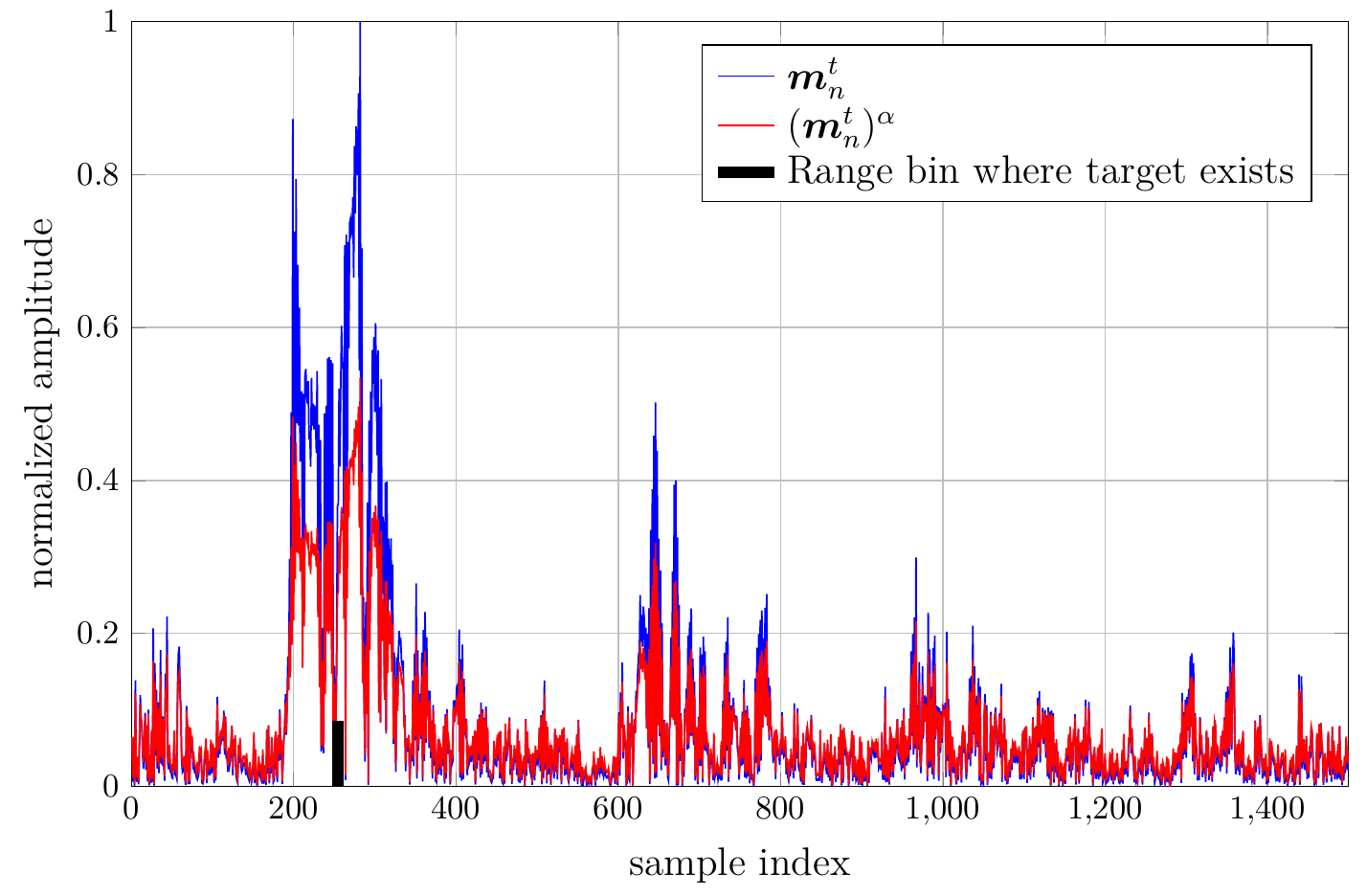}
\caption{Example of signal $\boldsymbol{m}_n^t$ along with the corresponding peak-suppressed samples  $(\boldsymbol{m}_n^t)^\alpha$ with $\alpha=0.75$.}
\label{fig:current_scan}
\end{figure}

The waveform ${\boldsymbol\epsilon}_n^t$ contains the echoes of both moving targets and non-static or static objects, together with noise. 
Echoes from static objects can be canceled out effectively by applying a clutter removal algorithm. 
In \ac{UWB} \acp{RSN}, the \ac{MTI} clutter removal technique turns out to be effective, especially when applied after noncoherent demodulation, owing to the possibility of performing very accurate \ac{ToA} estimation and therefore to accurately align the waveforms ${\boldsymbol\epsilon}_n^t$ at consecutive scans \cite{chiani2018sensor}. 
\ac{MTI} is implemented by calculating a reference waveform $\boldsymbol{\epsilon}_n^{t,\text{av}}$ by averaging the waveforms received within a time window, of $2\kappa$ scans, preceding the current scan, i.e.,
\begin{equation}\label{eq:er_energy}
 {\boldsymbol\epsilon}_n^{t,\text{av}}=\frac{1}{2\kappa} \sum_{p=1}^{2\kappa}{\boldsymbol\epsilon}_n^{t-\kappa-p}.
\end{equation}
Then, clutter removal is performed as ${\boldsymbol{m}}_n^{t} = \lVert {\boldsymbol\epsilon}_n^{t}-{\boldsymbol\epsilon}_n^{t,\mathrm{av}} \rVert$, where $\lVert\cdot\rVert$ is the  element-wise absolute value operator. The signal ${\boldsymbol{m}}_n^t$ is expected to contain target echoes along with noise and non-static clutter residue.
An example of real waveform ${\boldsymbol{m}}_n^{t}$ collected in the measurement campaign described in Section~\ref{sec:case_study1}, is presented in Fig.~\ref{fig:current_scan}.

%%%%%%%%%%%%%%%%%%%%%%%%%%%%%%%%%%%%%%%%%%%%%%%%%%%%%%%%%%%%%%%%%
\subsection{Voting for the Surveillance Area}\label{subsec:voting}
%%%%%%%%%%%%%%%%%%%%%%%%%%%%%%%%%%%%%%%%%%%%%%%%%%%%%%%%%%%%%%%%%
The surveillance area is divided into $N_x \times N_y$ grid cells, of size $\Delta_x \times \Delta_y$.
We use the notation $(i_x,i_y) \in \{1,\dots,N_x\} \times \{1,\dots,N_y\}$ to denote the pair of indexes identifying a cell in the score map. Moreover, we use $(x,y)$ to denote the actual position of a point in the map. The position of the center of cell $(i_x,i_y)$ is $(x_{\mathrm{c}}(i_x,i_y),y_{\mathrm{c}}(i_x,i_y))=((i_{x}-1/2)\Delta_x,(i_{y}-1/2)\Delta_y)$.

Processing at the \ac{FC} starts from the vectors ${\boldsymbol{m}}_n^t$, $n = 1,\dots,\numsens$, collected from the $\numsens$ sensors. 
The $j$-th element of ${\boldsymbol{m}}_n^t$ is denoted by $m_n^t(j)$, $j=1,\dots,\samplesperscan$. 
Each sample in ${\boldsymbol{m}}_n^t$ votes for a subset of the grid cells, as follows.
We denote by $\tau_n(j)$ the \ac{ToA} corresponding to sample $m_n^t(j)$ (a multiple of the sampling time $T_c$), defined as the delay with respect to the time of transmission. 
\begin{equation}\label{eq:tao}
   \tau_n(j)=\begin{cases}
        \tau_n^\text{m}(j),  & \text{monostatic \ac{RSN}}\\
        \tau_n^\text{b}(j),  & \text{multistatic \ac{RSN}.}
       \end{cases}
\end{equation}
The $\tau_n^\text{m}(j)$ and $\tau_n^\text{b}(j)$ means time delay in monostatic \ac{RSN} and multistatic \ac{RSN} respectively.

\indent In a monostatic \ac{RSN}, each sensor synchronizes on the time of transmission of the \ac{UWB} pulse. 
In this case, the sample ${m}_n^t({j})$ votes for all cells that are crossed by the circle of equation 
$$
\sqrt{(x - x_n)^2+(y - y_n)^2} = \frac{c\, \tau_n^\text{m}(j)}{2}
$$
where $(x_n,y_n)$ is the position of sensor $n$ and $c$ is the speed of light, i.e., the circle centered in the sensor and having radius $c\, \tau_n^\text{m}(j) / 2$. 

\indent In a multistatic \ac{RSN}, each sensor synchronizes on the \ac{ToA} of the \ac{UWB} pulse received along the direct path from the transmitter to the sensor; $\tau_n^\text{b}(j)$  means the excess delay with respect to the \ac{ToA} of the direct pulse.
In this case, denoting by $(x_{\mathrm{T}},y_{\mathrm{T}})$ the position of the transmitter and by $L_n$ the distance between the transmitter and sensor $n$, the sample $m_n^t(j)$ votes for all cells that are crossed by an ellipse with foci in $(x_n,y_n)$ and $(x_{\mathrm{T}},y_{\mathrm{T}})$ and major axis $L_n + c\, \tau_n^\text{b}(j)$, namely, the ellipse of equation 
$$
\sqrt{(x-x_{\mathrm{T}})^2+(y-y_{\mathrm{T}})^2} + 
  \sqrt{(x-x_n)^2+(y-y_n)^2} 
  = L_n+c\,\tau_n^\text{b}(j) \, .
$$
The voting algorithm starts by defining the amplitude of the circle or ellipse associated with $m_n^t(j)$ as 
\begin{equation}\label{eq:vote}
{V}_n^t({j})=\frac{({m}_n^t(j))^\alpha}{\frac{1}{\samplesperscan}\sum_{l=1}^{\samplesperscan} (m_n^t({l}))^\alpha}
\end{equation}
where $0<\alpha<1$. A total of $\samplesperscan$ circle or ellipse amplitudes are thus obtained from each vector ${\boldsymbol{m}}_n^t$. 
The score (or \emph{vote}) received by each of the $N_x \times N_y$ grid cells is defined as
\begin{equation}\label{eq:score_map_sensor}
  S_n^t(i_x,i_y) = \max_{j \in \{ 1,\dots,\samplesperscan\}} \{V_n^t(j)I_n(j,i_x,i_y)\}
\end{equation}
where the indicator function $I_n(j,i_x,i_y)$ equals $1$ if the \mbox{$j$-th} circle or ellipse passes through the cell $(i_x,i_y)$ and $0$ otherwise.
Depending on the sampling time $T_c$ and on the cell dimension, one grid cell may be voted by several circles or ellipses; according to \eqref{eq:score_map_sensor}, the cell score is the largest received vote. At the end of this voting process, we obtain $\boldsymbol{S}_n^t=\{S_n^t(i_x,i_y)\}_ {1\leq i_x\leq N_x,1\leq i_y\leq N_y}$ as the score map associated with $\boldsymbol{m}_n^t$.
The exponent $\alpha$ applied in \eqref{eq:vote} is used to mitigate strong clutter residues, avoiding excessively large cell scores from clutter.

The score maps constructed from all vectors $\boldsymbol{m}_n^t$, $n = 1, \dots, \numsens$, are first multiplied to obtain an overall score map $\bm{\Sigma}^t$, namely, 
\begin{equation}\label{eq:overall_score_map}
\bm{\Sigma}^t = \bigg\{\Sigma^t(i_x,i_y) = \prod_{n=1}^{\numsens} S_n^t(i_x,i_y)\bigg\}_{1\leq i_x\leq N_x,1\leq i_y\leq N_y}.
\end{equation}
Then, score values that are below a threshold are forced to zero, yielding the final score map at scan $t$, $\bm{S}^t = \left\{ S^t(i_x,i_y) \right\}_{1\leq i_x\leq N_x,1\leq i_y\leq N_y}$ where
\begin{equation}\label{eq:M_to_S}
   S^t(i_x,i_y)=\begin{cases}
        \Sigma^t(i_x,i_y),  & \text{if} \,\,\,\Sigma^t(i_x,i_y) > \eta^t_{\mathrm{score}}\\
        0,  & \text{otherwise.}
       \end{cases}
\end{equation}
The score threshold in \eqref{eq:M_to_S} is adaptive and proportional to the average score map. Specifically, we have
\begin{align}\label{eq:etat_threshold}
    \eta^t_{\mathrm{score}} = \frac{\beta}{N_x N_y} \sum_{{(i_x,i_y)}} \Sigma^t(i_x,i_y)
\end{align}
for some positive $\beta$. As a main feature of the proposed approach, the threshold $\eta^t_{\mathrm{score}}$ may be kept relatively low (compared with track-after-detect approaches), which is very helpful in detecting weak targets. The high number of generated false alarms are then dealt with in the subsequent stages of the processing, in which the temporal dimension is introduced. In some scenarios (especially some indoor ones) a clutter removal step may also be applied to the score maps generated by \eqref{eq:M_to_S}. The clutter removal procedure is described in Appendix~\ref{app:clutter}.
%%%%%%%%%%%%%%%%%%%%%%%%%%%%%%%%%%%%%%%%%%%%%%%%%%%%%%%%%%%%%%%%%%%%%%%%
\subsection{3D Region Growing and 3D Opening Operation}\label{subsec:3d}
%%%%%%%%%%%%%%%%%%%%%%%%%%%%%%%%%%%%%%%%%%%%%%%%%%%%%%%%%%%%%%%%%%%%%%%%
The scan maps $\boldsymbol{S}^t$ generated through the voting process are stacked one over another, yielding a \ac{3D} data structure $\boldsymbol{M}^t = [M^t(i_x,i_y,k)]$, $1 \leq i_x \leq N_x$, $1 \leq i_y \leq N_y$, $1 \leq k \leq w$, where $M^t(i_x,i_y,k)=S^k(i_x,i_y)$. 
Hereafter, we denote by $\boldsymbol{M}^{k_1:k_2}$ the \ac{3D} structure composed of all layers of $\boldsymbol{M}^{t}$ between scan period $k_1$ and scan period $k_2$, with $k_1$ and $k_2$ included. 
We refer to each element of $\boldsymbol{M}^t$ as a \emph{cell} and to $M^t(i_x,i_y,k)$ as the score of cell $(i_x,i_y,k)$. 
Moreover, we say that two cells $(i_x,i_y,k)$ and $(j_x,j_y,h)$ are neighboring cells when all of the following conditions are true: $\lVert i_x-j_x\rVert \leq 1$, $\lVert i_y-j_y \rVert \leq 1$, $ \lVert k-h\rVert \leq 1$. As such, every cell has at most $26$ neighbors.

The data structure $\boldsymbol{M}^t$ is processed in a sliding window fashion with time window size $W$. 
In particular, the top (i.e., most recent) $W$ layers of $\boldsymbol{M}^t$, $\boldsymbol{M}^{(t-W+1):t}$, are processed by applying two operations borrowed from digital image processing, namely, region growing and opening operation \cite{wan2003symmetric}. 
These two operations are applied on $\boldsymbol{M}^{(t-W+1):t}$ every $s$ scan periods, for some $1 \leq s \leq W$, i.e., at scan periods $t = W + q \cdot s$, $q \in \mathbb{N}$. 
For example, if $W=4$ and $s=2$ then region growing and open operation are applied on $\boldsymbol{M}^{1:4}$, $\boldsymbol{M}^{3:6}$ $\boldsymbol{M}^{5:8}$, and so on.\footnote{If $s=W$ then every layer of $\boldsymbol{M}^t$ is processed once. If instead $1 \leq s < W$ then the time windows are partially overlapped and every layer of $\boldsymbol{M}^t$ is processed multiple times.}
Region growing and opening operation are described in the following. More details (e.g., the pseudocode of \ac{3D} region growing) are provided in Appendix~\ref{app:3DRG}.

Region growing consists of grouping cells to form cell clusters called \emph{regions}.
Starting from a set of ``seed'' cells, regions are progressively grown by including in each of them new neighboring pixels that meet certain criteria. 
Region growing is performed in a \ac{3D} fashion, as described hereafter.

In the beginning, a number of cells in $\boldsymbol{M}^{(t-W+1):t}$ are declared as seed cells.
Seed cells are evenly selected and their set is $\mathcal{C}_{\mathrm{seed}}^{(t-W+1):t} = \{(a d_x, b d_y, t-W + c d_t)\}$, where $d_x$, $d_y$, and $d_t$ are given positive integers and $a=1,\dots, \lfloor N_x / d_x \rfloor$, $b=1,\dots,\lfloor N_y / d_y \rfloor$, $c=1, \dots, \lfloor W / d_t \rfloor$.
At the beginning, all seed cells are declared to be active. 
Seed cells are then processed in order.\footnote{The order in which seed cells are processed is irrelevant, as the final set of generated regions is the same irrespective of the seed cell ordering.} 
If a seed cell is active and its score is nonzero, a new region $\boldsymbol{\mathcal{R}}$ is initialized, having the seed cell as its sole elements.
All neighbors of the seed cell are then considered and any of them having a nonzero score is included in $\boldsymbol{\mathcal{R}}$.
Next, all newly included cells are processed. The neighbors of each of them are considered and, if any such neighbor is not yet in $\boldsymbol{\mathcal{R}}$ and has a nonzero score, it is included in $\boldsymbol{\mathcal{R}}$. This processing is iterated until no new cell can be included in the region. Importantly, if during some iteration the region grown from a seed cell includes another seed cell, this latter cell is declared to be inactive. This way, no new region will be initialized for it.

Not all of the built regions are further processed. In order for a region to become a confirmed one, the two conditions
\begin{align}
  \sum_{(i_x,i_y,k) \in \boldsymbol{\mathcal{R}}} M^t(i_x,i_y,k) \geq \gamma^t_{\rm{score}} \label{eq:region_score_condition} \\
  |\boldsymbol{\mathcal{R}}| \geq \gamma^t_{\rm{num}} \label{eq:region_card_condition}
\end{align}
must be simultaneously fulfilled, where $|\cdot|$ denotes cardinality.
This means that, to regard a region as potentially corresponding to a target, we require that it possesses a minimum score and a minimum cardinality.
For all regions not fulfilling both \eqref{eq:region_score_condition} and \eqref{eq:region_card_condition}, the score of all cells is forced to zero ($M^t(i_x,i_y,k)=0$). 
Moreover, the score of any cell not included in any of the obtained region is also forced to zero, which is effective in removing isolated cells due to noise and clutter.
Investigating the optimum values of the two threshold $\gamma^t_{\mathrm{score}}$ and $\gamma^t_{\mathrm{num}}$ analytically turns to be a hard task, as they depend on several factors that include the target size, radar cross-section, radar parameters, measurement noise, and grid cell area. 
In practice, suitable values of these thresholds can be found experimentally.
The \ac{3D} region growing procedure is formalized in Appendix~\ref{app:3DRG}.

At the end of \ac{3D} region growing, we obtain a set $\{\boldsymbol{\mathcal{R}}_1^t, \boldsymbol{\mathcal{R}}_2^t, \dots, \boldsymbol{\mathcal{R}}_{R^t}^t\}$ of $R^t$ regions.
Note that $R^t$ may be considered as a preliminary estimate of the number of targets at scan $t$, although no detection is performed at this stage. 
The $W$ \ac{2D} layers of any region $\boldsymbol{\mathcal{R}}_i^t$ are usually irregular, especially due to the effect of clutter. Hence, a \ac{3D} opening operation \cite{maintz20013d,grinzato1998corrosion,gonzalez2008digital} is performed to smooth the contour of $\boldsymbol{\mathcal{R}}_i^t$, break narrow isthmuses, and eliminate thin protrusions. Opening operation is effective in improving the localization accuracy of targets corresponding to regions for which a detection is declared at the end of the whole processing, as well as in reducing false alarms.
 
The opening of region $\boldsymbol{\mathcal{R}}_i^t$ by a structuring element (or kernel) $\bm{B}$ returns a new region $\bm{A}_i^t$ defined as
\begin{equation}\label{eq:opening}
    \bm{A}_i^t = (\bm{\mathcal{R}}_i^t \ominus \bm{B}) \oplus \bm{B}
\end{equation}
where $\bm{\mathcal{R}}_i^t \ominus \bm{B}$ denotes erosion of $\bm{\mathcal{R}}_i^t$ by the structuring element $\bm{B}$ and $(\bm{\mathcal{R}}_i^t \ominus \bm{B}) \oplus \bm{B}$ denotes dilation of $(\bm{\mathcal{R}}_i^t \ominus \bm{B})$, again by $\bm{B}$. The erosion and dilation operations (in a \ac{3D} setting) are defined as follows. 

Let $\bm{F}$ be a \ac{3D} region and $\bm{B}$ be a \ac{3D} structuring element, one cell of which is regarded as the origin of $\bm{B}$. Then, $\bm{F} \ominus \bm{B}$ is the set of all cells $(i_x,i_y,k)$ such that the structural element with origin in $(i_x,i_x,k)$ is included in $\bm{F}$. Formally:
\begin{equation}\label{eq:18}
\bm{F} \ominus \bm{B} = \{ (i_x,i_y,k) | \bm{B}_{i_x,i_y,k} \subseteq \bm{F} \} 
\end{equation}
where $\bm{B}_{i_x,i_y,k}$ is the structural element with origin in $(i_x,i_y,k)$.  Moreover, $\bm{F} \oplus \bm{B}$ is the set of all cells $(i_x,i_y,k)$ such that the structural element with origin in $(i_x,i_x,k)$ has a nonzero intersection with $\bm{F}$. Formally:
\begin{equation}\label{eq:218}
\bm{F} \oplus \bm{B} = \{ (i_x,i_y,k) | \bm{B}_{i_x,i_y,k} \cap \bm{F} \ne \emptyset \} .
\end{equation}
The employed \ac{3D} structuring element is composed of $7$ cells, with a central cell playing the role of the origin and the $6$ cells adjacent to its $6$ facets (regarding a \ac{3D} cells as a parallelepiped). After application of opening operation to each region $\bm{\mathcal{R}}_i^t$, we obtain a new region set $\{{\boldsymbol{{A}}}_1^t, {\boldsymbol{{A}}}_2^t, \dots, {\boldsymbol{{A}}}^t_{{R^t}}\}$ that is further processed for generation of points.

\begin{figure*}[tbp]
\centering
\subfigure[]{
\includegraphics[width=0.31\columnwidth]{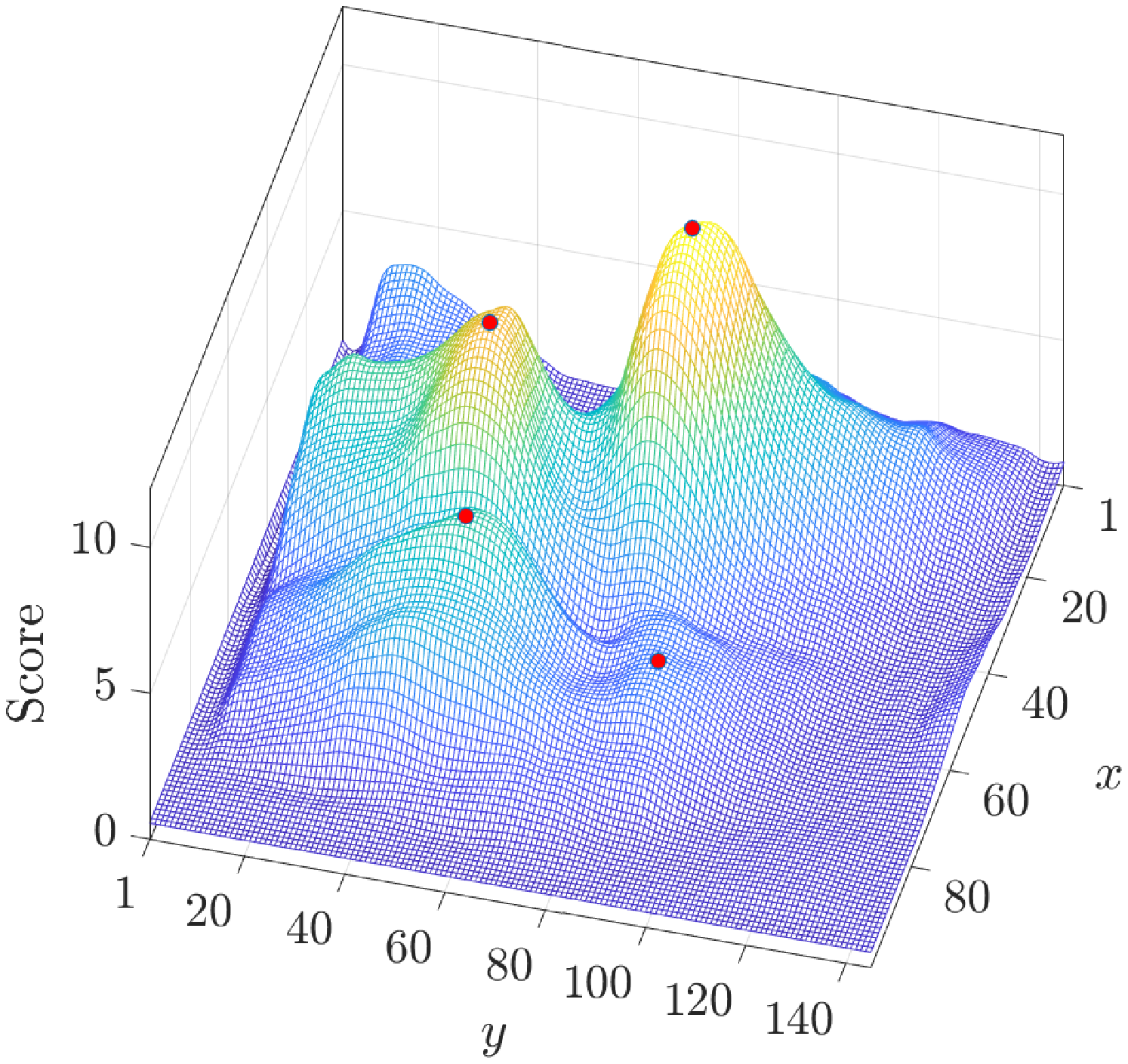}
\label{fig:smoothed_sm_a}}
\subfigure[ ]{
\includegraphics[width=0.31\columnwidth]{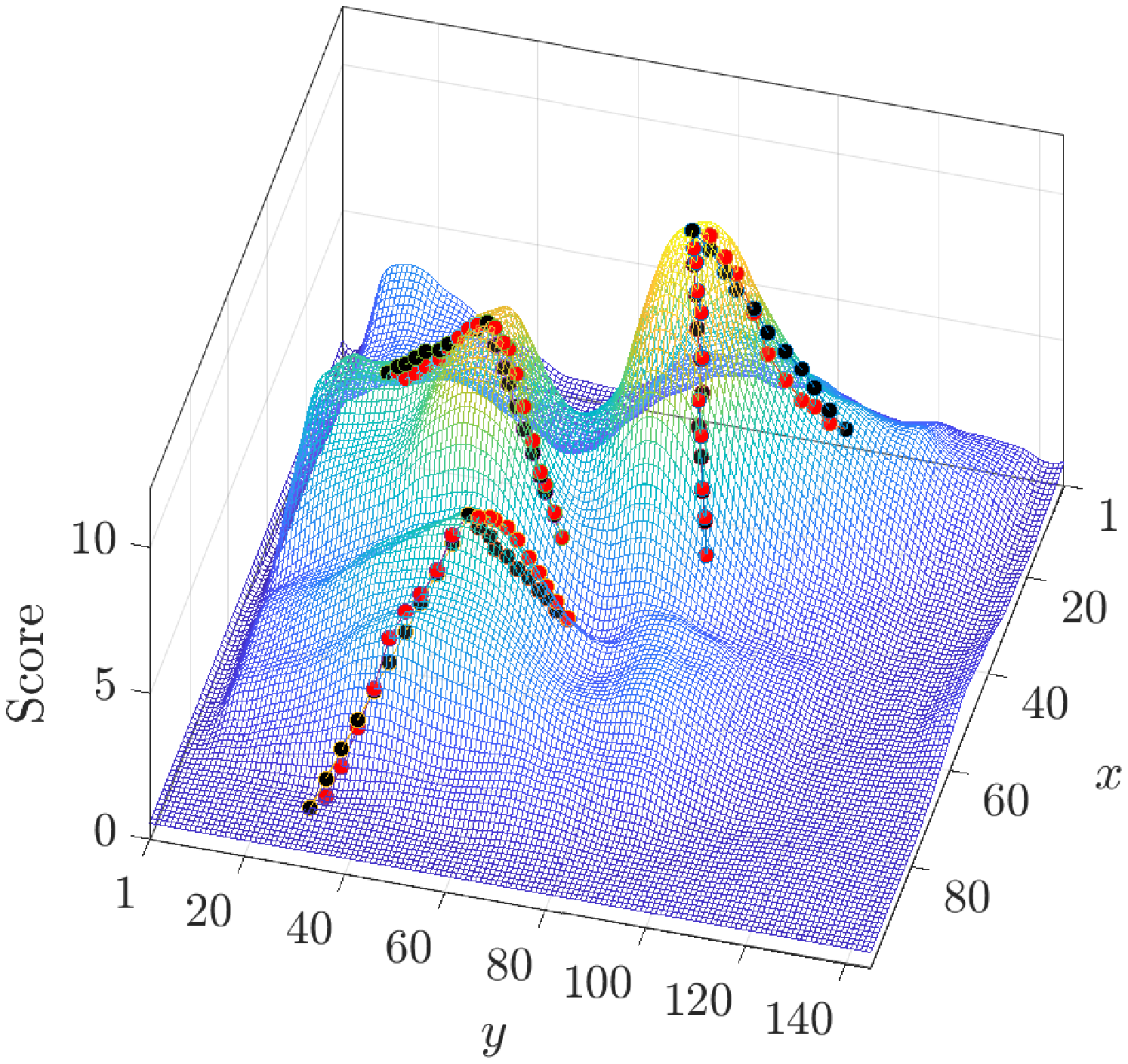}
\label{fig:smoothed_sm_b}}
\subfigure[ ]{
\includegraphics[width=0.31\columnwidth]{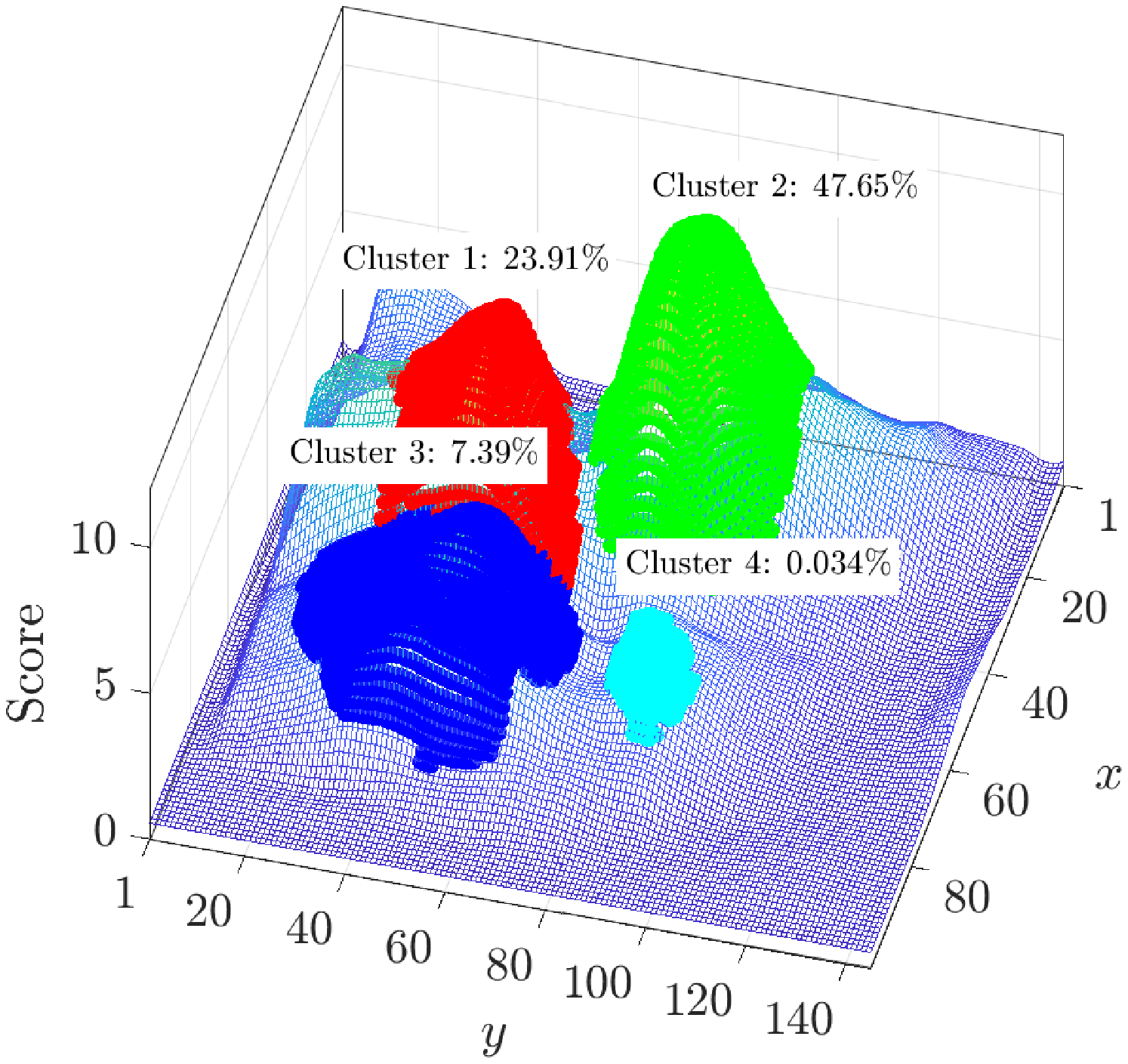}
\label{fig:smoothed_sm_c}}
\caption{(a) Example of surface $\mathcal{S}'$ and centers of local maximum cells. (b) Example of application of the criterion to defined clusters ($N=10$). Black: points $Q_h$. Red: points $\tilde{Q}_h$. (c) Portions of $\mathcal{S}'$ corresponding to the four clusters.}
\label{fig:part_method}
\end{figure*}

\subsection{Points Generation}\label{subsec:detection}
We aim at generating a set of points, also referred to as \emph{measurements}, out of the $R^t$ regions constructed via region growing and opening operation. 
The starting point is a \ac{3D} data structure whose $k$-th layer, $k=t-W+1,\dots,t$, is a \ac{2D} score map denoted by $\hat{m}^k(i_x,i_y)$. 
Each \ac{2D} score map $\hat{m}^k(i_x,i_y)$ is processed separately. Note that
this score map is contributed by layer $k$ of all \ac{3D} regions $\boldsymbol{A}_i^t$ identified in the previous step. 

A \ac{2D} image smoothing operation is first performed on $\hat{m}^k(i_x,i_y)$, yielding a smoothed score map $m^k(i_x,i_y)$. Specifically, we perform
\begin{equation}\label{eq:smooth}
m^k(i_x,i_y) = \frac{1}{(2w+1)^2}\sum_{p=-w}^{w}\sum_{q=-w}^{w} \hat{m}^k(i_x+p,i_y+q)
\end{equation}
where $w$ is the width of the smoothing window. Next, we search for cells being local maxima in $m^k(i_x,i_y)$. A cell is said to be a local maximum when its score is larger than the score of all of its $8$ nearest neighbors in the \ac{2D} layer.
 
Once all local maxima in $m^k(i_x,i_y)$ have been identified, we grow a \ac{2D} region (hereafter referred to as a \emph{cluster} associated with $(i_x,i_y)$) around each of them, according to the following strategy. 
Let $(i_{x,1},i_{y,1})$ and $(i_{x,2},i_{y,2})$ be a local maximum cell and a cell under test, respectively. 
Let $(x_{\mathrm{c},1},y_{\mathrm{c},1})$ and $(x_{\mathrm{c},2},y_{\mathrm{c},2})$ be the positions of the centers of cells $(i_{x,1},i_{y,1})$ and $(i_{x,2},i_{y,2})$, respectively. 
Moreover, let $m^k_1=m^k(i_{x,1},i_{y,1})$ and $m^k_2=m^k(i_{x,2},i_{y,2})$.  

Consider the two points $P_1=(x_{\mathrm{c},1},y_{\mathrm{c},1},m^k_1)$ and $P_2=(x_{\mathrm{c},2},y_{\mathrm{c},2},m^k_2)$, draw the \ac{3D} segment having $P_1$ and $P_2$ as endpoints, and find $N-1$ equally-spaced points on this segment. 
Letting the $h$-th such point be $Q_h=(\chi_h,\psi_h,c^k_h)$, $h=1,\dots,N-1$, we have 
\begin{equation}\label{eq:segment_points}
(\chi_h,\psi_h,c^k_h)=\frac{N-h}{N}(x_{\mathrm{c},1},y_{\mathrm{c},1},m^k_1)+\frac{h}{N}(x_{\mathrm{c},2},y_{\mathrm{c},2},m^k_2).
\end{equation}

The smoothed score map may be represented as a \ac{3D} surface $\mathcal{S}$.
Note that the $z$-axis value of points in $\mathcal{S}$ is cell-wise constant.
A second surface $\mathcal{S}'$ can be obtained from $\mathcal{S}$ through a \ac{2D} interpolation operation. 
Each point in $\mathcal{S}'$ is in the form $(x,y,f^k(x,y))$, where $f^k(x,y)$ is constructed as follows. Let $(a_1,b_1)$, $(a_1,b_2)$, $(a_2,b_1)$, and $(a_2,b_2)$ be the coordinates of the centers of the four cells closest to $(x,y)$, where $a_1<x<a_2$ and $b_1<y<b_2$, and let $m^k_{a_1,b_1}$, $m^k_{a_1,b_2}$, $m^k_{a_2,b_1}$, and $m^k_{a_2,b_2}$ be the scores of the four corresponding cells. Then 
\begin{align}
  f^k(x,y) &= 
\frac{a_2-x}{a_2-a_1}\frac{b_2-y}{b_2-b_1}m^k_{a_1,b_1} + \frac{a_2-x}{a_2-a_1}\frac{y-b_1}{b_2-b_1}m^k_{a_1,b_2}\notag \\ 
&+ \frac{x-a_1}{a_2-a_1}\frac{b_2-y}{b_2-b_1}m^k_{a_2,b_1}
  +\frac{x-a_1}{a_2-a_1}\frac{y-b_1}{b_2-b_1}m^k_{a_2,b_2} .
\end{align}
Note that, when $(x,y)$ coincides with any of the four centers, $f^k(x,y)$ is equal to the score of the corresponding cell.

For each of the above-defined points $Q_h$, we consider a second point $\tilde{Q}_h = (\chi_h, \psi_h, f^k_h)$ where $f^k_h=f^k(\chi_h, \psi_h)$. We then say that the cell under test $(i_{x,2},i_{y,2})$ belongs to the subset associated with $(i_{x,1},i_{y,1})$ if and only if 
\begin{align}\label{eq:22}
f^k_h \geq c^k_h \qquad \forall\,\, h=1,\dots,N-1\,.
\end{align}
Operatively, for each local maximum cell, we start by testing its $8$ neighbors, including the neighbor in the cluster if \eqref{eq:22} is fulfilled. We then proceed by testing each neighbor of newly included cells (if not already tested), and we go ahead until no further cells can be included in the cluster.

\emph{Discussion}: Condition~\eqref{eq:22} imposes that all of the $N-1$ points $Q_h$, lying on the segment connecting the centers of a local maximum cell and a cell under test, falls ``below'' the surface $\mathcal{S}'$. 
The net result is that the portion of $\mathcal{S}'$ corresponding to a cluster has an approximately convex shape. 
This is exemplified in Fig.~\ref{fig:part_method}, obtained with real measurements. 
The surface $\mathcal{S}'$ is first depicted\footnote{The surface $\mathcal{S}'$ is here shown only for illustration purposes. In the proposed method, only $N-1$ $z$-axis values of $\mathcal{S}'$ shall be computed for each cell under test.} along with markers corresponding to the centers of the four local maximum cells; examples of points $Q_h=(\chi_h,\psi_h,c^k_h)$ (black) and $\tilde{Q}_h = (\chi_h, \psi_h, f^k_h)$ (red) are then shown, for three local maximum cells and two test cells per maximum; finally, the portions of $\mathcal{S}'$ corresponding to the four clusters are highlighted (in red, green, blue, and cyan). 
The proposed approach to consider local maxima and growing clusters around them according to \eqref{eq:22} is motivated by practical considerations. 
In our experiments, the image of a human target at this stage of the processing is typically a group of contiguous cells with a pronounced maximum in terms of score, a pronounced difference between the score of inner cells and the score of cells on the border (this is in part due to the smoothing operation \eqref{eq:smooth}), and an approximately convex shape of the portion of the surface obtained via interpolation. 
Thus, the described processing aims at identifying groups of cells exhibiting these features.

The cluster grown around each local maximum cell is intrinsically associated with (layer $k$ of) one of the regions $\boldsymbol{A}_i^t$. 
In fact, since at the beginning of the points generation step the score of every cell not belonging to any region $\boldsymbol{A}_i^t$ is equal to zero, every local maximum belongs to (layer $k$ of) some region. 
To make this connection explicit, we denote the identified clusters at layer $k$ by $U_{i,j}^k$, where $i$ is the index of the region and $j$ is the index of the cluster associated with the region. 
Exactly one point is extracted from each cluster $U_{i,j}^k$; moreover, only points characterized by a large enough \emph{score} become useful for tracklet generation, as follows. The mean score and score variance in $U_{i,j}^k$ are defined as 
\begin{align}
\bar{m}_{i,j}^k = \frac{1}{|U_{i,j}^k|}\sum_{(i_x,i_y) \in U_{i,j}^k} m_{i,j}^k(i_x,i_y) \qquad \mathrm{and} \qquad
v_{i,j}^k = \frac{1}{|U_{i,j}^k|}\sum_{(i_x,i_y) \in U_{i,j}^k}(m(i_x,i_y)-\bar{m}_{i,j}^k)^2
\end{align}
respectively. 
Clusters with large mean score, score variance, and cardinality are more likely to be generated by a target. (A large variance arises from the above-mentioned difference between the score of inner cells and the score of cells on the border of a cluster.)
Hence, the point corresponding to $U_{i,j}^k$ is constructed as $\boldsymbol{z}_{i,j}^k = (x_{i,j}^k,y_{i,j}^k,k,p_{i,j}^k)$, where 
\begingroup
\allowdisplaybreaks
\begin{equation}\label{eq:measurement}
\begin{aligned}
x_{i,j}^k = \frac{\sum_{(i_x,i_y) \in U_{i,j}^k} x_{\mathrm{c}}(i_x,i_y) \cdot m(i_x,i_y)}{\sum_{(i_x,i_y) \in U_{i,j}^k} m(i_x,i_y)} & \qquad \qquad
y_{i,j}^k = \frac{\sum_{(i_x,i_y) \in U_{i,j}^k} y_{\mathrm{c}}(i_x,i_y)\cdot m(i_x,i_y)}{\sum_{(i_x,i_y) \in U_{i,j}^k} m(i_x,i_y)} \\
p_{i,j}^k =&\,\, \bar{m}_{i,j}^k \cdot v_{i,j}^k \cdot |U_{i,j}^k|\,.
\end{aligned}
\end{equation}
\endgroup
Note that $(x_{i,j}^k,y_{i,j}^k)$ is the cluster centroid, while $p_{i,j}^k$ assumes the meaning of score of $\boldsymbol{z}_{i,j}^k$. 
Measurements whose score is below a threshold are removed.

At the end of the process, a collection of points is generated for each region $\boldsymbol{A}_i^t$, $i=1,\dots,R^t$. This is expressed as $\mathcal{Z}_i^{(t-W+1):t} = \{ \mathcal{Z}_i^{t-W+1}, \mathcal{Z}_i^{t-W+2}, \dots, \mathcal{Z}_i^t \}$ where, for all $k \in \{t-W+1, \dots, t\}$, we have denoted by $\mathcal{Z}_i^k = \{\boldsymbol{z}_{i,j}^k | j = 1, \dots, N_i^k \}$ the subset of points corresponding to layer $k$ of $\boldsymbol{A}_i^t$. The final set of points is
\begin{align}\label{eq:window_points}
\mathcal{Z}^{(t-W+1):t} = \{ \mathcal{Z}^{t-W+1}, \mathcal{Z}^{t-W+2}, \dots, \mathcal{Z}^t \} 
\end{align}
where $\mathcal{Z}^k = \bigcup_{i=1}^{R^t} \mathcal{Z}_i^k$. 

\subsection{Tracklets Generation}\label{subsec:tracklet_gen}

Points generated by targets, collected in subsequent scans, tend to concentrate around tracklets, i.e., segments in the \ac{3D} space\footnote{The \ac{3D} space dimensions are represented by the $x$ coordinate, $y$ coordinate, and time.} whose directions correspond to the kinematic states of the targets. For this reason, successive points within a time window can be associated to obtain tracklets. 

We exploit partially overlapping time windows, each containing points of $W$ scans; the points of the most recent window are given by \eqref{eq:window_points}. Tracklets are detected by the points in each time window independently.
In this work, we employed the tracklet generation technique proposed in \cite{yan2019detection}, adapted to a human target case.
In particular, since the direction of a human target can change in one or two scans a narrower time window, i.e., a small enough $W$, should be used in our setting. 
Note that several tracklets may be generated in a time window due to false alarm points.

The $l$-th tracklet at scan $t$ is denoted by $\bm{\mathcal{T}}_l^t$ and is a collection of $W$ points belonging to $\mathcal{Z}^{t-W+1}$, $\mathcal{Z}^{t-W+2}$, $\dots$, and $\mathcal{Z}^t$, respectively.
To keep the computational complexity under control, only tracklets with a large enough score are forwarded to the next step, while the others are discarded. To further limit complexity, depending on the application it is also possible to set a maximum number of detected tracklets per time window. 
The tracklet score is defined considering both the score of the corresponding points and an error measure for each point. 
More specifically, let the $W$ points $\boldsymbol{z}^{t-W+1}, \dots, \boldsymbol{z}^t$ be considered to form a possible tracklet. 
Then, the tracklet score is $(\prod_{k=t-W+1}^t p^k) / (\prod_{k=t-W+1}^t d^k)$ where $p^k$ is the measurement score defined in the third equation of \eqref{eq:measurement} and $d^k$ is an error metric for $\boldsymbol{z}^k=(x^k,y^k,k,p^k)$. 
This is calculated as the Euclidean distance between $(x^k,y^k)$ and $(\Dot{x}{}^k,\Dot{y}{}^k)$, where $(\Dot{x}{}^k,\Dot{y}{}^k,k)$ is obtained by linear fitting of the $W-1$ points $(x^i,y^i,i)$, $i \neq k$, with a least-squares criterion.

\subsection{Tracklets Association, Outlier Removal and Trajectory Smoothing}\label{subsec:tracklets_assoc}

The tracklets originated from the same target should be associated (i.e., joined) with each other to obtain the whole target trajectory \cite{chong2000architectures}. 
In this work we have employed the \ac{MHT} tracklet association algorithm proposed in \cite{yan2019detection}, with minor adjustments.
Since adjacent windows are partially overlapping, tracklets including the same points are regarded as more likely to be originated from the same target. The algorithm also attempts to identify and remove false alarm tracklets to possibly let each target be represented by a unique point in every scan.

After tracklet association, at the generic scan period $t$ we obtain a set of active trajectories $\{\bGamma\}^t$, whose time-variant cardinality $N^t = | \{ \bGamma \}^t |$ is the number of detected targets at scan $t$. 
Each such detected target, at some scan $k \leq t$, is represented by a point belonging to one of these trajectories, along with its score and its error metric. Denoting by $\boldsymbol{\xi}^k(\bGamma)$ the point of trajectory $\bGamma$ at scan $k$, the structure of this point is $\boldsymbol{\xi}^k(\bGamma) = (x^k,y^k,k;p^k,d^k)$. 

Position errors along the trajectory may still be too large, due to outliers and measurement noise. Therefore, outliers removal and trajectory smoothing are performed to obtain the final smoothed trajectory. 
Outlier removal is based again on the metric $d_k$. Point $\boldsymbol{\xi}^k(\bGamma)$ is regarded as an outlier whenever the error metrics of points along trajectory $\bGamma$ fulfill 
\begin{align}\label{eq:outlier_removal}
    d^k \geq \frac{\nu}{2 D} \left( \sum_{n=-D}^{-1} d^{k+n} + \sum_{n=1}^{D} d^{k+n} \right)
\end{align}
where $D = \lfloor W/2 \rfloor$ and $\nu > 0$ is a system design parameter. 
The above search for outliers is performed in a sliding window fashion, with a window size of $W+1$ scans (apart from the initial and final scan periods). Then the point $\boldsymbol{\xi}^k(\bGamma)$ is replaced by the smoothed point $\hat{\boldsymbol{\xi}}^k(\bGamma) = (\hat{x}^k, \hat{y}^k,k ; p^k, \hat{d}^k)$, obtained by linear fitting the other $W$ points in the time window, with a least-squares criterion.

\section{Experimental Results}\label{sec:experimental}

An experimental campaign was performed to assess the effectiveness of the proposed processing chain. 
The performance analysis was carried out both by computer simulations (synthetic data) and by employing actual \ac{UWB} waveforms collected in real environments.

Concerning computer simulations, two sets of simulations were performed. In the first set, the performance of a monostatic \ac{UWB} \ac{RSN} with $\numsens=4$ sensors was analyzed, both with two tracks and with three tracks; in all cases the tracks include maneuvering as well as walking-stop-walking targets.   
In the second set of simulations, a multistatic \ac{UWB} \ac{RSN} with $\numsens=4$ sensors and one transmitter was considered. Again, two scenarios with two and three targets were analyzed.

Regarding the analysis with actual \ac{UWB} waveforms, two case studies were investigated. In the first case study we used a monostatic \ac{RSN} to detect and track one human target in an outdoor environment (specifically, a balcony); in the second case study we used a multistatic \ac{RSN} to detect two human targets in an indoor environment.
We employed Humatics \ac{UWB} nodes (P440 and P410) and a laptop acting as the \ac{FC}. The nodes operate in the $[3.1, 4.8]\,\mathrm{GHz}$ band, and have two antenna ports (for transmission and reception), a power interface, and a control~port. In all simulation and case studies, the sampling frequency at the receiver is $f_s=16.48\,\mathrm{GHz}$ (sampling time $T_s=61\,\mathrm{ps}$), the scan period is $0.45\,\mathrm{s}$ and the cell dimensions are $\Delta_x=10\,\mathrm{cm}$ and $\Delta_y=10\,\mathrm{cm}$.

The value of $\alpha$ in \eqref{eq:vote} was set to $0.75$ in all experiments, and the maximum number of detected tracklets per time window was set to $5$. The proposed approach involves a number of thresholds, whose values often need to be set by manual tuning. We used $\beta=2$ in the adaptive threshold $\eta^t_{\mathrm{score}}$ in \eqref{eq:etat_threshold}. Moreover, good results were obtained by setting the adaptive threshold $\gamma^t_{\mathrm{score}}$ in \eqref{eq:region_score_condition} equal to $\eta^t_{\mathrm{score}}$, i.e., $\gamma^t_{\mathrm{score}}=\eta^t_{\mathrm{score}}$.
Table~\ref{tab:value of parameters} summarizes the values of the several parameters used in the experiments. A singe value indicates that the corresponding parameter was did not change in all experiments. The third column of the table provides a brief description of each parameter along with its influence to the processing performance.

\begin{table}[!t]
  \centering
  \fontsize{8}{8.5}\selectfont
  \begin{threeparttable}
  \caption{Numerical values of the parameters involved in the processing}
  \label{tab:value of parameters}
    \begin{tabular}{p{3.2cm}
    p{4.1cm} 
    p{7.8cm}
    }
   \toprule Parameter & Value  & Description\cr
    \midrule
    \multirow{4}{*}{$(N_x,N_y,K)$} &
    Experiment 1:  $(150,150,40)$
    & - $(N_x,N_y)$ proportional to the surveillance area and inversely\\ & Experiment 2: $(150,150,40)$
   &  proportional to grid cell size. \\ & Case study 1: $(120,80,124)$ & \multirow{2}{*}{- $K$ inversely proportional to the scan period.}\\ 
    & Case study 2: $(120,60,40)$& \cr
    \hline 
    \multirow{4}{*}{Number of sensors, $\numsens$} & Experiment 1: $\numsens=4$ &\multirow{2}{*}{- Monostatic \ac{RSN} in simulation set 1 and case study 1.} \\ & Experiment 2: $\numsens=4$ (+ 1 Tx) & \\& Case study 1: $\numsens=3$ & \multirow{2}{*}{- Multistatic \ac{RSN} in simulation set 2 and case study 2.}\\& Case study 2: $\numsens=4$ (+ 1 Tx) \cr
     \hline
     \multirow{4}{*}{Pulse integration factor, $\pulseintegfactor$} & Experiment 1: $4096$ & \multirow{1}{*}{- A larger $N_s$ improves the SNR, beneficial to weak target} \\ & Simulation 2: $20800$ & \multirow{1}{*}{detection.}\\ & Case study 1: $4096$& \multirow{1}{*}{- A smaller $N_s$ enables a lower scan period, beneficial to}\\ 
    & Case study 2: $20800$ &\multirow{1}{*}{maneuvering target detection.}\cr
    \hline
    Samples per scan, $\samplesperscan$ &$1500$ & $\samplesperscan$ is proportional to coverage range of sensor.\cr
    $\kappa$ in \eqref{eq:er_energy} \ &$10$ & - A smaller $\kappa$ reduces complexity in static clutter removal, but brings a performance deterioration for slowly moving targets.\cr
     $\alpha$ in \eqref{eq:vote} \ &$0.75$ & - $\alpha \in (0,1)$ helps suppressing strong clutter; a smaller $\alpha$ should be applied in strongly cluttered environments.\cr
    $\beta$ in \eqref{eq:etat_threshold} & $2$ & - A larger $\beta$ helps decreasing the false alarms in a single scan, but tends to hinder detection of weak targets.\cr
     $(W,s)$ &$(4,2)$ & - Larger $W$ and $s$ are beneficial to weak target detection but deteriorate tracking performance of maneuvering targets.\cr
     $(d_x,d_y)$ & $(5,5)$ & - Related to the size of target region; larger $(d_x,d_y)$ reduces complexity but tend to deteriorate detection performance.\cr
     $\gamma^t_{\rm{score}}$ in \eqref{eq:region_score_condition} & $\eta^t_{\mathrm{score}}$ in \eqref{eq:etat_threshold} & - $\gamma^t_{\rm{score}}$ and $\eta^t_{\mathrm{score}}$ are related to the SNR of target echoes; lower values favor weak target detection but yield more false alarms.\cr
     $\gamma^t_{\rm{num}}$ in \eqref{eq:region_card_condition} & $150$ & - Proportional to the target size; a smaller $\gamma^t_{\rm{num}}$ is beneficial detection of smaller targets but brings more false alarms.\\
     $w$ in \eqref{eq:smooth}& $3$& - Related to the target size and distribution of noise; a larger $w$ is beneficial to suppress the ghost echoes but increases complexity and hinders partition of regions of closely-spaced targets. \cr
     $N$ in \eqref{eq:segment_points}& $10$ & - A larger $N$ is beneficial to target region partition but increases complexity.\cr
     $\nu$ in \eqref{eq:outlier_removal}& $3$ & - A smaller $\nu$ is beneficial to outlier removal but may deteriorate localization accuracy when target is highly maneuvering.\\
    \bottomrule
    \end{tabular}
    \end{threeparttable}
    \label{tab:positioanl_error}
\end{table}

\subsection{Computer Simulations}\label{sec:case_study0}

As mentioned above, in order to numerically assess the performance of the proposed processing chain, we performed two sets of simulations, hereafter referred to as ``Experiment 1'' and ``Experiment 2'' and corresponding to a monostatic and a multistatic \ac{UWB} \ac{RSN}, respectively. In both sets of simulations, the signals produced by the \ac{UWB} sensors, representing the input of the \ac{FC}, were generated according to distributions extracted from real data, as described hereafter.

Measurements were first collected with Humatics \ac{UWB} radios both in a monostatic configuration in an outdoor environment and in a multistatic configuration in an indoor environment (hence, the two sets of measurements reflect the above-mentioned case studies), using human targets in all cases. 
Then, for each collected \ac{UWB} waveform, the corresponding signal ${\boldsymbol{m}}_n^{t}$ was obtained by applying the in-sensor processing described in Section~\ref{subsec:in_sensor_processing}. This way, two sets of signals ${\boldsymbol{m}}_n^{t}$ were built, one relevant to the monostatic configuration (in outdoor environment) and one to the multistatic configuration (in indoor environment). Next, separately for each of the two signal sets, the distribution of the generic element of ${\boldsymbol{m}}_n^{t}$ was obtained, both for those range bins (elements of ${\boldsymbol{m}}_n^{t}$) in which a target echo exists and for those ones associated with clutter echoes and noise only. During each simulation, given the positions of the targets at scan $t$ and given a sensor $n$, each sample of ${\boldsymbol{m}}_n^{t}$ was generated randomly by applying the appropriate distribution.\footnote{The statistical distributions according to which the signals ${\boldsymbol{m}}_n^{t}$ are generated throughout the simulations are therefore tied to the specific in-sensor processing described in Section~\ref{subsec:in_sensor_processing}. This is not an issue because this in-sensor processing has been specifically designed for \ac{UWB} signals and because the same distributions are employed to generate the signals ${\boldsymbol{m}}_n^{t}$ at the input of all \ac{FC} algorithms whose detection and tracking performance we seek to compare.} 

A method for points generation in presence of \ac{UWB} signals has been developed in \cite{chiani2018sensor}. To benchmark the performance of the proposed method measured via numerical simulations, the point generation method of \cite{chiani2018sensor} has been employed and the points generated by this method have been further processed for track detection by two different techniques, namely, the \ac{PHD} filter from \cite{granstrom2012extended} and the \ac{TBD} technique from \cite{YAN2021107821}, both approaches targeting tracking of extended targets.

\begin{figure}[!tb]
    \centering
    \subfigure[]{\includegraphics[width=0.4\columnwidth]{{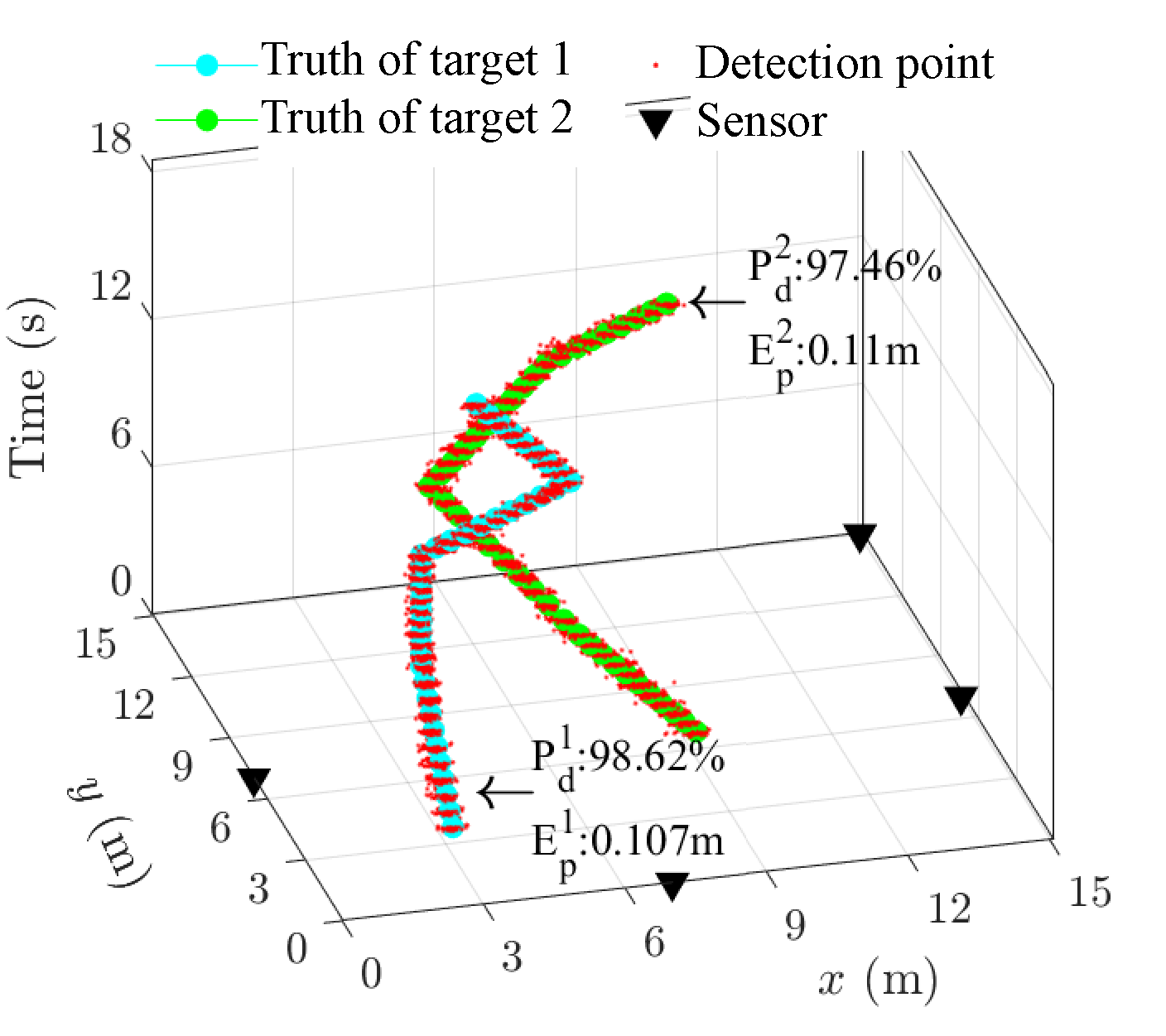}}}
    \subfigure[]{\includegraphics[width=0.4\columnwidth]{{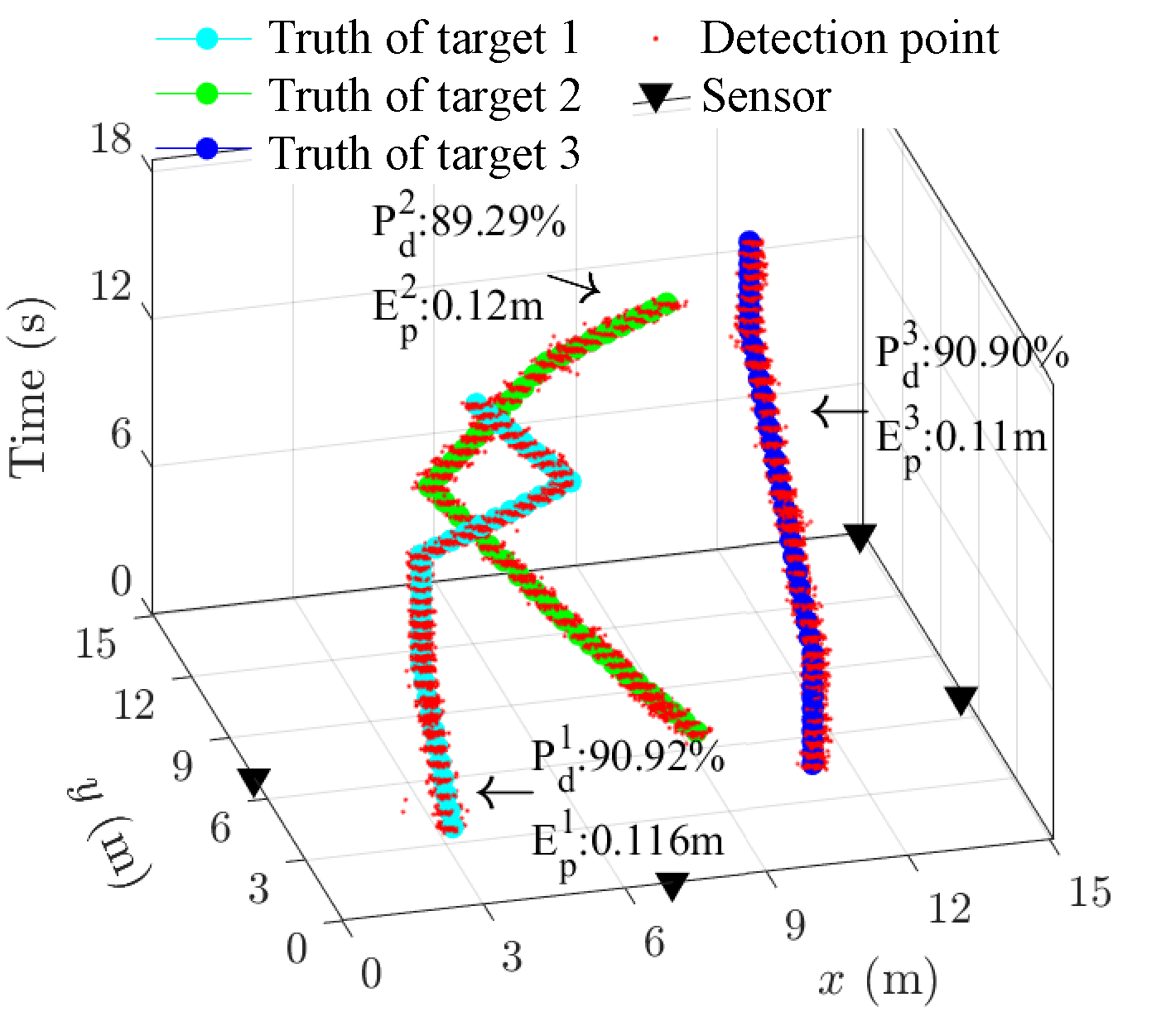}}}
    \subfigure[]{\includegraphics[width=0.4\columnwidth]{{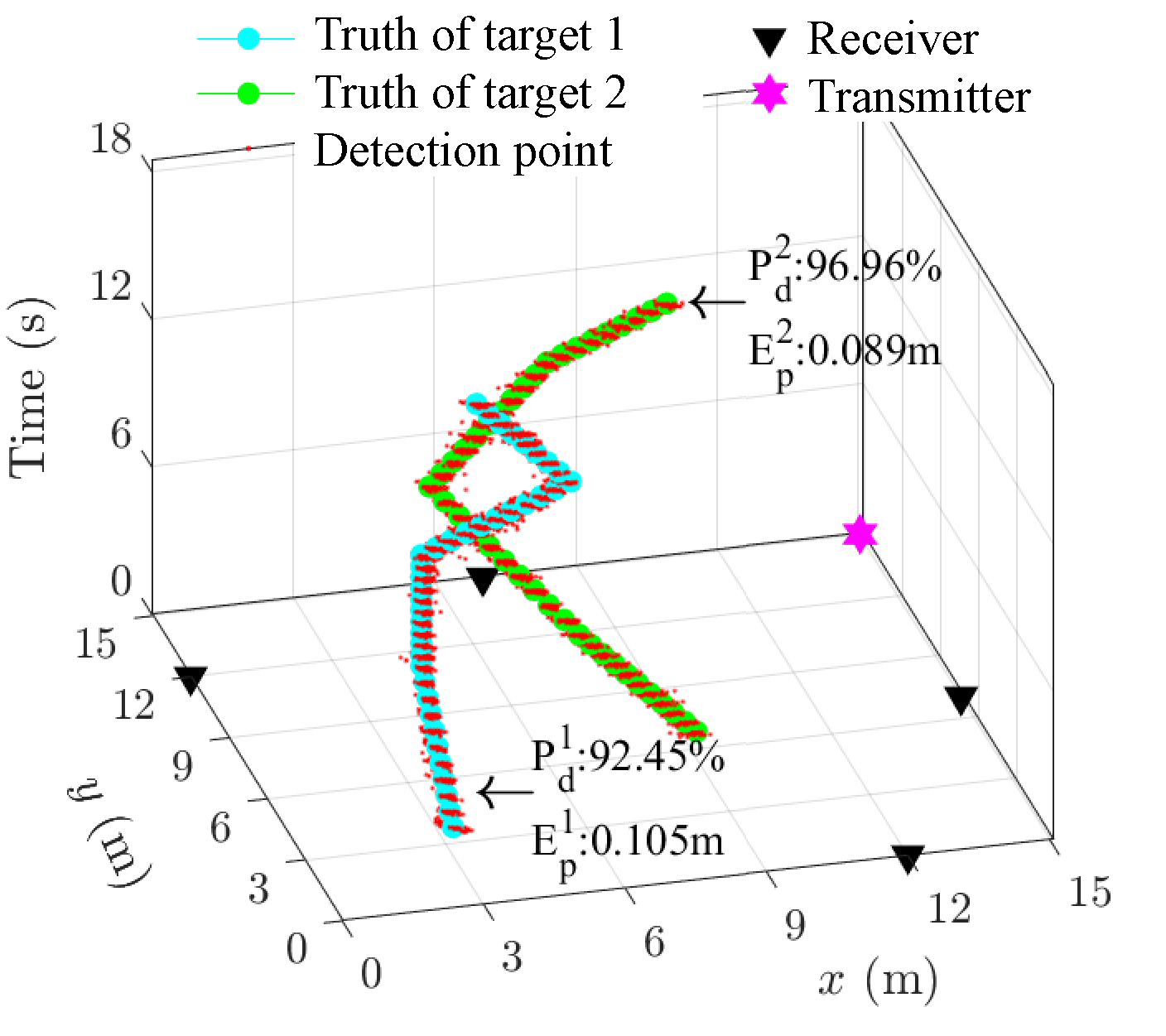}}}
   \subfigure[]{\includegraphics[width=0.4\columnwidth]{{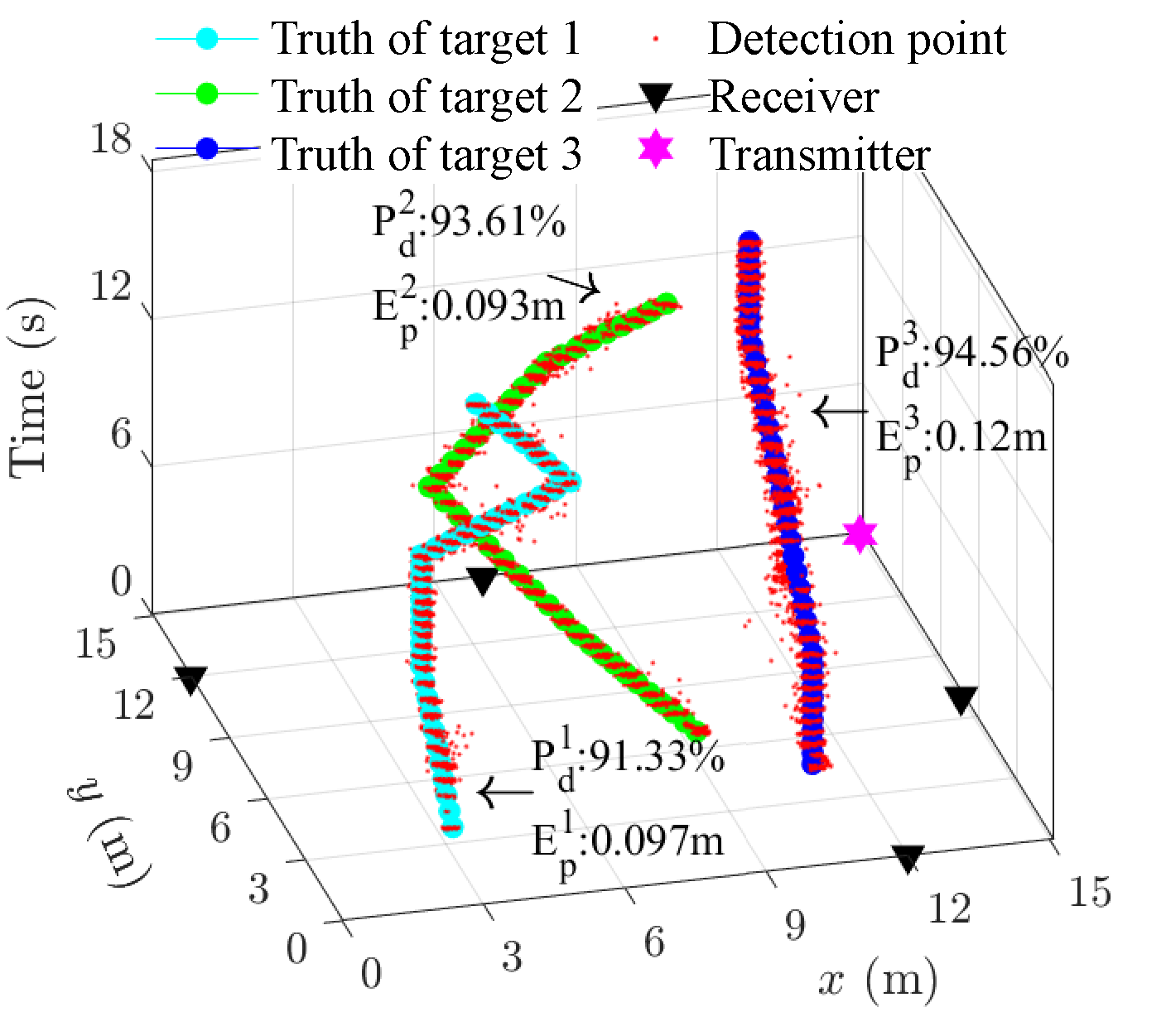}}}
    \caption{Results of the proposed method through 100 numerical simulations. (a) Experiment 1 (monostatic \ac{RSN}) -- Test 1 (two targets); (b) Experiment 1 -- Test 2 (three targets); (c) Experiment 2 (multistatic \ac{RSN}) -- Test 1; (d) Experiment 2 -- Test 2.}
    \label{fig:Res_pro_simu}
\end{figure}

\begin{figure}[!tb]
    \centering
    \subfigure[]{\includegraphics[width=0.4\columnwidth]{{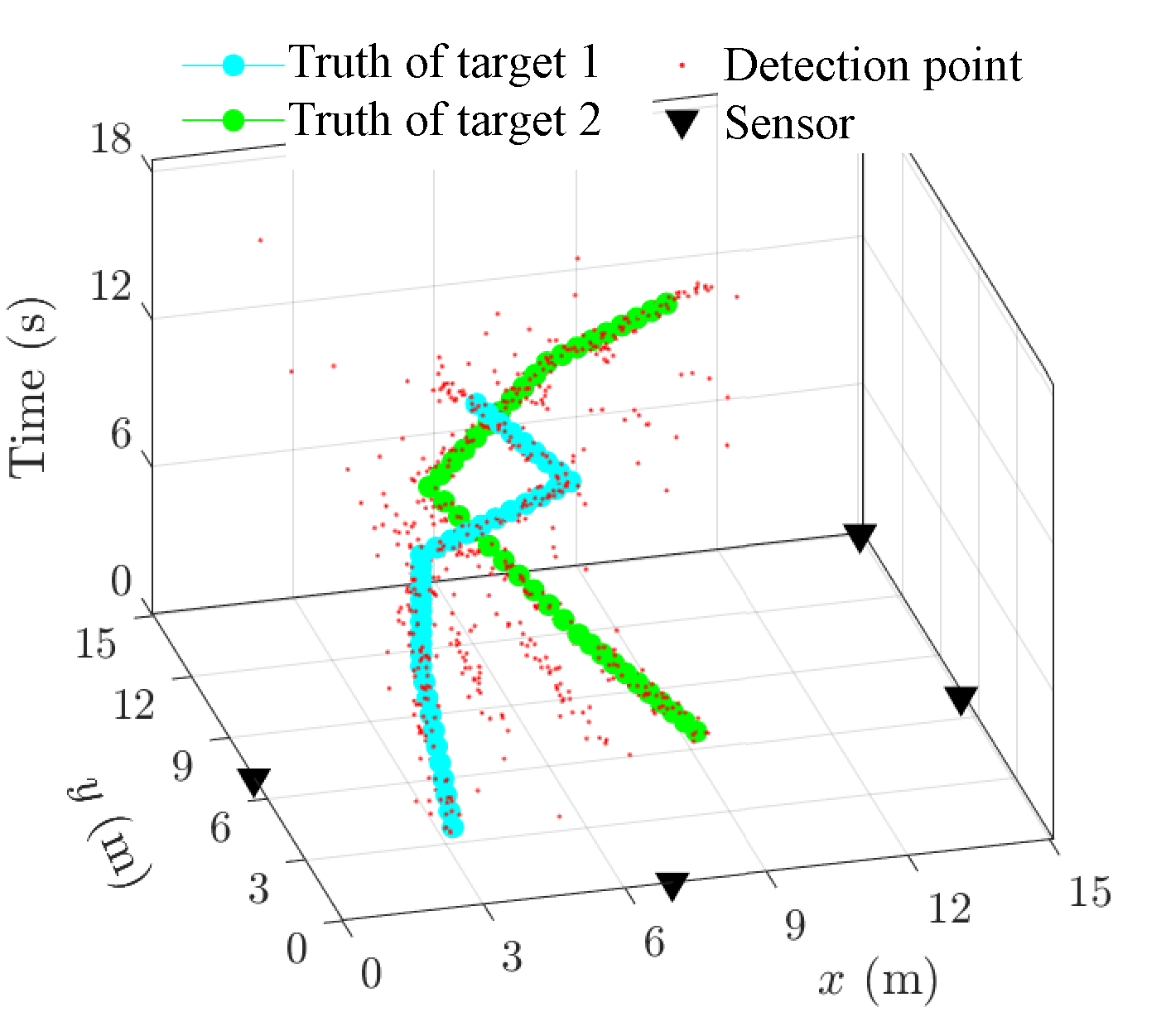}}}
    \subfigure[]{\includegraphics[width=0.4\columnwidth]{{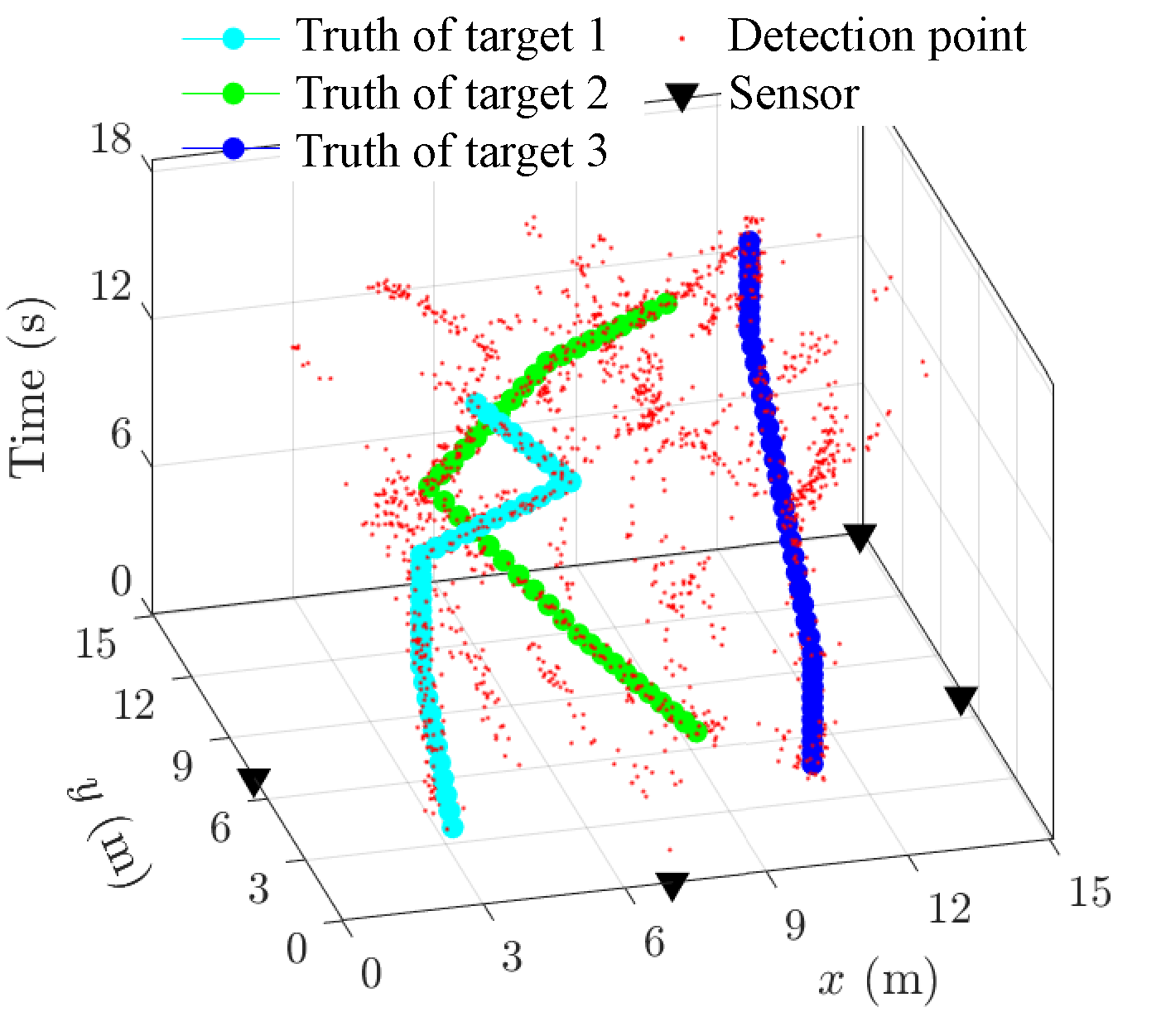}}}
    \subfigure[]{\includegraphics[width=0.4\columnwidth]{{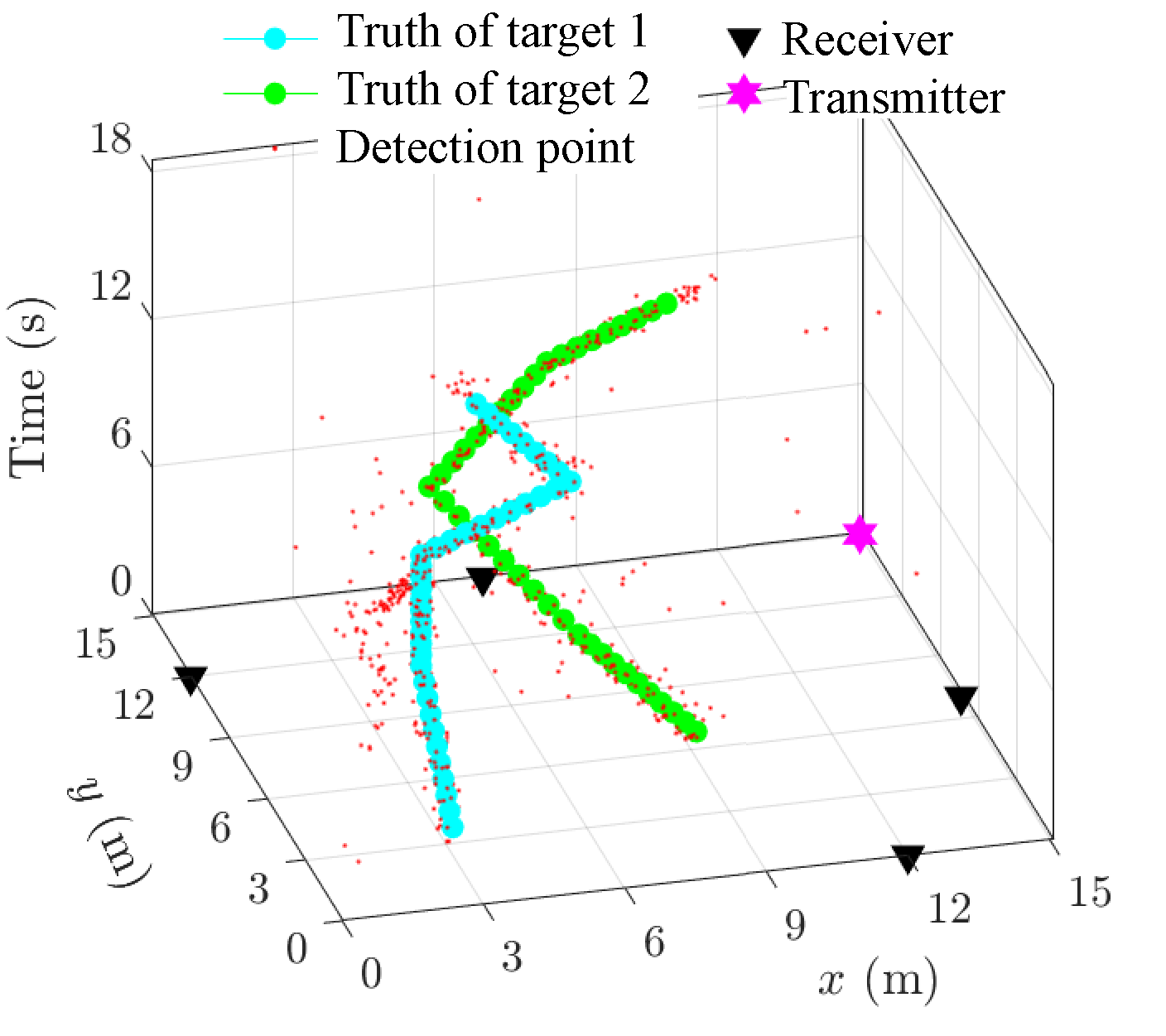}}}
   \subfigure[]{\includegraphics[width=0.4\columnwidth]{{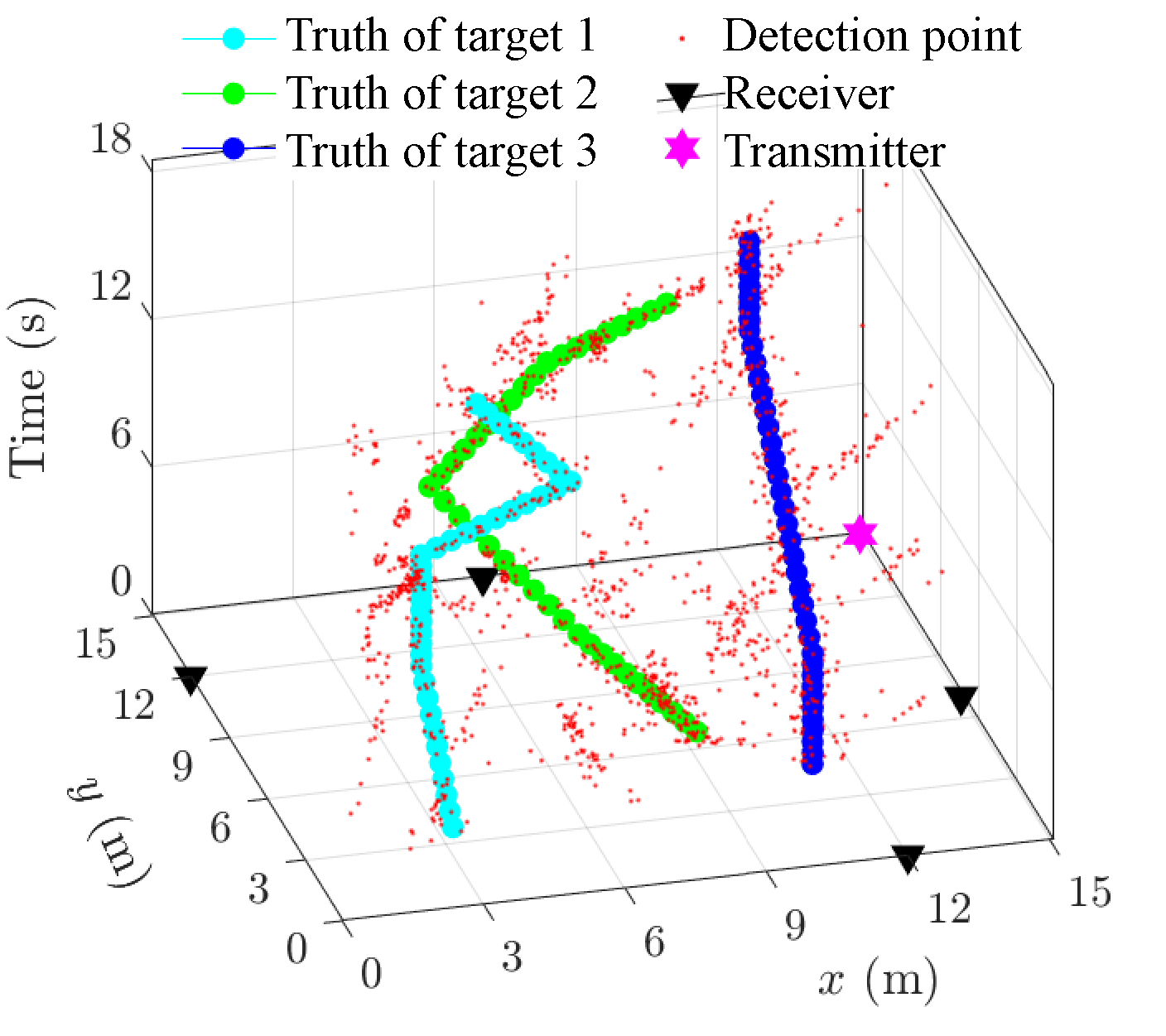}}}
    \caption{Detection points generated by the method \cite{chiani2018sensor} through 10 numerical simulations. (a) Experiment 1 --  Test 1; (b) Experiment 1 -- Test 2; (c) Experiment 2 -- Test 1; (d) Experiment 2 -- Test 2.}
    \label{fig:mea_simu}
\end{figure}

The detected tracks of a monostatic and of a multistatic \ac{RSN} (Experiment~1 and Experiment~2, respectively) using the proposed method, both in presence of two targets (``Test~1'') and of three targets (``Test~2''), are presented in Fig.~\ref{fig:Res_pro_simu}. Each subfigure shows the target ground truths together with the final detection points produced by the proposed method (red points), after trajectory smoothing. In particular, the superposition of all detection points generated through 100 simulations is illustrated. 
Both in presence of two targets and of three targets, the target detection rate is higher than 90\% with an average positional error lower than $12\,\mathrm{cm}$. 
The points generated by the detection approach in  \cite{chiani2018sensor} are shown in Fig.~\ref{fig:mea_simu}. We can see that a relatively large number false of alarm points arise, obscuring the target points.
Compared with the case of two targets, in the three target case the target detection rate is lower and more false alarm points are generated.
A comparison between Fig.~\ref{fig:Res_pro_simu} and Fig.~\ref{fig:mea_simu} allows appreciating the remarkably better performance offered by the proposed method with respect to a ``points generation only'' approach in presence of weak \ac{UWB} echoes, especially in cluttered environments with multiple (e.g., three) targets.

\begin{figure}[!tb]
    \centering
    \subfigure[]{\includegraphics[width=0.4\columnwidth]{{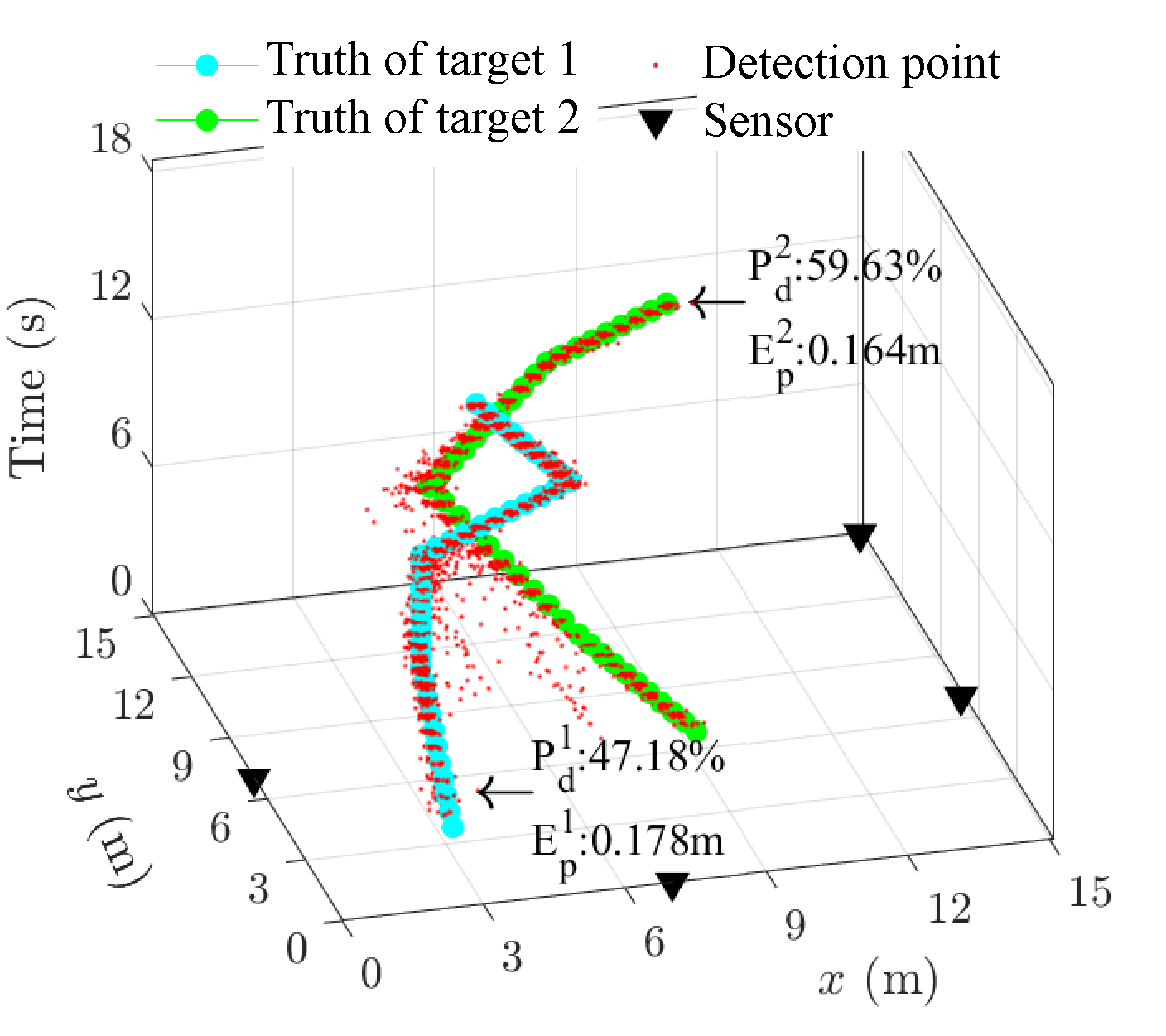}}}
    \subfigure[]{\includegraphics[width=0.4\columnwidth]{{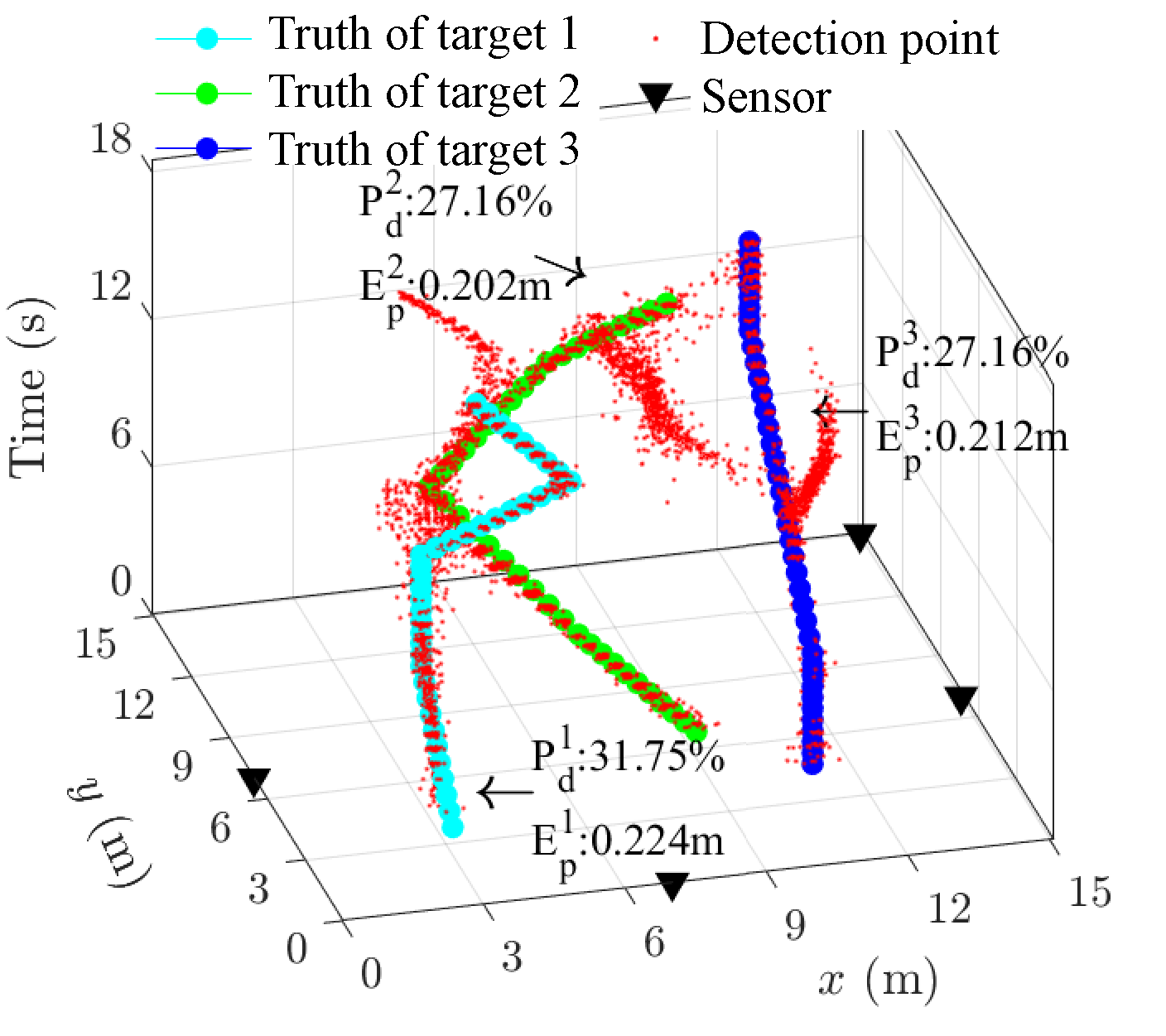}}}
    \subfigure[]{\includegraphics[width=0.4\columnwidth]{{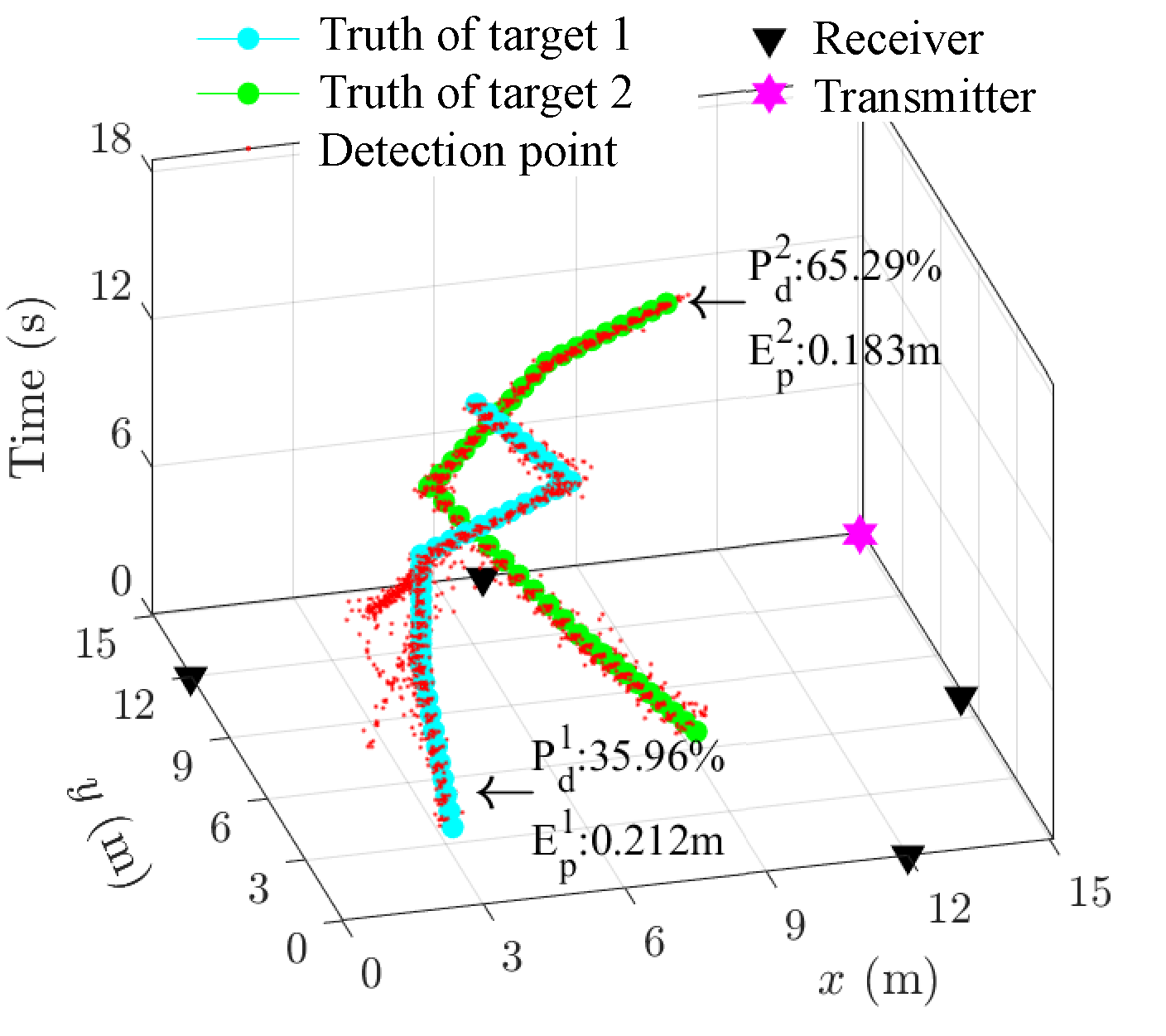}}}
   \subfigure[]{\includegraphics[width=0.4\columnwidth]{{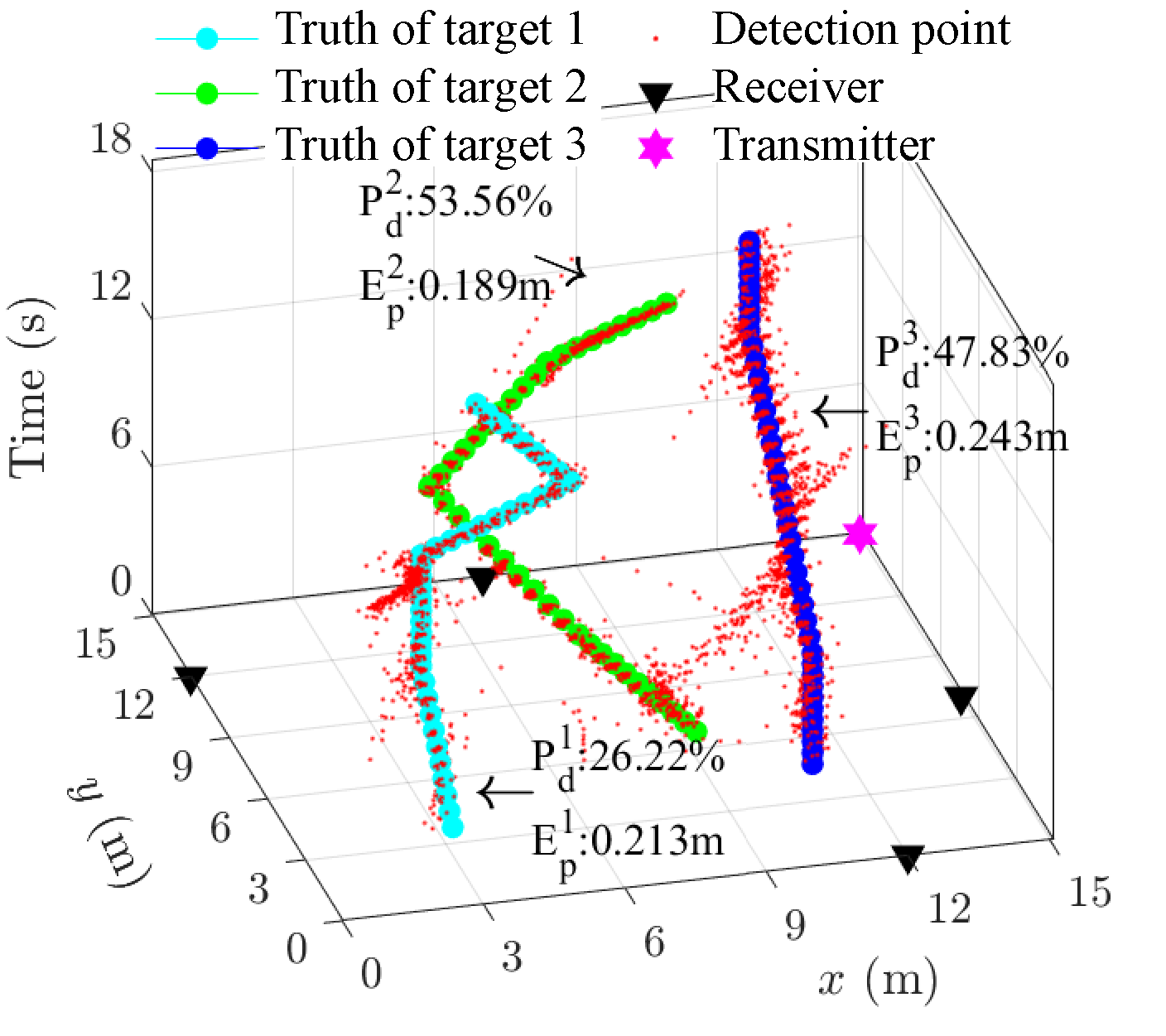}}}
    \caption{Results of the \ac{TBD} method in \cite{YAN2021107821} fed with the points generated by the detection technique in \cite{chiani2018sensor} through 100 numerical simulations. (a) Experiment 1 -- Test 1; (b) Experiment 1 -- Test 2; (c) Experiment 2 -- Test 1; (d) Experiment 2 -- Test 2.}
    \label{fig:Res_TBD_simu}
\end{figure}

Fig.~\ref{fig:Res_TBD_simu} shows the detected tracks returned by the  \ac{TBD} algorithm recently proposed in \cite{YAN2021107821}, using as input the points generated by the detection technique in \cite{chiani2018sensor}.
A lower detection rate is achieved and and more false tracks are generated.
In \cite{YAN2021107821}, a \ac{TBD} approach based on a 3D Projection was developed and shown to be able to outperform other methods, including the \ac{3D} Hough transform \ac{TBD} algorithm proposed in \cite{yan2019detection} and the \ac{PHD} filter based algorithm proposed in \cite{granstrom2012extended}. The fact that the method here developed performs better than the one of \cite{YAN2021107821} indicates that it can outperform other tracking approaches in presence of \ac{UWB} waveforms.

The results of Experiment~1 and Experiment~2 (both Test~1 and Test~2 for each of them) are summarized in Table~\ref{tab:Simulation_C3T2}, Table~\ref{tab:Simulation_C3T3}, Table~\ref{tab:Simulation_C4T2}, and Table~\ref{tab:Simulation_C4T3}, including the detection rate and the positional error of each target, the number of average false alarm tracks per scan, the \ac{OSPA} distance \cite{ristic2010performance}, and the average running time over 100 Monte Carlo experiments.
As it can be observed from the reported data, the detection rate of the proposed method is remarkably higher than that of the benchmark methods, with less false alarm tracks. As a consequence, the proposed method achieves a lower \ac{OSPA} distance. In contrast, the average running time of proposed method is higher than that of the other two methods. This fact mainly comes as a consequence of the voting (Sec.~\ref{subsec:voting}) and the \ac{3D} image processing (Sec.~\ref{subsec:3d}) stages. Importantly, we can trade-off the computational complexity and the performance of the proposed method by adjusting the size of grid cell, defined by $\Delta_x$ and $\Delta_y$. Besides $\Delta_x=\Delta_y=\Delta=0.1\,\mathrm{m}$, tests we carried out for $\Delta=0.05\,\mathrm{m}$ and $\Delta=0.2\,\mathrm{m}$, corresponding to a smaller and a larger grid cell, respectively. As from the tabular data, using a larger $\Delta$, such as $\Delta=0.2\,\mathrm{m}$, yields considerable savings in complexity at the price of a slight deterioration in performance, which however remains generally better that that of the other techniques.  
A smaller $\Delta$, such as $\Delta=0.05\,\mathrm{m}$ achieves very marginal gains over $\Delta=0.1\,\mathrm{m}$ while imposing unacceptable complexity.

\begin{figure}[!tb]
    \centering
    \subfigure[]{\includegraphics[width=0.48\columnwidth]{{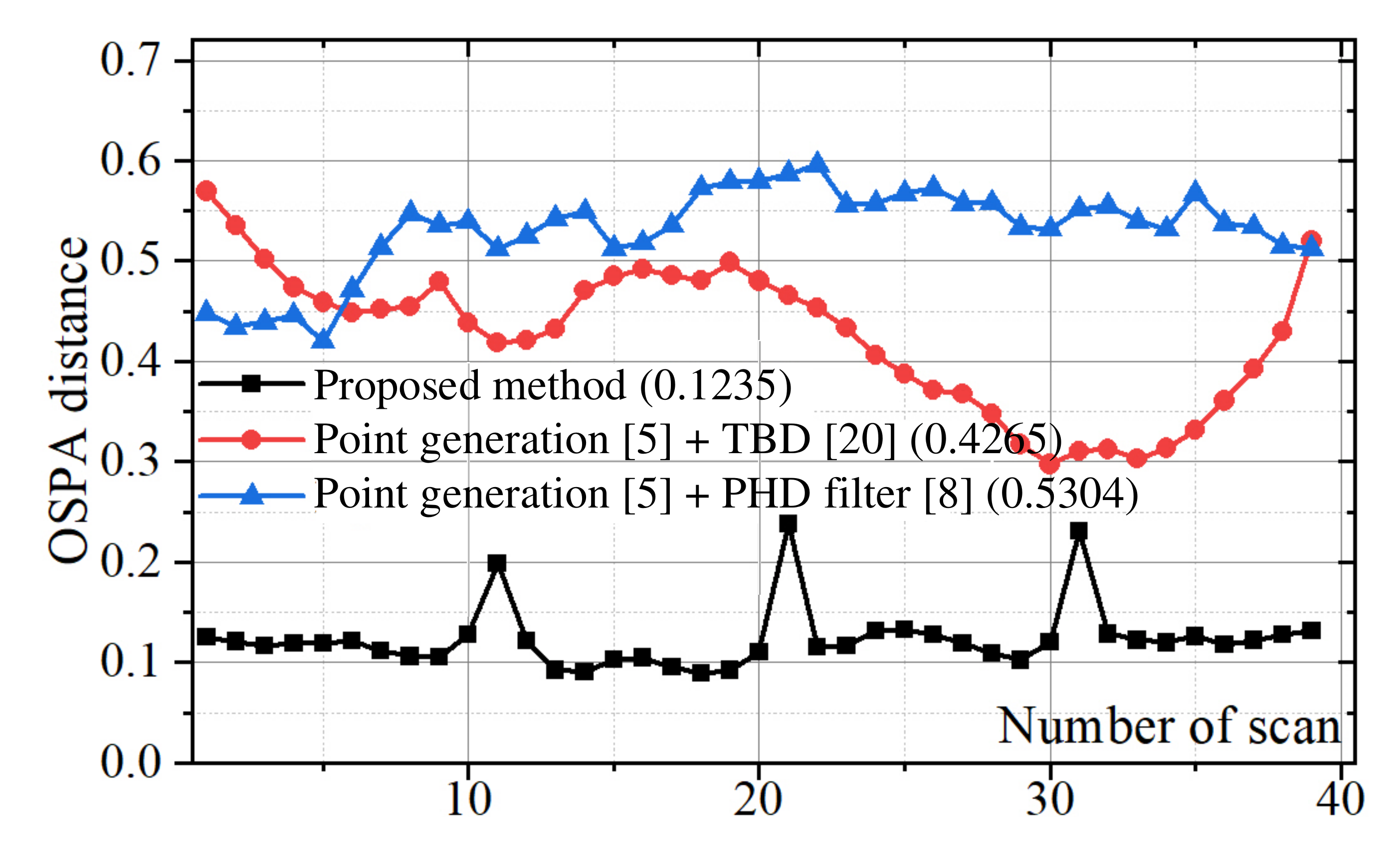}}}
    \subfigure[]{\includegraphics[width=0.48\columnwidth]{{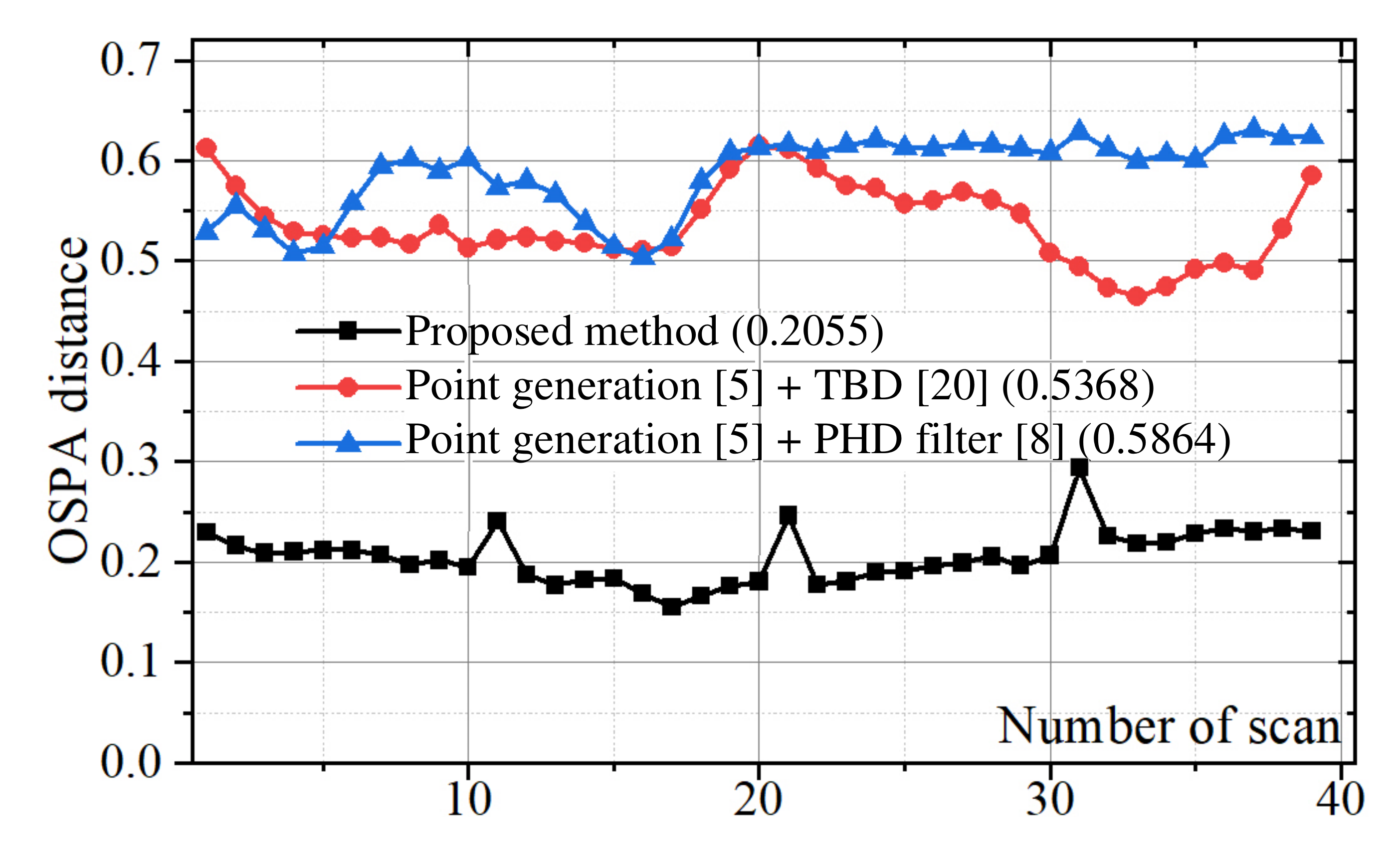}}}
    \subfigure[]{\includegraphics[width=0.48\columnwidth]{{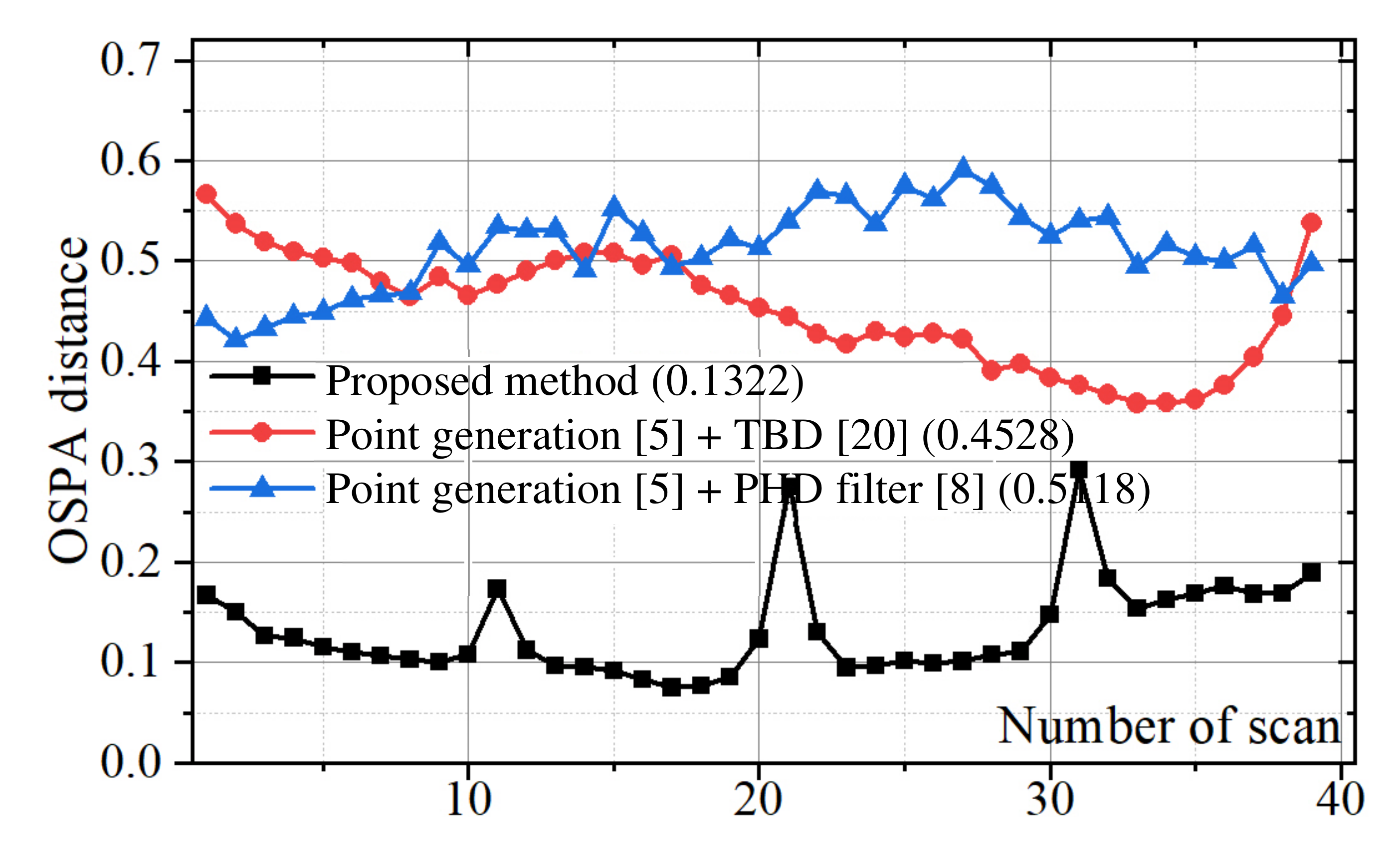}}}
   \subfigure[]{\includegraphics[width=0.48\columnwidth]{{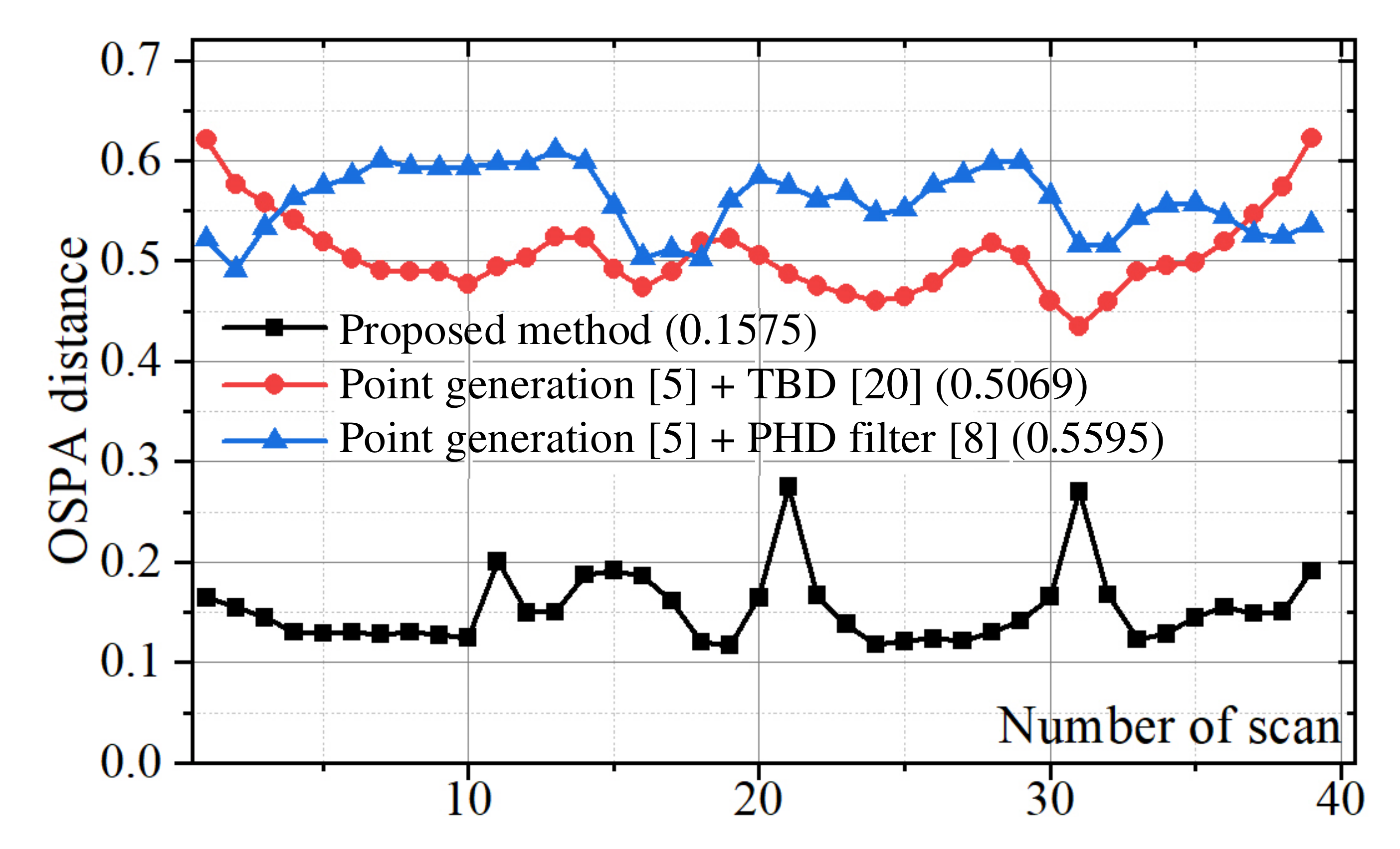}}}
    \caption{ \ac{OSPA} distance of methods. (a) Experiment 1 test 1. (b) Experiment 1  test 2 . (c) Experiment 2 test 1. (d) Experiment 2  test 2.}
    \label{fig:Simu_OSPA}
\end{figure}

Fig.~\ref{fig:Simu_OSPA} illustrates the \ac{OSPA} distance through different scans for the proposed method and the two methods used as benchmarks; the applied \ac{OSPA} metric employs the Euclidean distance with a cut-off parameter $0.7$ meters and an order parameter equal to $1$. In all cases (Experiments 1 and 2, Tests 1 and 2), the proposed method achieves an \ac{OSPA} distance lower than the one achieved by competing approaches. 
As we can observe in Fig.~\ref{fig:Simu_OSPA}, the \ac{OSPA} distance values achieved by the proposed method at the $11^{\text{th}}$, $21^{\text{th}}$, and $31^{\text{th}}$ scans are higher than those in the other scans. This fact can be attributed to the least-square criterion adopted in trajectory smoothing: A larger positional error is unavoidable when the target is maneuvering because of the mismatch between the smoothing model and actual target motion state \cite{chen2020novel}. The smoothing model trades off positional error between target in straight line and in maneuvering. A smaller scan period is helpful to decrease the positional error in those scans in which the target is maneuvering; this can be achieved by reducing pulse integration factor $N_s$, but at the expense of decreasing the target detection rate.

\renewcommand{\arraystretch}{1}
\begin{table*}[htp]
  \centering
  \fontsize{8.5}{8}\selectfont
  \begin{threeparttable}
  \caption{Results of Experiment 1 -- Test 1 (monostatic \ac{RSN}, 2 targets).}
  \label{tab:Simulation_C3T2}
     \begin{tabular}{wl{4.4cm}<{\centering} p{2cm}<{\centering}p{2cm}<{\centering}p{1cm}<{\centering}p{1cm}<{\centering}p{1cm}<{\centering}p{1cm}<{\centering}}
    \toprule
     & $P_d^{1}$\tnote{(a)} \qquad  $E_p^1$\tnote{(b)}
     & $P_d^2$ \qquad $E_p^2$
     & $N_{\text{FA}}$\tnote{(c)}
     & \ac{OSPA}
     & Time
\cr
    \midrule% 
    Proposed method ($\Delta=0.1\,\mathrm{m}$) & $98.62\%$ \quad $0.107$ & $97.46\%$ \quad $0.114$ & $0.0087$ & $0.123$ & $21.87\,\mathrm{s}$ \cr
    Point generation \cite{chiani2018sensor} + \ac{TBD} \cite{YAN2021107821} & $47.18\%$ \quad  $0.178$ & $59.63\%$ \quad $0.164$ & $0.0318$ & $0.426$ & $12.45\,\mathrm{s}$\cr
    Point generation \cite{chiani2018sensor} + \ac{PHD} filter \cite{granstrom2012extended} & $49.40\%$ \quad  $0.329$ & $63.01\%$ \quad $0.334$ & $0.9810$ & $0.531$ & $07.44\,\mathrm{s}$\cr
   Proposed method ($\Delta=0.2\mathrm{m}$) & $98.28\%$ \quad $0.167$ & $94.87\%$ \quad $0.155$ & $0.0052$ & $0.175$ & $15.87\,\mathrm{s}$\cr
   Proposed method ($\Delta=0.05\,\mathrm{m}$) & $98.72\%$ \quad $0.105$ &$95.98\%$ \quad $0.110$ & $0.0089$ & $0.164$ & $61.51\,\mathrm{s}$\cr
    \bottomrule
    \end{tabular}
    \begin{tablenotes}
        \footnotesize
        \item[(a)]{$P_d^i$: detection rate of target $i$.}
        \item[(b)]{$E_p^i$: average positional error of target $i$.} 
        \item[(c)]{$N_{\text{FA}}$: average number of false alarms per scan.} 
      \end{tablenotes}
    \end{threeparttable}
\end{table*}

\renewcommand{\arraystretch}{1}
\begin{table*}[htp]
  \centering
  \fontsize{8.5}{8}\selectfont
  \begin{threeparttable}
  \caption{Results of Experiment 1 -- Test 2 (monostatic \ac{RSN}, 3 targets).}
  \label{tab:Simulation_C3T3}
     \begin{tabular}{wl{4.4cm}<{\centering} p{2cm}<{\centering}p{2cm}<{\centering}p{2cm}<{\centering}p{1cm}<{\centering}p{1cm}<{\centering}p{1cm}<{\centering}}
    \toprule
     & $P_d^1$ \qquad $E_p^1$ 
     & $P_d^2$ \qquad $E_p^2$
     & $P_d^3$ \qquad $E_p^3$
     & $N_{\text{FA}}$
     & \ac{OSPA}
     & Time
\cr
    \midrule% 
    Proposed method ($\Delta=0.1\,\mathrm{m}$) & $90.92\%$ \quad $0.116$ & $89.29\%$ \quad  $0.122$ & $90.89\%$ \quad $0.118$ & $0.0280$ & $0.206$ & $22.38\,\mathrm{s}$\cr
    Point generation \cite{chiani2018sensor} + \ac{TBD} \cite{YAN2021107821} & $31.75\%$ \quad $0.224$ & $46.86\%$ \quad $0.202$ & $27.16\%$ \quad $0.212$ & $0.4010$ & $0.537$ & $13.86\,\mathrm{s}$\cr
    Point generation \cite{chiani2018sensor} + \ac{PHD} filter \cite{granstrom2012extended} & $53.45\%$ \quad $0.314$ & $61.47\%$ \quad  $0.366$ & $51.10\%$ \quad $0.392$ & $3.5780$ & $0.587$ & $07.84\,\mathrm{s}$\cr
   Proposed method ($\Delta=0.2\,\mathrm{m}$) & $86.67\%$ \quad $0.228$ & $88.60\%$ \quad  $0.197$ & $87.56\%$ \quad $0.197$ & $0.0220$ & $0.244$ & $16.12\,\mathrm{s}$\cr
   Proposed method ($\Delta=0.05\,\mathrm{m}$) & $92.92\%$ \quad $0.114$ & $86.29\%$ \quad $0.121$ & $92.90\%$ \quad $0.109$ & $0.0028$ & $0.208$ & $71.85\,\mathrm{s}$\cr
    \bottomrule
    \end{tabular}
    \end{threeparttable}
\end{table*}

\renewcommand{\arraystretch}{1}
\begin{table*}[htp]
  \centering
  \fontsize{8.5}{8}\selectfont
  \begin{threeparttable}
  \caption{Results of Experiment 2 -- Test 1 (multistatic \ac{RSN}, 2 targets).}
  \label{tab:Simulation_C4T2}
     \begin{tabular}{wl{4.4cm}<{\centering} p{2cm}<{\centering}p{2cm}<{\centering}p{1cm}<{\centering}p{1cm}<{\centering}p{1cm}<{\centering}p{1cm}<{\centering}}
    \toprule
     & $P_d^1$ \qquad $E_p^1$ 
     & $P_d^2$ \qquad $E_p^2$
     & $N_{\text{FA}}$
     & \ac{OSPA}
     & Time
\cr
    \midrule% 
    Proposed method ($\Delta=0.1\,\mathrm{m}$) & $96.64\%$ \quad $0.105$ & $92.45\%$ \quad $0.089$ & $0.0045$ & $0.132$ & $19.02\mathrm{s}$ \cr
    Point generation \cite{chiani2018sensor} + \ac{TBD} \cite{YAN2021107821} & $35.96\%$ \quad $0.212$ & $65.29\%$ \quad $0.183$ & $0.0650$ & $0.453$ & $15.89\mathrm{s}$\cr
    Point generation \cite{chiani2018sensor} + \ac{PHD} filter \cite{granstrom2012extended} & $51.13\%$ \quad $0.331$ & $65.75\%$ \quad $0.341$ & $0.7540$ & $0.513$ & $12.21\,\mathrm{s}$\cr
   Proposed method ($\Delta=0.2\,\mathrm{m}$) & $95.89\%$ \quad $0.157$ & $92.82\%$ \quad $0.150$ & $0.0091$ & $0.210$ & $16.49\,\mathrm{s}$\cr
   Proposed method ($\Delta=0.05\,\mathrm{m}$) & $93.84\%$ \quad $0.110$ & $91.02\%$ \quad $0.094$ & $0.0094$ & $0.180$ & $64.05\,\mathrm{s}$\cr
    \bottomrule
    \end{tabular}
    \end{threeparttable}
\end{table*}

\renewcommand{\arraystretch}{1}
\begin{table*}[htp]
  \centering
  \fontsize{8.5}{8}\selectfont
  \begin{threeparttable}
  \caption{Results of Experiment 2 -- Test 2 (multistatic \ac{RSN}, 3 targets).}
  \label{tab:Simulation_C4T3}
     \begin{tabular}{wl{4.4cm}<{\centering} p{2cm}<{\centering}p{2cm}<{\centering}p{2cm}<{\centering}p{1cm}<{\centering}p{1cm}<{\centering}p{1cm}<{\centering}}
    \toprule
     & $P_d^1$ \qquad $E_p^1$ 
     & $P_d^2$ \qquad $E_p^2$
     & $P_d^3$ \qquad $E_p^3$
     & $N_{\text{FA}}$
     & \ac{OSPA}
     & Time
\cr
    \midrule% 
    Proposed method ($\Delta=0.1\,\mathrm{m}$) & $91.33\%$ \quad $0.097$ & $93.61\%$ \quad $0.093$ & $94.56\%$ \quad $0.121$ & $0.0092$ & $0.157$ & $20.25\,\mathrm{s}$\cr
    Point generation \cite{chiani2018sensor} + \ac{TBD} \cite{YAN2021107821} & $26.22\%$ \quad $0.213$ & $53.56\%$ \quad $0.189$ & $47.83\%$ \quad $0.243$ & $0.1620$ & $0.507$ & $16.76\,\mathrm{s}$\cr
    Point generation \cite{chiani2018sensor} + \ac{PHD} filter \cite{granstrom2012extended} & $53.42\%$ \quad $0.334$ & $56.75\%$ \quad $0.362$ & $62.35\%$ \quad $0.340$ & $2.6200$ & $0.559$ & $11.37\,\mathrm{s}$\cr
   Proposed method ($\Delta=0.2\,\mathrm{m}$) & $93.64\%$ \quad $0.199$ & $92.61\%$ \quad  $0.182$ & $93.64\%$ \quad $0.193$ & $0.0190$ & $0.214$ & $16.39\,\mathrm{s}$\cr
   Proposed method ($\Delta=0.05\,\mathrm{m}$) & $92.89\%$ \quad $0.129$ & $96.71\%$ \quad $0.096$ & $94.23\%$ \quad $0.110$ & $0.0097$ & $0.178$ & $72.95\,\mathrm{s}$\cr
    \bottomrule
    \end{tabular}
    \end{threeparttable}
\end{table*}

\subsection{Case Study 1}\label{sec:case_study1}

\begin{figure}[t]
\centering
\subfigure[]{
\includegraphics[width=0.5\columnwidth]{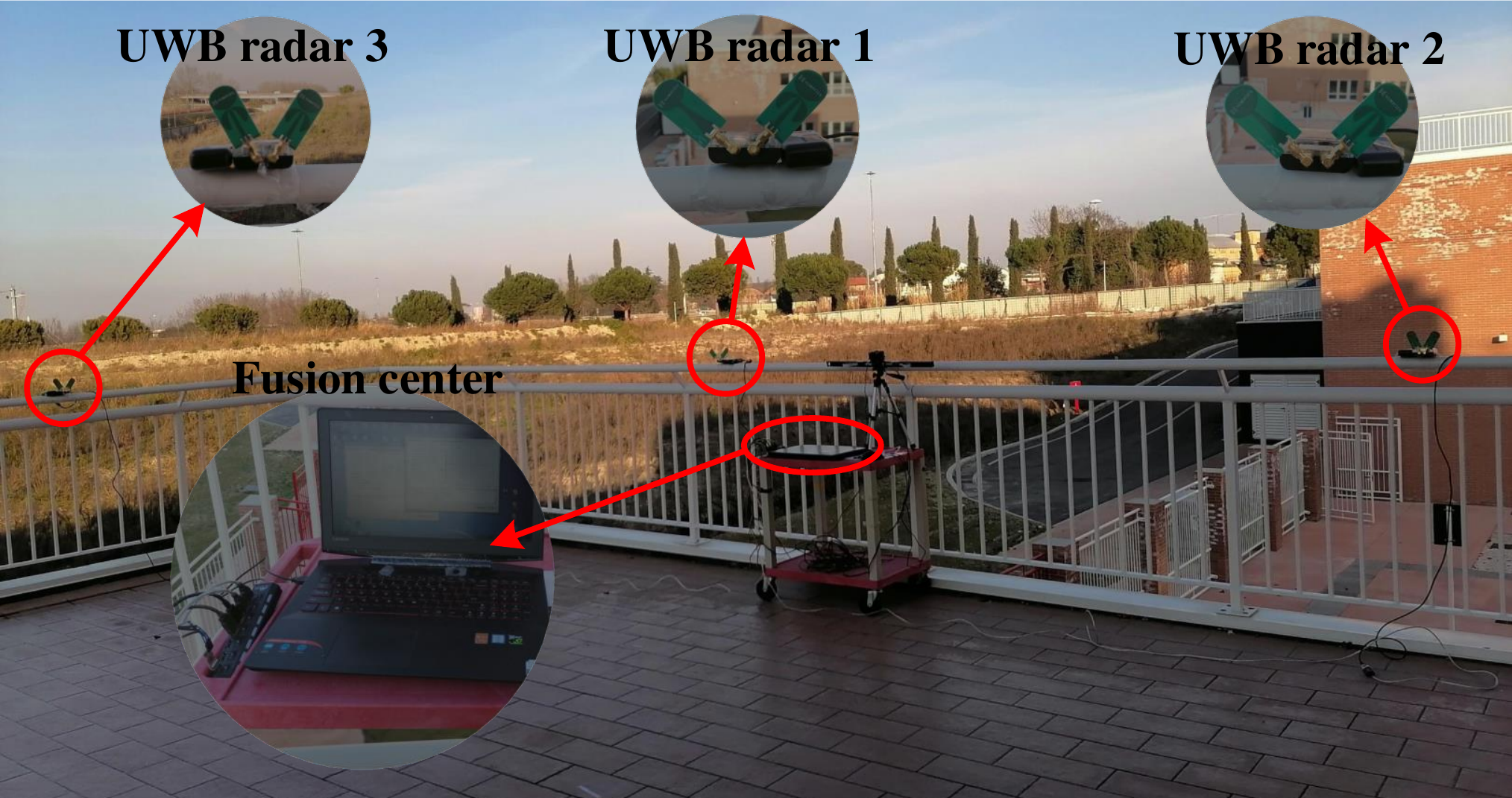}}
\subfigure[]{
\includegraphics[width=0.5\columnwidth]{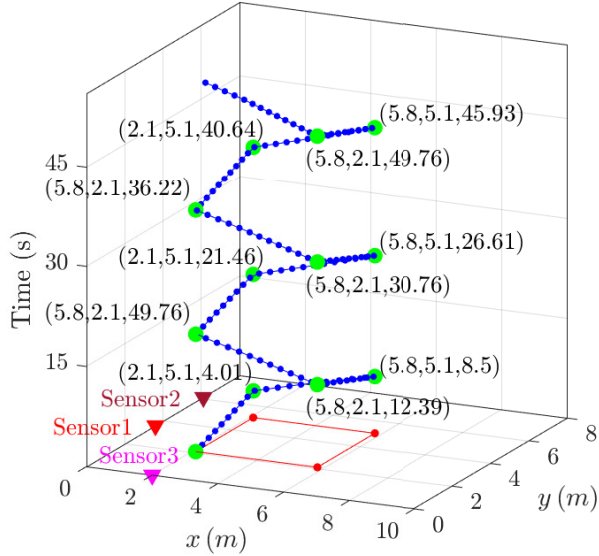}}
\caption{(a) Picture of the experimental setup for case study 1. (b) Positions of the sensors and ground truth.} \label{fig:scenario}
\end{figure}

In the first experiment, the goal was to detect a human target, of weight $80\,\mathrm{kg}$ and height $170\,\mathrm{cm}$, walking over a balcony. The experimental scenario is portrayed in Fig.~\ref{fig:scenario}(a). 
The \ac{RSN} was composed of $\numsens=3$ \ac{UWB} sensors, each configured as a monostatic radar.
Fig.~\ref{fig:scenario}(b) illustrates, on the $x\!-\!y$ plane, both the sensor positions and the target trajectory, a rectangular path of size $3.7\,\mathrm{m}\times 3\,\mathrm{m}$ covered three times in about $56\,\mathrm{s}$ with an approximately constant speed of $0.717\,\mathrm{m/s}$. 
The evolution of the target trajectory in the \ac{3D} $x\!-\!y\!-\!t$ space is also shown. 
The total number of scans is $K=124$.

\begin{figure}[]
\centering
\subfigure[]{
\includegraphics[width=0.5\columnwidth]{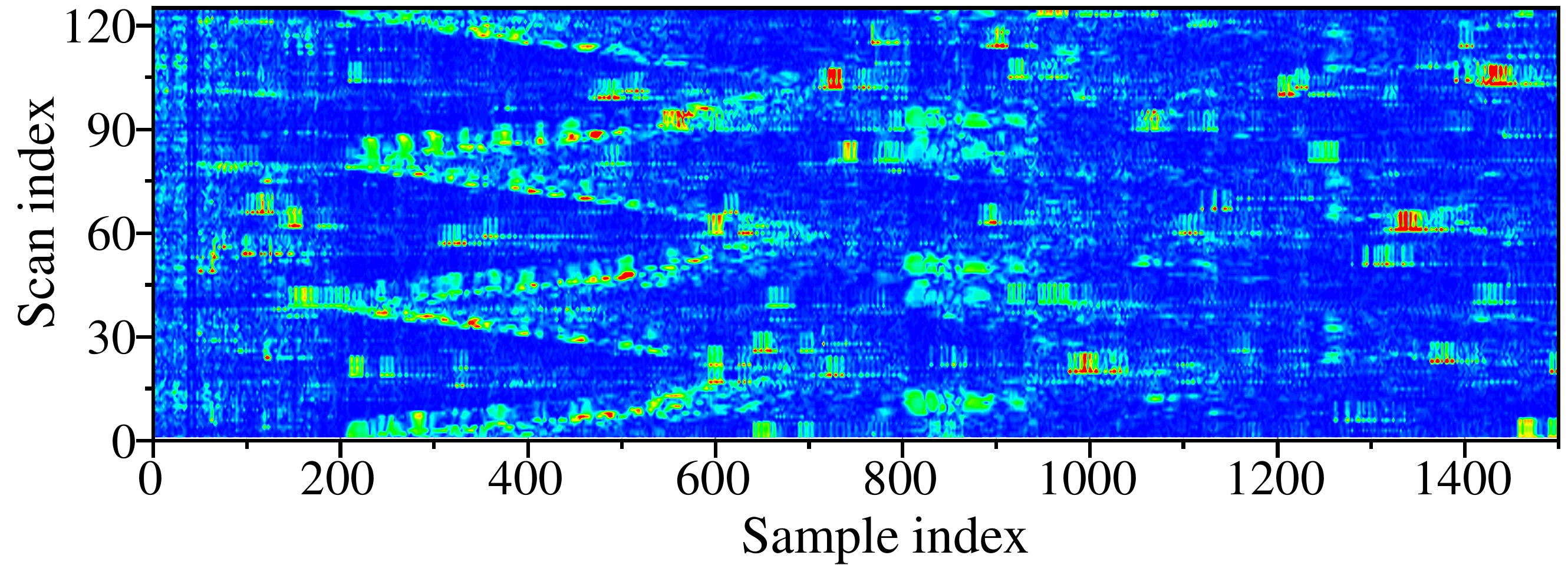}}
\subfigure[]{
\includegraphics[width=0.5\columnwidth]{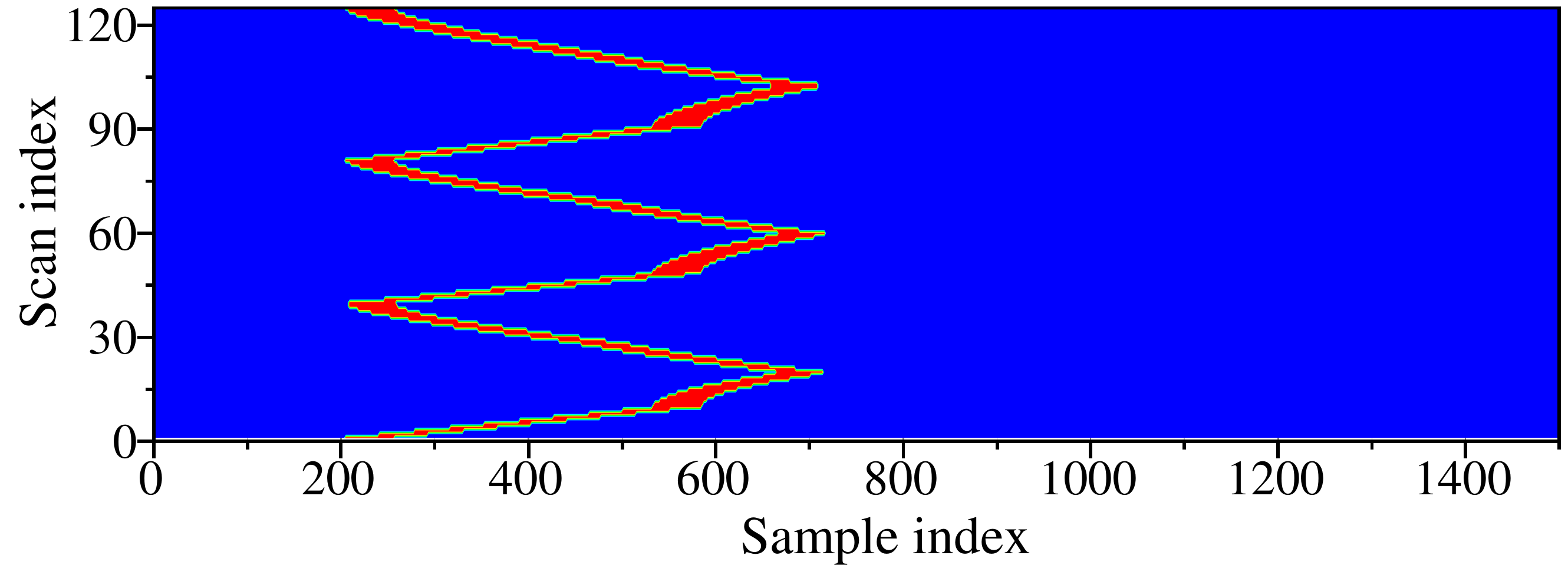}}
\quad
\caption{(a) Illustration of the signals ${\boldsymbol{m}}_n^{t}$ for $n=3$ and $1 \leq t \leq 124$. (b) Actual target \ac{ToA}.}\label{fig:case1_mea}
\end{figure}

The signals ${\boldsymbol{m}}_n^{t}$ (i.e., the result of the in-sensor processing) are represented in Fig~\ref{fig:case1_mea}(a) for the sensor $n=3$ located in position $(2,0)$. 
Each row corresponds to one of the $K=124$ scan periods. 
In each row $t$, $1 \leq t \leq 124$, all samples $m_n^t(j)$ are depicted by means of colors, where blue represents zero and red the largest sample value; every two consecutive samples correspond to a difference of $61\,\mathrm{ps}$ in terms of \ac{ToA} and of $0.91\,\mathrm{cm}$ in terms of range.
The actual target \ac{ToA} is depicted in Fig.~\ref{fig:case1_mea}(b).
It is possible to observe how signals incoming to the \ac{FC} are affected by residual clutter and some strong echoes, not due to the target but even stronger (by a factor up to $3$) than the target's ones. 
We found that these echoes originated by cars going along a large street close to the balcony where the experiment was performed. 
Very similar figures are obtained for sensors $n=1$ and $n=2$.

\begin{figure}[]
    \centering
    \subfigure[]
    {
        \includegraphics[width=0.4\columnwidth]{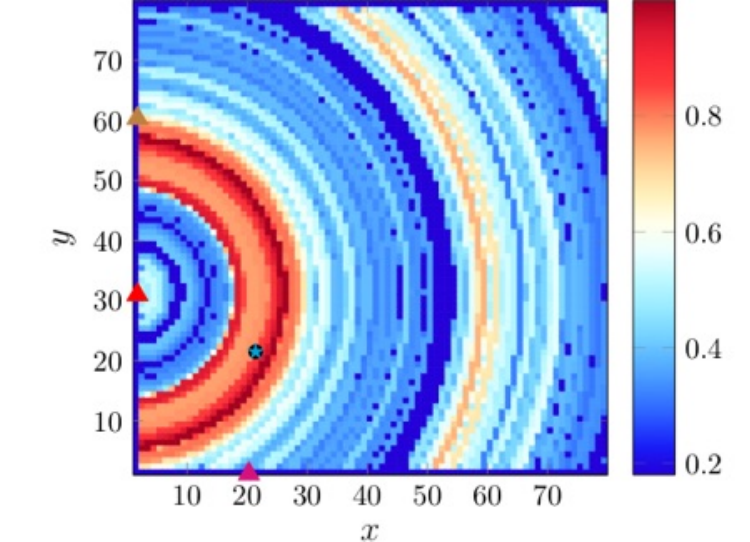}
    }
    \subfigure[]
    {
        \includegraphics[width=0.4\columnwidth]{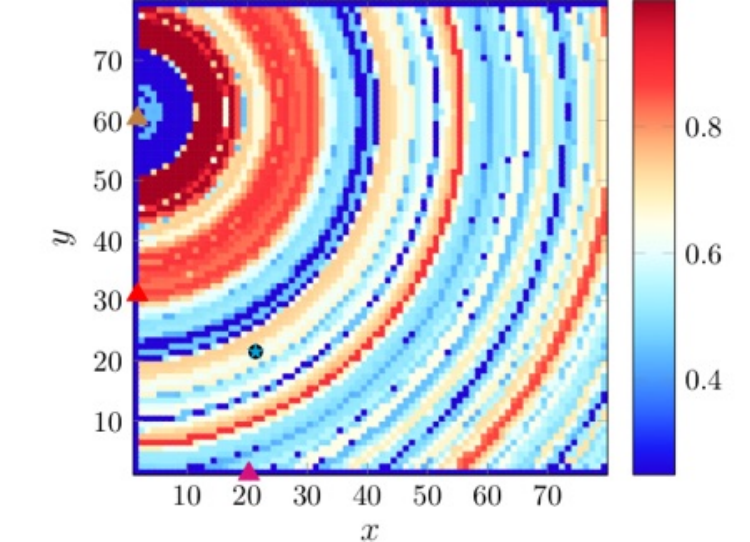}
    }
    \subfigure[]
    {
        \includegraphics[width=0.4\columnwidth]{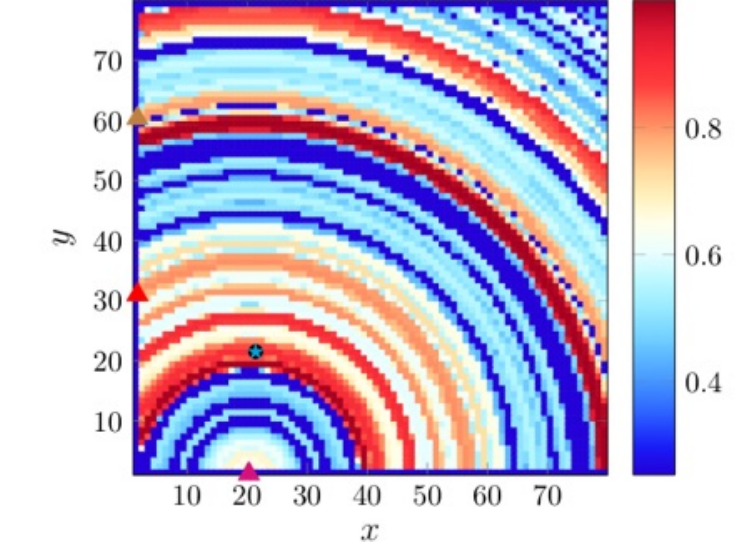}
    }
        \subfigure[]
    {
        \includegraphics[width=0.4\columnwidth]{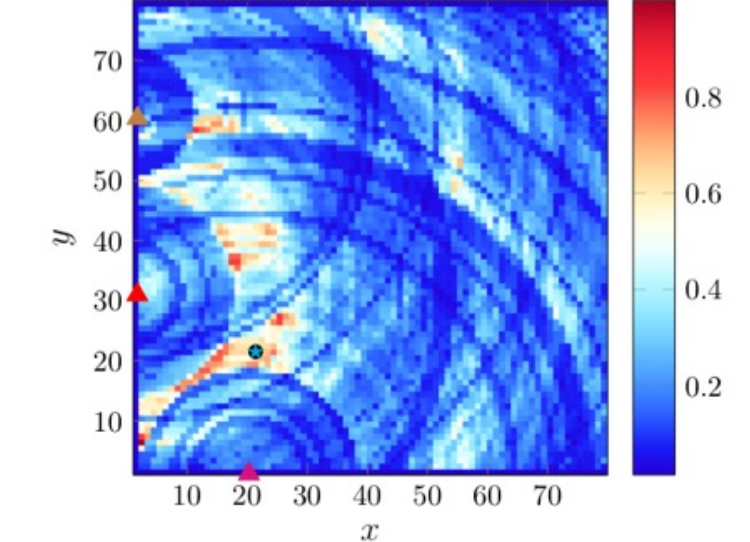}
    }
    \caption{Score maps in the first scan period ($t=1$). (a) Score map $\boldsymbol{S}_1^t$ for the sensor in position $(0,3.1\,\mathrm{m})$. (b) Score map $\boldsymbol{S}_2^t$ for the sensor in position $(0,6.1\,\mathrm{m})$. (c) Score map $\boldsymbol{S}_3^t$ for the sensor in position $(2\,\mathrm{m},0)$. (d) Overall score map $\boldsymbol{S}^t$.}
    \label{fig:case1_waves_and_colormap}
\end{figure}

\begin{figure}[!tb]
\centering
\subfigure[]{
\includegraphics[width=0.31\columnwidth]{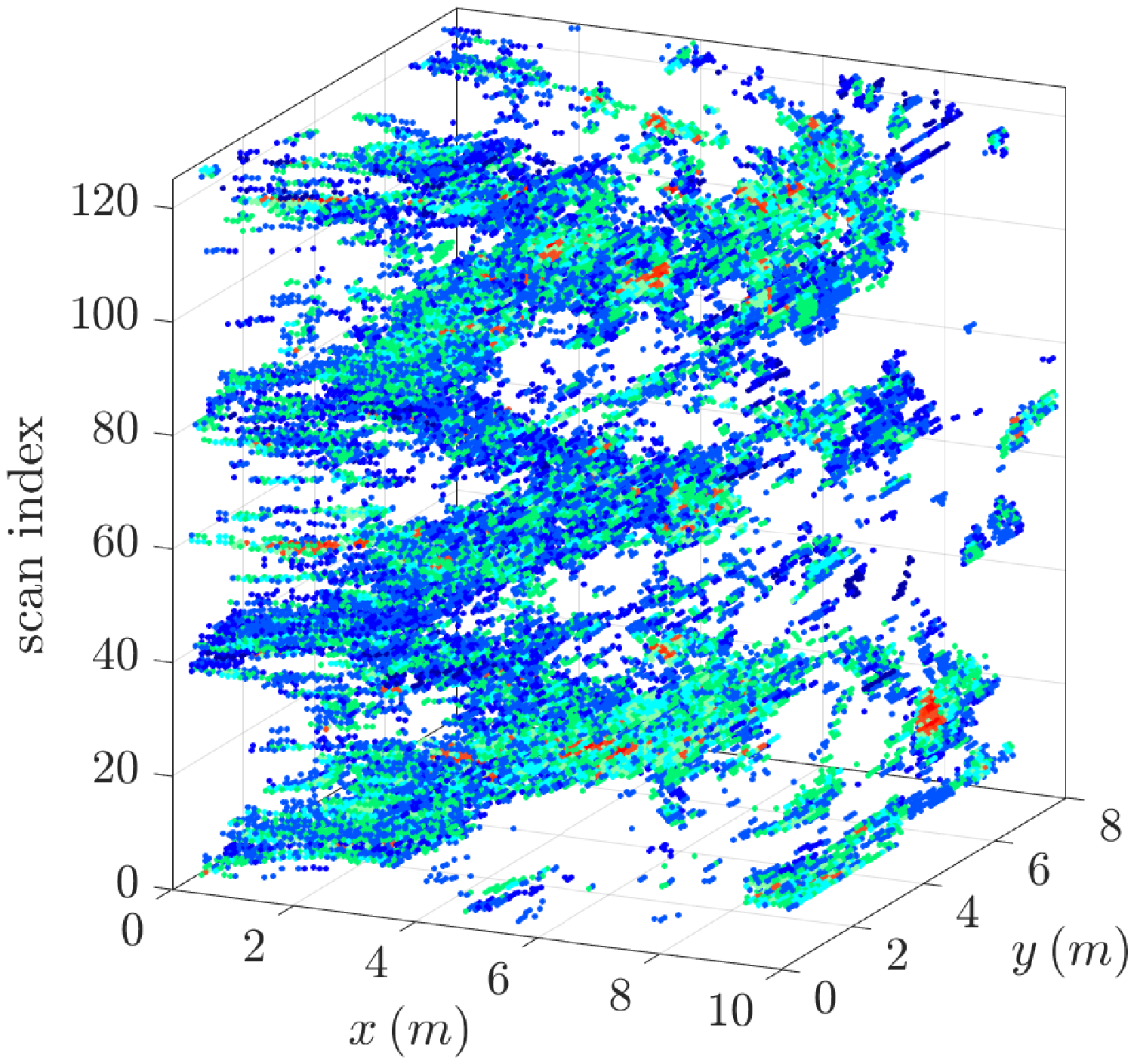}
\label{fig:6a}}
\subfigure[ ]{
\includegraphics[width=0.31\columnwidth]{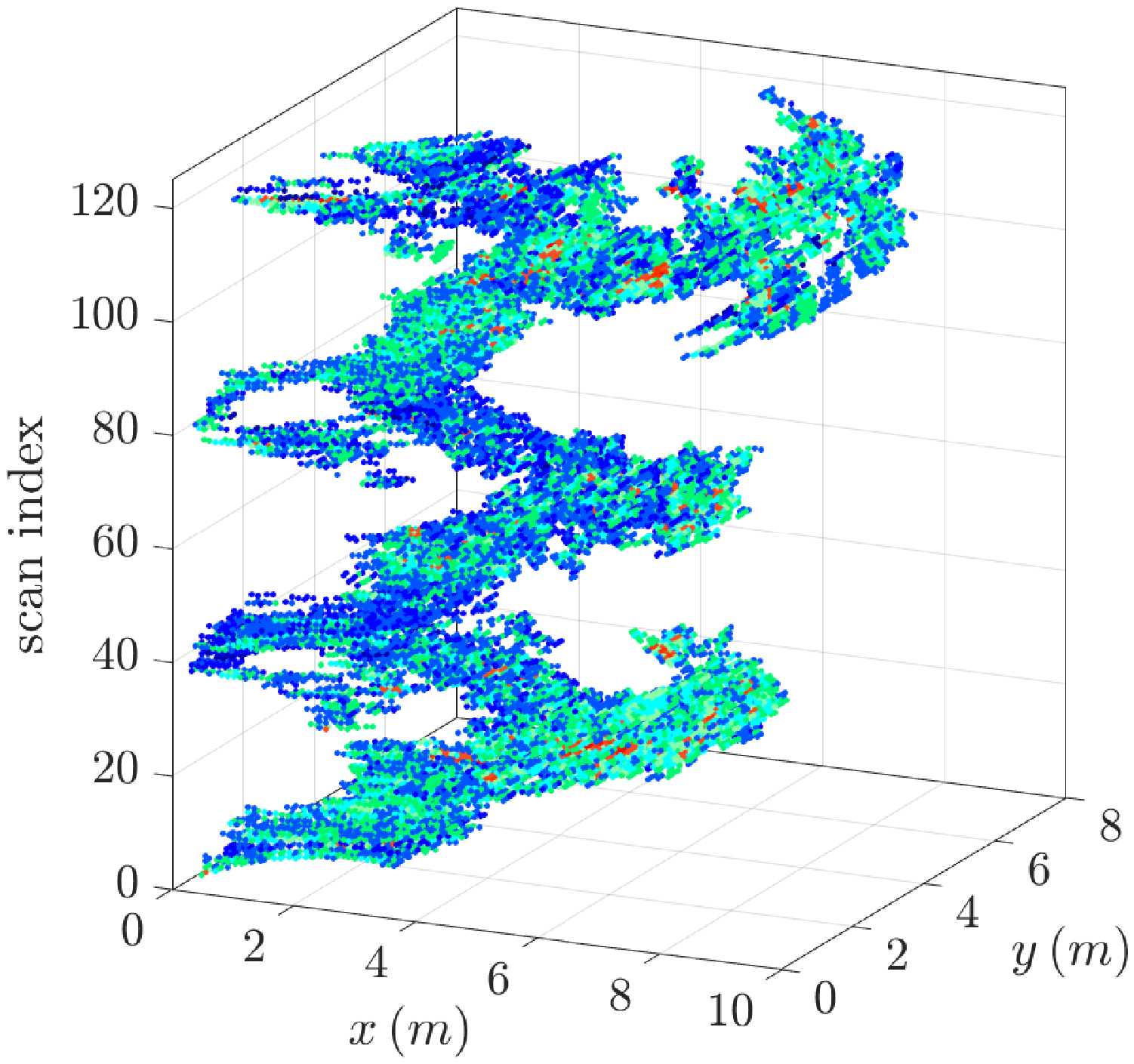}
\label{fig:6b}}
\subfigure[ ]{
\includegraphics[width=0.31\columnwidth]{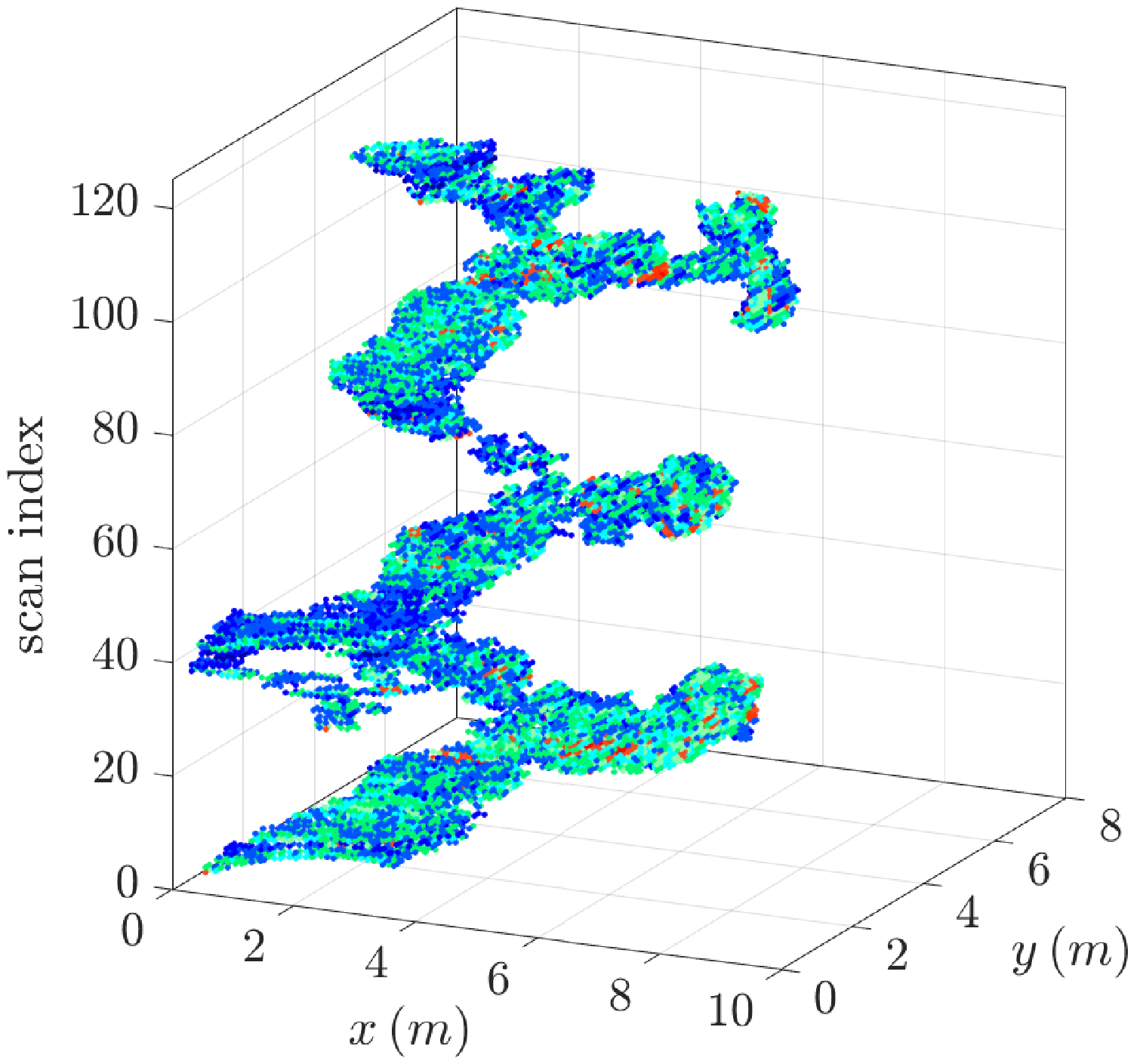}
\label{fig:6c}}
\subfigure[ ]{
\includegraphics[width=0.31\columnwidth]{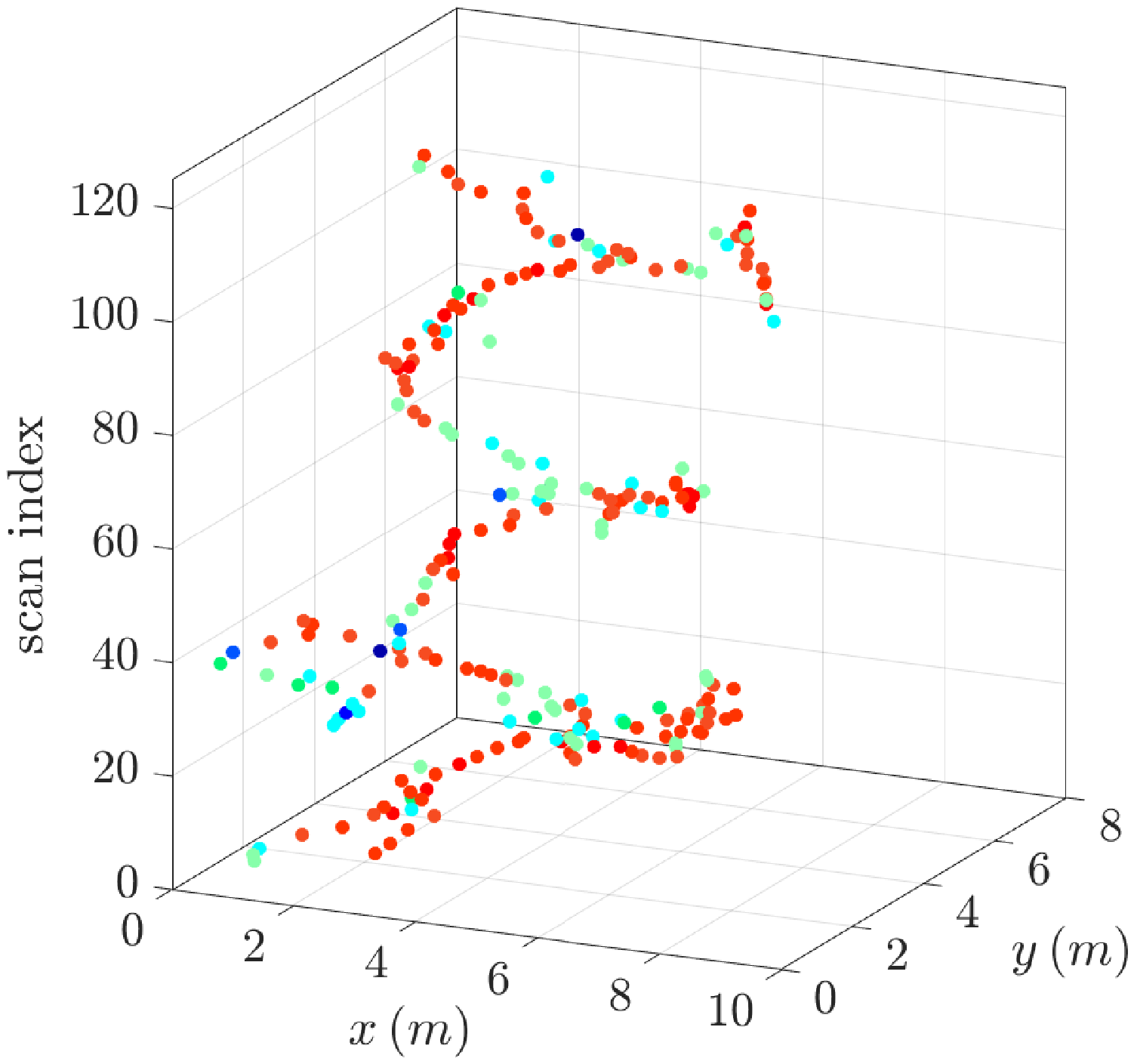}
\label{fig:6d}}
\subfigure[ ]{
\includegraphics[width=0.31\columnwidth]{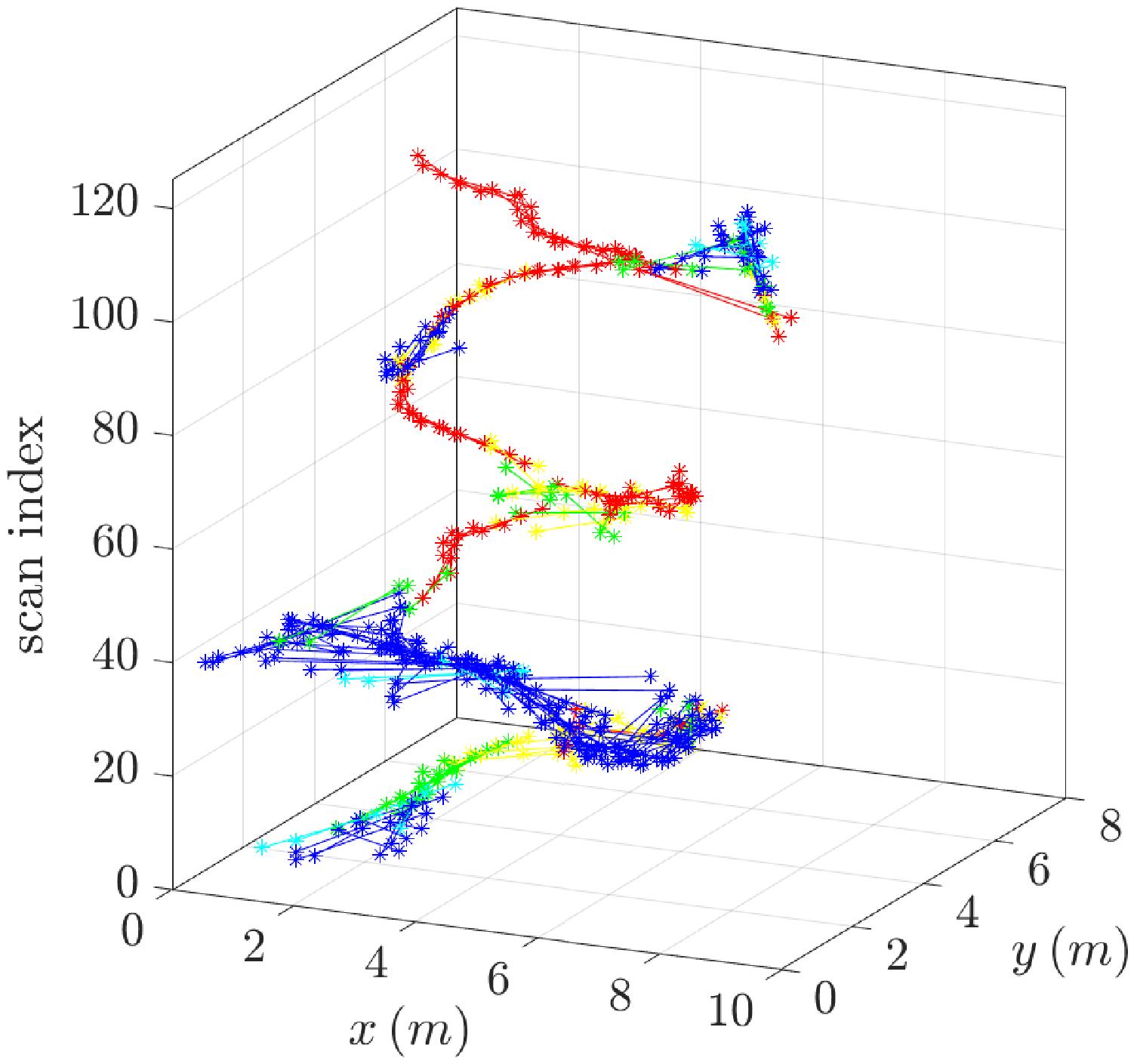}
\label{fig:6e}}
\subfigure[ ]{
\includegraphics[width=0.31\columnwidth]{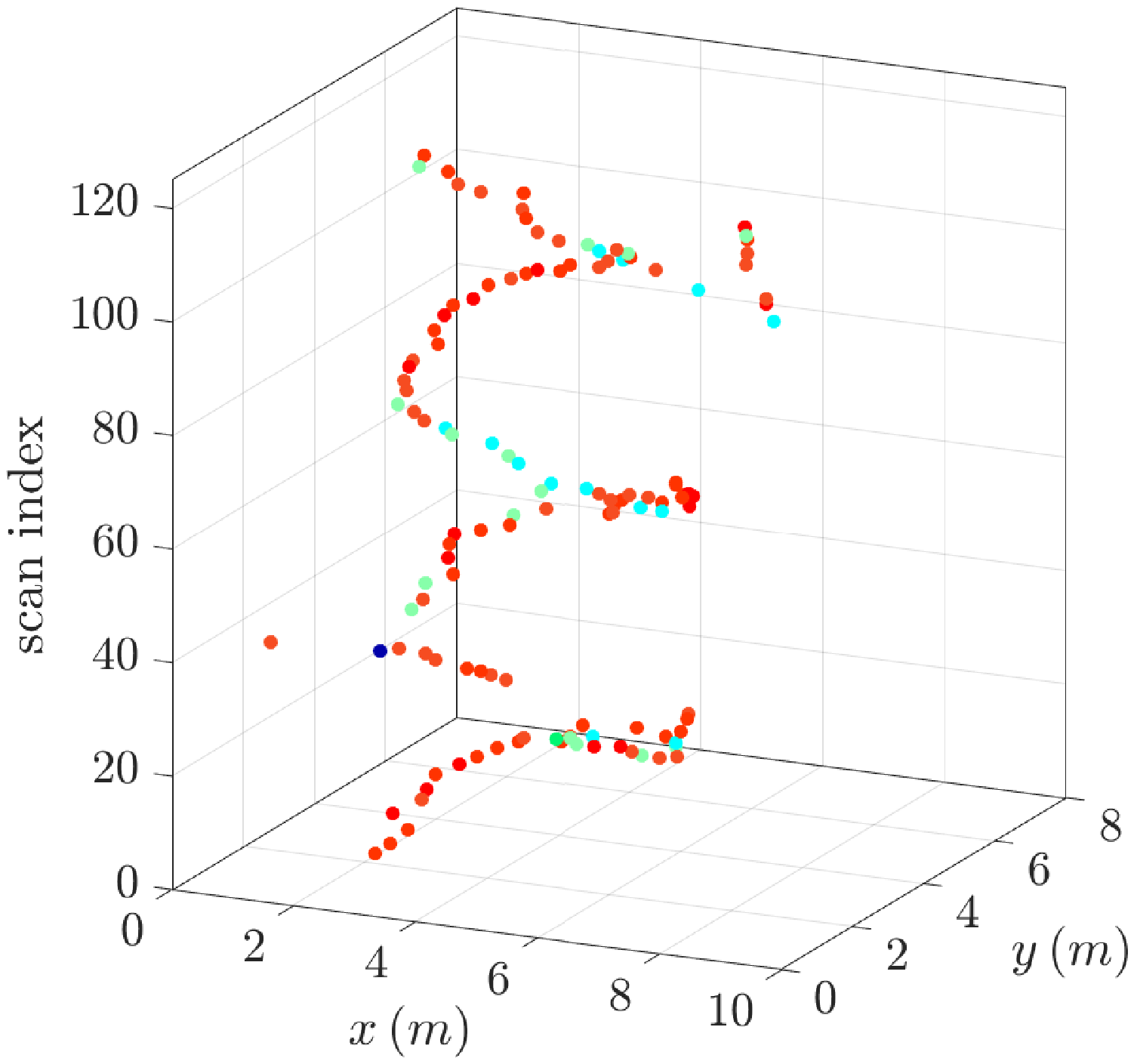}
\label{fig:6f}}
\caption{(a) Example of 3D data structure applying voting, i.e., \eqref{eq:M_to_S} only. Red and blue colors denote the largest and the smallest score, respectively. (b) The same data structure obtained by applying \ac{3D} region growing. (c) The same data structure obtained by applying also \ac{3D} opening operation. (d) The points obtained with the proposed generation method, and (e) the corresponding tracklets. (f) Trajectory points after tracklet association (before outlier removal and trajectory smoothing).}
\label{fig:cloud_point}
\end{figure}

Fig.~\ref{fig:current_scan} shows the original samples $m_n^t(j)$, $j=1,\dots,N_\text{c}$, generated by one of the sensors along with the corresponding peak-suppressed samples, $(m_n^t(j))^{\alpha}$,  with $\alpha=0.75$. 
The range bin where the target exists is marked by a thick black line and it reveals how the amplitude of target echoes may not be the largest, although it is larger than that of most non-target range bins.

An example of score maps $\boldsymbol{S}_1^t$, $\boldsymbol{S}_2^t$, and $\boldsymbol{S}_3^t$, one for each receiver, is illustrated through color maps in Fig.~\ref{fig:case1_waves_and_colormap}(a), Fig.~\ref{fig:case1_waves_and_colormap}(b), and Fig.~\ref{fig:case1_waves_and_colormap}(c), respectively. 
The sensor positions are marked with triangles, the actual position of the target is indicated by a cyan star and, in each map, all scores are normalized to the peak one for visualization purposes. These score maps are the ones obtained during the first scan period ($t=1$). The corresponding overall (normalized) score map, obtained via \eqref{eq:overall_score_map} and \eqref{eq:M_to_S}, is shown in Fig.~\ref{fig:case1_waves_and_colormap}(d). Note that the overall score map predicts a potential target in the correct position.
However, the presence of a relatively-strong residual clutter prevents from performing a correct detection in a single scan, leading to a considerable number of false alarms for a low detection threshold or to excessive misdetections of weak targets if the detection threshold is increased. In a classical track-after-detect configuration, this makes the tracking filter task very problematic, which justifies our \ac{TBD} approach.

Fig.~\ref{fig:cloud_point}(a), \ref{fig:cloud_point}(b), and \ref{fig:cloud_point}(c) show the result of the \ac{3D} processing proposed in Section~\ref{subsec:3d}. 
In particular, Fig.~\ref{fig:cloud_point}(a) illustrates how the \ac{3D} data structure $\boldsymbol{M}^t$ would appear in the last scan period ($t=124$) by simply stacking score maps $S(i_x,i_y)$ obtained by \eqref{eq:M_to_S} and \eqref{eq:etat_threshold} (cells with a null score are left blank). 
Several isolated cells with a nonzero score, originated by noise and clutter, can be observed. 
Fig.~\ref{fig:cloud_point}(b) shows the result of application of region growing only with $W=4$ and $s=2$. Only one region was found in all time windows ($R^t=1$ for all $t = W + q \cdot s$, $q=0,\dots,59$). As it is possible to appreciate comparing Fig.~\ref{fig:cloud_point}(b) with Fig.~\ref{fig:cloud_point}(a), most isolated cells are now removed and the generated \ac{3D} region resembles the ground truth depicted in Fig.~\ref{fig:scenario}(b). An even closer match with the ground truth is obtained after application of \ac{3D} opening operation, as depicted in Fig.~\ref{fig:cloud_point}(c). The points obtained in each scan through the proposed point generation method (Section~\ref{subsec:detection}) are depicted in Fig~\ref{fig:cloud_point}(d). 
These points are processed in each time window by the employed tracklet generation algorithm (Section~\ref{subsec:tracklet_gen}). 
In each time window, we employed the tracklet generation by enumeration. 
The non-discarded detected tracklets are presented in Fig~\ref{fig:cloud_point}(e), where the red color indicates that only one tracklet was generated in the time window and, on the opposite side, the blue color indicates that $5$ tracklets (the largest number) were generated.
Fig~\ref{fig:cloud_point}(f) shows the unique point obtained in each scan after tracklet association. 
Compared with the Fig~\ref{fig:cloud_point}(d), we can see how several false alarm points have been removed.

\begin{figure}[tbp]
\centering
{\includegraphics[width=0.4\columnwidth]{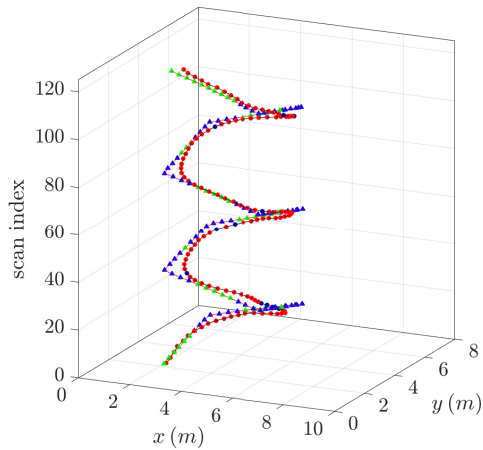}}
\caption{The ground truth trajectory (triangles) and the smoothed one (circles). Green triangles indicate straight portions of the ground truth; blue triangles
indicate a turning target.}\label{fig:case1_final_track}
\end{figure}

The smoothed \ac{3D} trajectory after outlier removal and smoothing is presented in Fig.~\ref{fig:case1_final_track} (red circles) together with the ground truth (green and blue triangles). 
Comparing it with the points in Fig.~\ref{fig:cloud_point}(f), we can see how the positional error is decreased after smoothing. 
The projection of the smoothed trajectory on the $x\!-\!y$ plane has an ``oval'' shape as an effect of smoothing, while the actual one is rectangular. 
Therefore, positioning errors close to the corners tend to be larger.

\begin{figure}[t]
\centering
\includegraphics[width=0.45\columnwidth]{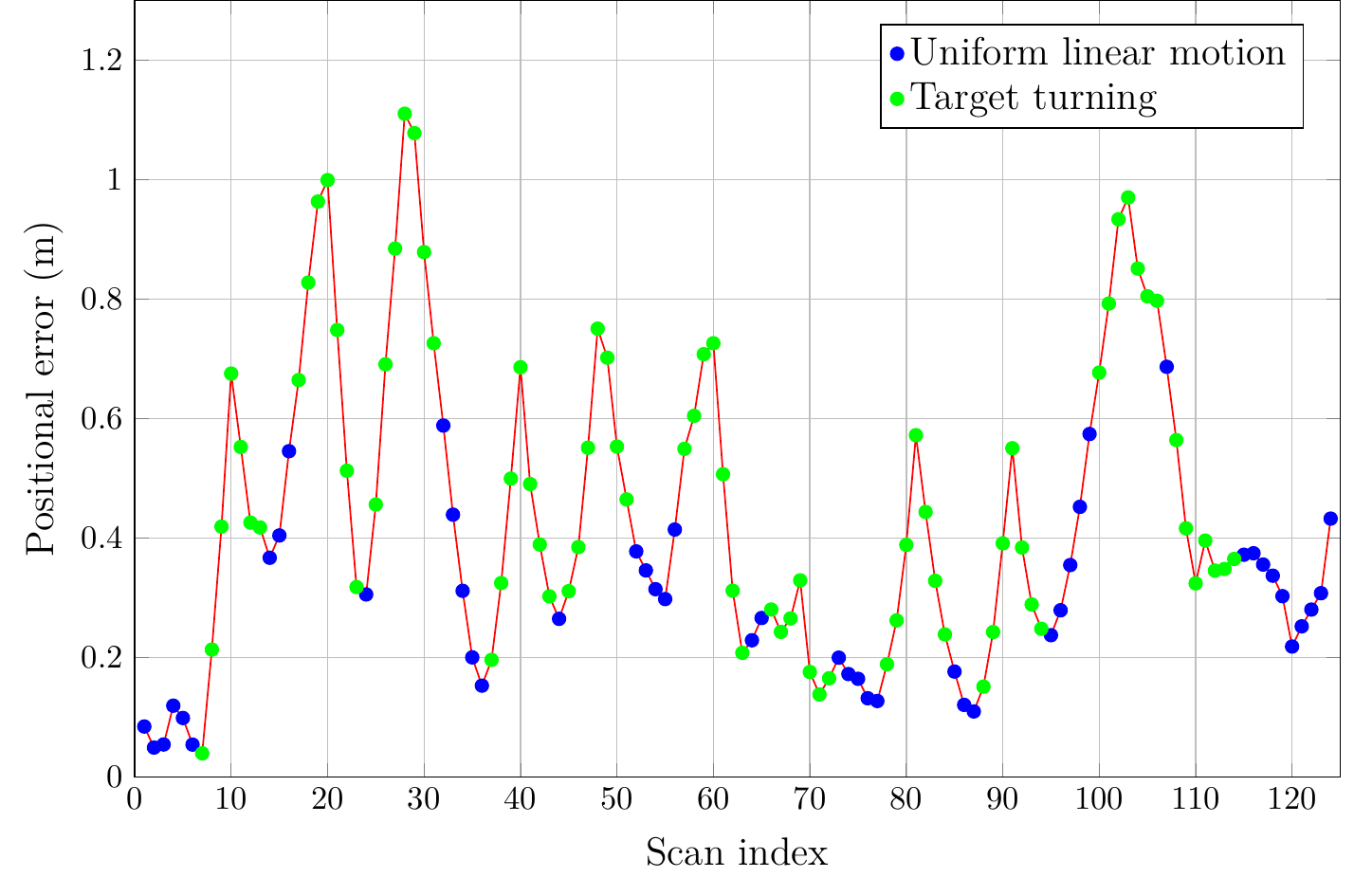} 
\caption{The positional error of the smoothed trajectory among the 124 scans.}
\label{fig:Case1_result_error}
\end{figure}

To account for both cardinality and localization errors, the tracking performance was assessed by \ac{OSPA}  distance \cite{ristic2010performance}.
Since the trajectory was detected in all $124$ scans without false alarms, the \ac{OSPA} distance corresponds to the positioning error (distance between the actual target position and the estimated one); this is shown in Fig.~\ref{fig:Case1_result_error} as a function of the scan index, where green and blue markers correspond to scan periods where the target is turning and where it moves with uniform linear motion, respectively.
The average positioning error over all scans is $0.419\,\mathrm{m}$. As expected, the performance is better during uniform linear motion (blue points), where the average error is $0.283\,\mathrm{m}$, than during target turning (green points) where it reaches $0.502\,\mathrm{m}$. Similar with the \ac{OSPA} distance in Fig.~\ref{fig:Simu_OSPA}, it is irreparable because of the mismatching between  smooth model and actual target motion state when the target is maneuvering.
However, the overall performance is remarkably good, considering that the average error is comparable with the target transverse size.

The effectiveness of the proposed points generation method (Section~\ref{subsec:detection}) followed by tracklet detection and association can be observed by repeating the processing with the very same data and turning all of these steps off. 
Specifically, instead of considering layer $k$ of the region $\boldsymbol{A}_i^t$, finding clusters $U_{i,j}^k$ out of it, extracting one point per cluster, and applying tracklet detection and association to them, we can imagine to extract one point from layer $k$ of $\boldsymbol{A}_i^t$, in particular its centroid, and to feed the smoothing filter with these points directly. 
This latter technique is here referred to as the ``simple method''.
Doing this, we would obtain an average positional error of $0.535\,\mathrm{m}$, slightly worse than the one of the proposed technique after points generation, equal to $0.514\,\mathrm{m}$.
The gap increases substantially after smoothing and outlier removal, which are almost ineffective on the points generated by the simple method. 
The overall average positioning error of smoothed trajectory is $0.499\,\mathrm{m}$; in particular, it equals $0.526\,\mathrm{m}$ during target turning and $0.455\,\mathrm{m}$ during uniform linear motion.
The superiority of the proposed approach is mainly due to clutter cells being entirely removed from the calculation of the target position. The comparison is summarized in Table~\ref{tab:performance_comparison2}.

\renewcommand{\arraystretch}{1}
\begin{table}[t]
  \centering
  \fontsize{8.5}{8}\selectfont
  \caption{Average positioning error.}
    \begin{tabular}{p{2.9cm}<{\centering} p{2.5cm}<{\centering} p{2.5cm}<{\centering} p{2.5cm}<{\centering} p{2.5cm}<{\centering}}
   \toprule 
   $\,$ & After point generation & After trajectory smoothing & Target turning & Uniform linear motion\\
    \midrule
    Simple method & $0.535$ & $0.499$ & $0.526$ & $0.455$\\
    Proposed method &$0.514$ & $0.419$ & $0.502$ & $0.283$\\
    \bottomrule
    \end{tabular}
    \label{tab:performance_comparison2}
\end{table}

\begin{figure}[!tb]
\centering
\subfigure[]{
\includegraphics[width=0.45\columnwidth]{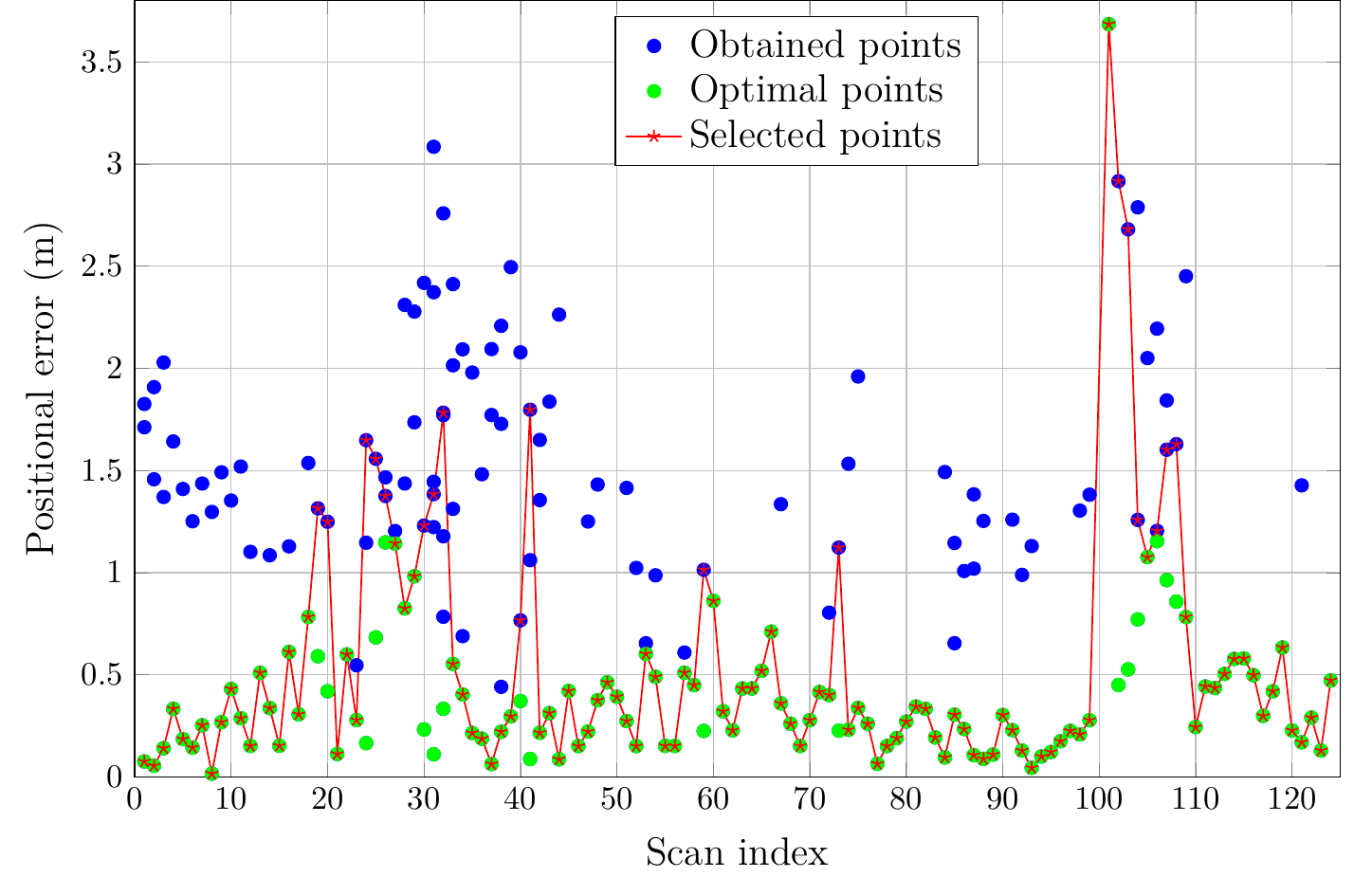}}
\quad
\subfigure[]{
\includegraphics[width=0.45\columnwidth]{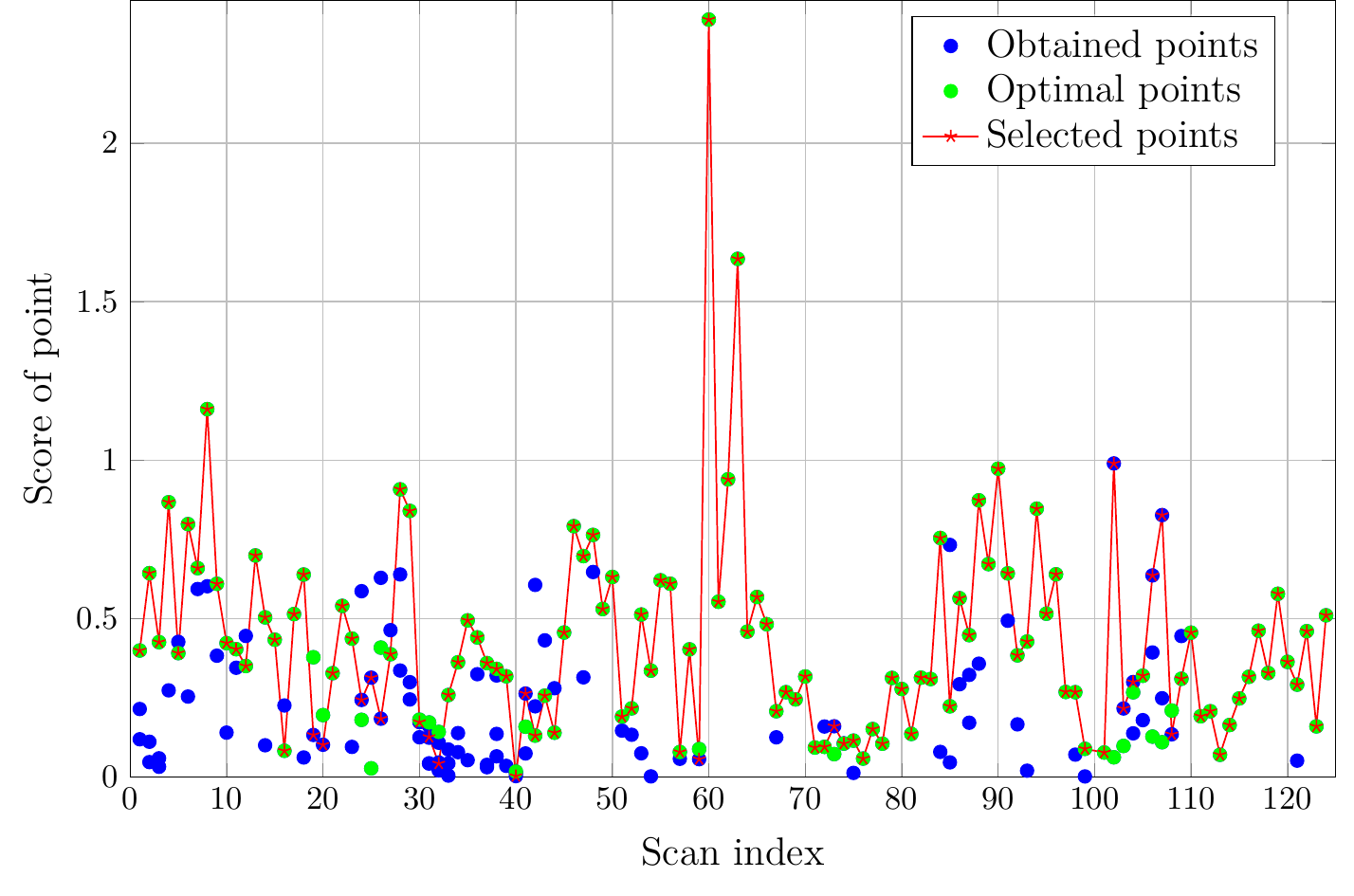}}
\quad
\caption{(a) The positional error of the points before trajectory smooth. (b) The score of the points obtained in target detection.  }\label{fig:Case1_result_analysis}
\end{figure}

To provide insights about the tracklet generation and association steps (Section~\ref{subsec:tracklet_gen} and Section~\ref{subsec:tracklets_assoc}), in Fig.~\ref{fig:Case1_result_analysis}(a) we show the positioning error of all points generated through all scan periods (blue and green markers -- recall from Section~\ref{subsec:detection} that multiple points per scan may be generated).
For each scan index, green markers are associated with the optimal (least error) point and blue markers with the other points.
The average positioning error of all generated points (green and blue) and optimal points (green only) are $0.909\,\mathrm{m}$ and $0.387\,\mathrm{m}$, respectively. 
The selected points, corresponding to the points in Fig.~\ref{fig:cloud_point}(f), are marked with a red star in the figure.
Out of all $124$ scan periods, the optimal point is selected as a trajectory points $105$ times, with an average positional error of the selected points as low as $0.535\,\mathrm{m}$ which reveals the benefits provided by tracklets generation and association in terms of localization accuracy. The score of the points generated through all scans is presented in Fig.~\ref{fig:Case1_result_analysis}(b). As we can see, in most scans the optimal points (green marks) exhibit the largest score and in the majority of scans the points with the largest score are the ones selected as the target points. 
This shows how the adopted score is beneficial in terms of trajectory detection.

The overall results of Case Study 1 are presented in Table~\ref{tab:Res_Case1}. The detection rate of the proposed method is remarkably higher than that of the other two methods. The detection rate of points generation \cite{chiani2018sensor} followed by the \ac{TBD} method of \cite{YAN2021107821} is slightly lower than the one achieved by the same points generation technique followed by \ac{PHD} filtering, because the \ac{TBD} technique in \cite{YAN2021107821} is tailored to straight-line tracklets. The number of false alarms with \ac{PHD} filtering is higher because false tracks can be suppressed by multi-scan detection. According to these results, the proposed method can outperform the two competing methods within a monostatic \ac{RSN} configuration.

\renewcommand{\arraystretch}{1}
\begin{table}[t]
  \centering
  \fontsize{8.5}{8}\selectfont
  \begin{threeparttable}
  \caption{Results of Case Study 1}
  \label{tab:Res_Case1}
     \begin{tabular}{p{4.4cm}<{\centering} p{3cm}<{\centering}p{3cm}<{\centering}p{2cm}<{\centering}p{1cm}}
    \toprule
     & $P_d^1$ \& $E_p^1$ 
     & $N_{\text{FA}}$
     & \ac{OSPA}
\cr
    \midrule% 
Proposed method ($\Delta$=0.1m)&124/124 \& 0.419 &0&0.419\cr
    Point generation \cite{chiani2018sensor} + \ac{TBD} \cite{YAN2021107821} &47/124\& 0.591&3/124 & 0.659\cr
    Point generation \cite{chiani2018sensor} + \ac{PHD} filter \cite{granstrom2012extended} &59/124 \& 0.583&19/124&0.652\cr
    \bottomrule
    \end{tabular}
    \end{threeparttable}
\end{table}
\begin{figure}[!tb]
\centering
\includegraphics[width=0.5\columnwidth]{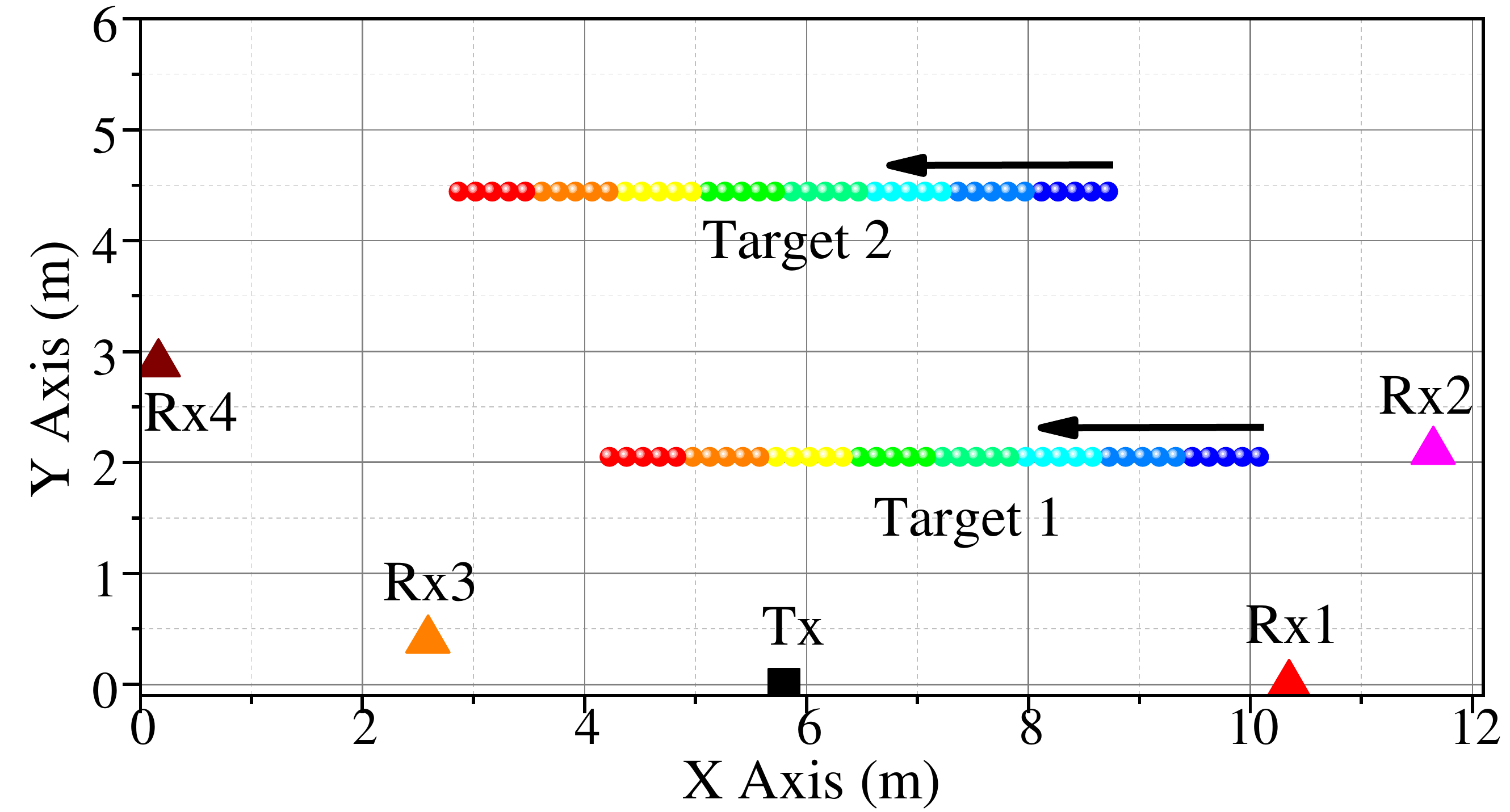} 
\caption{Case study 2 scenario. Tx: \ac{UWB} transmitter. Rx: \ac{UWB} sensor.}
\label{fig:Case2_scenario}
\end{figure}

%[width=0.6\columnwidth
\begin{figure}[!tb]
    \centering
    \includegraphics[width=0.5\columnwidth]{{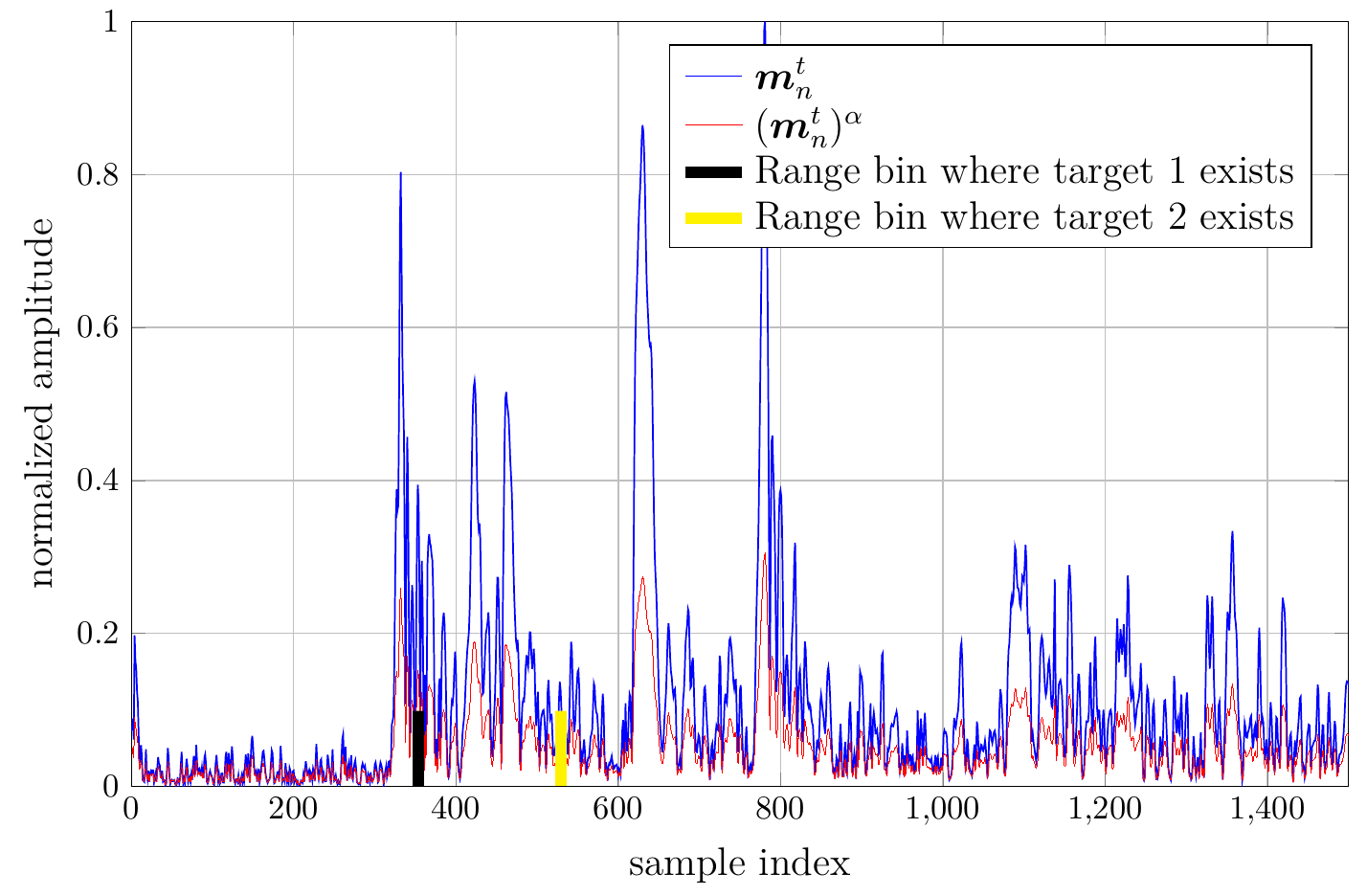}}
    \caption{Example of signal $\boldsymbol{m}_n^t$ collected during the second experiment.}
    \label{fig:my_label}
\end{figure}

\begin{figure}[!tb]
\centering
\subfigure[]{
\includegraphics[width=6cm]{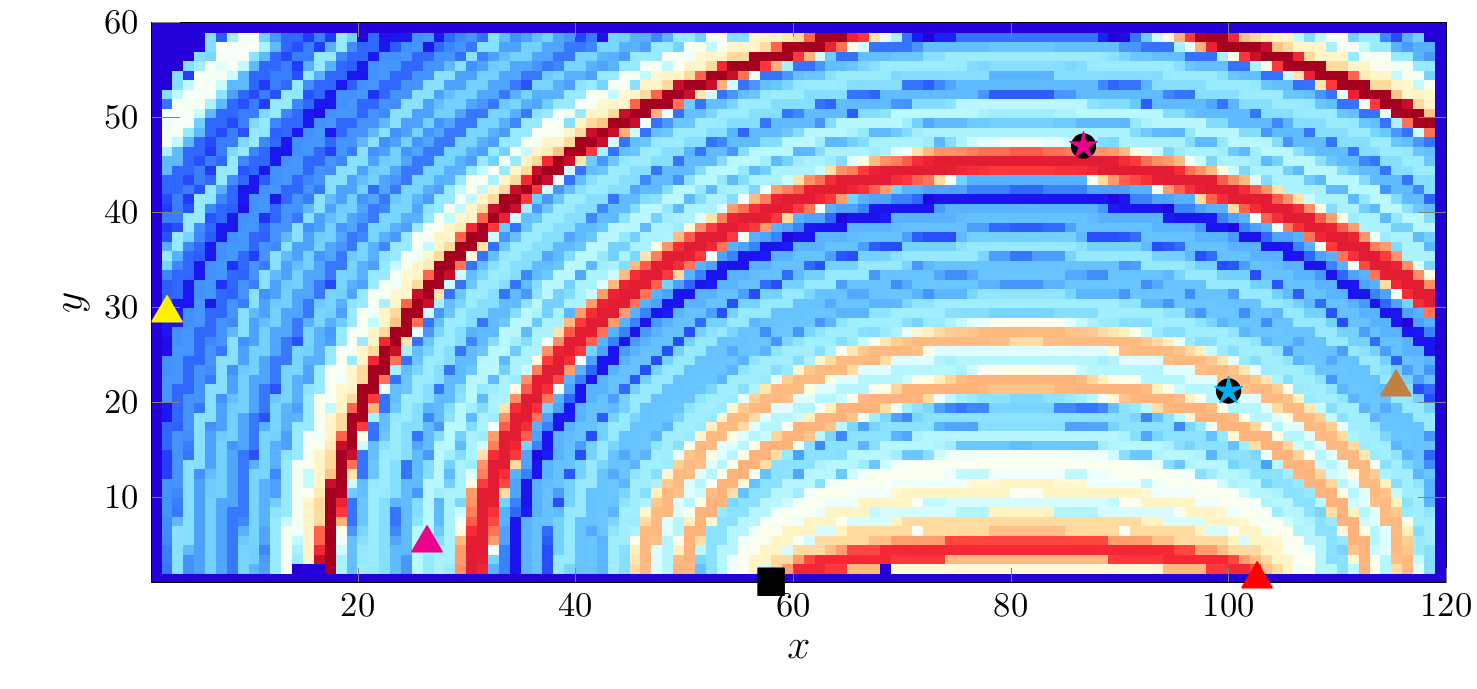}}
\quad
\subfigure[]{
\includegraphics[width=6cm]{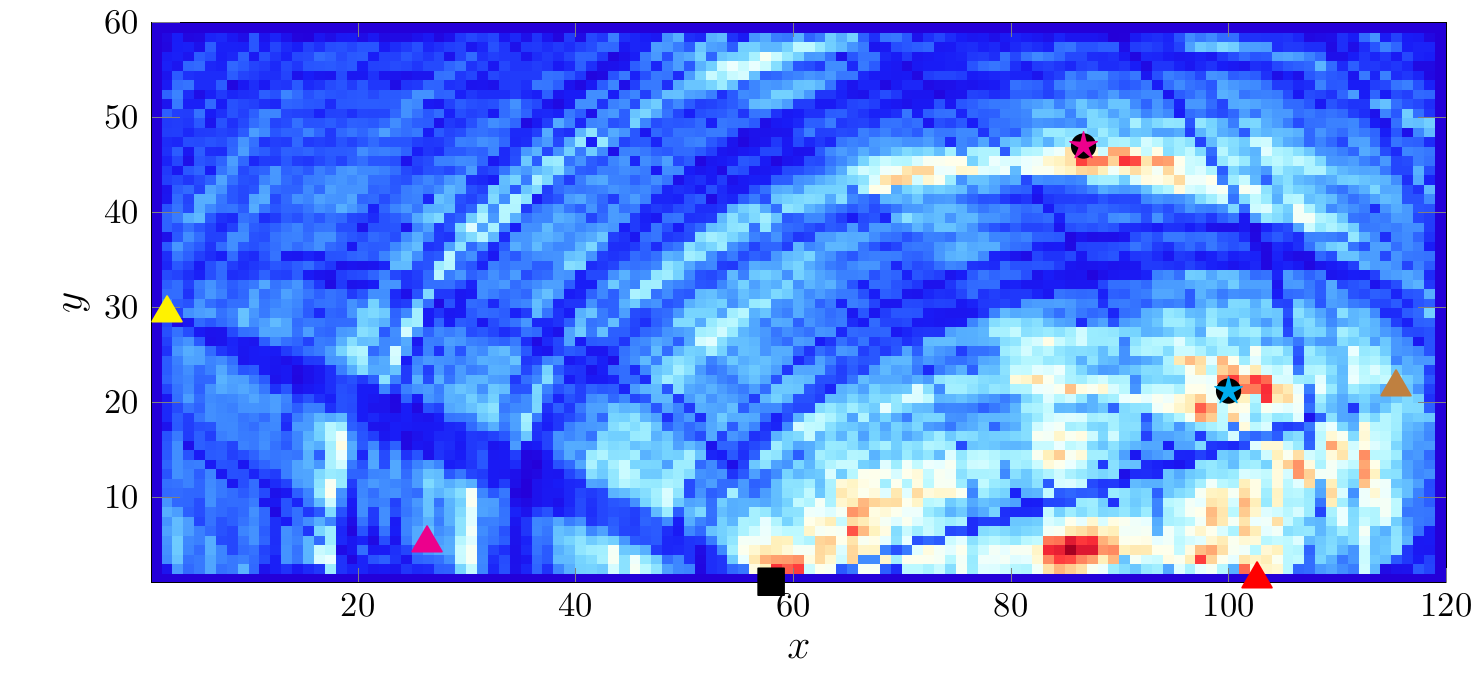}}
\quad
\subfigure[]{
\includegraphics[width=6cm]{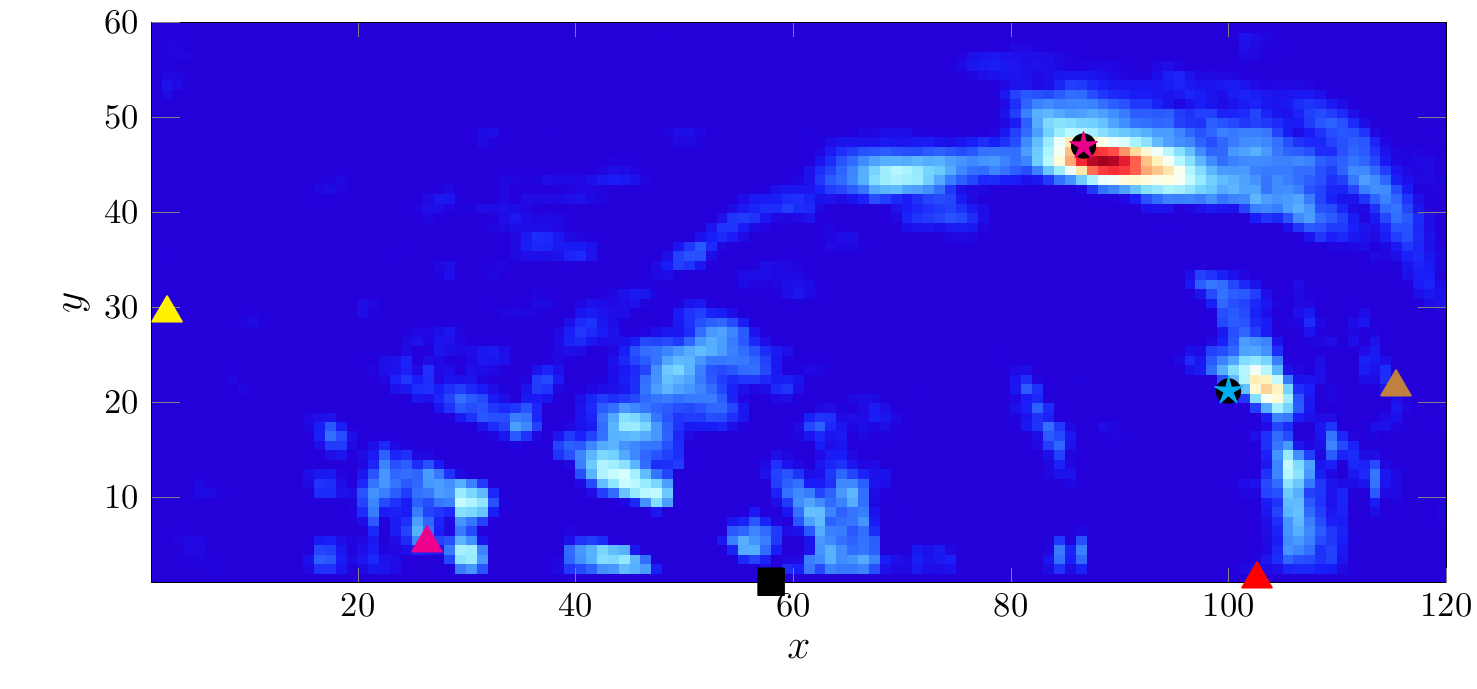}}
\caption{(a) Example of score map $\bm{S}_n^t$. (b) The overall score map $\bm{S}^t$ before clutter suppression. (c) The overall score map $\hat{\bm{S}}^t$ after clutter suppression.}\label{fig:Case2_map_sensor}
\end{figure}

\begin{figure}[!t]
\centering
\subfigure[]{
\includegraphics[width=0.30\columnwidth]{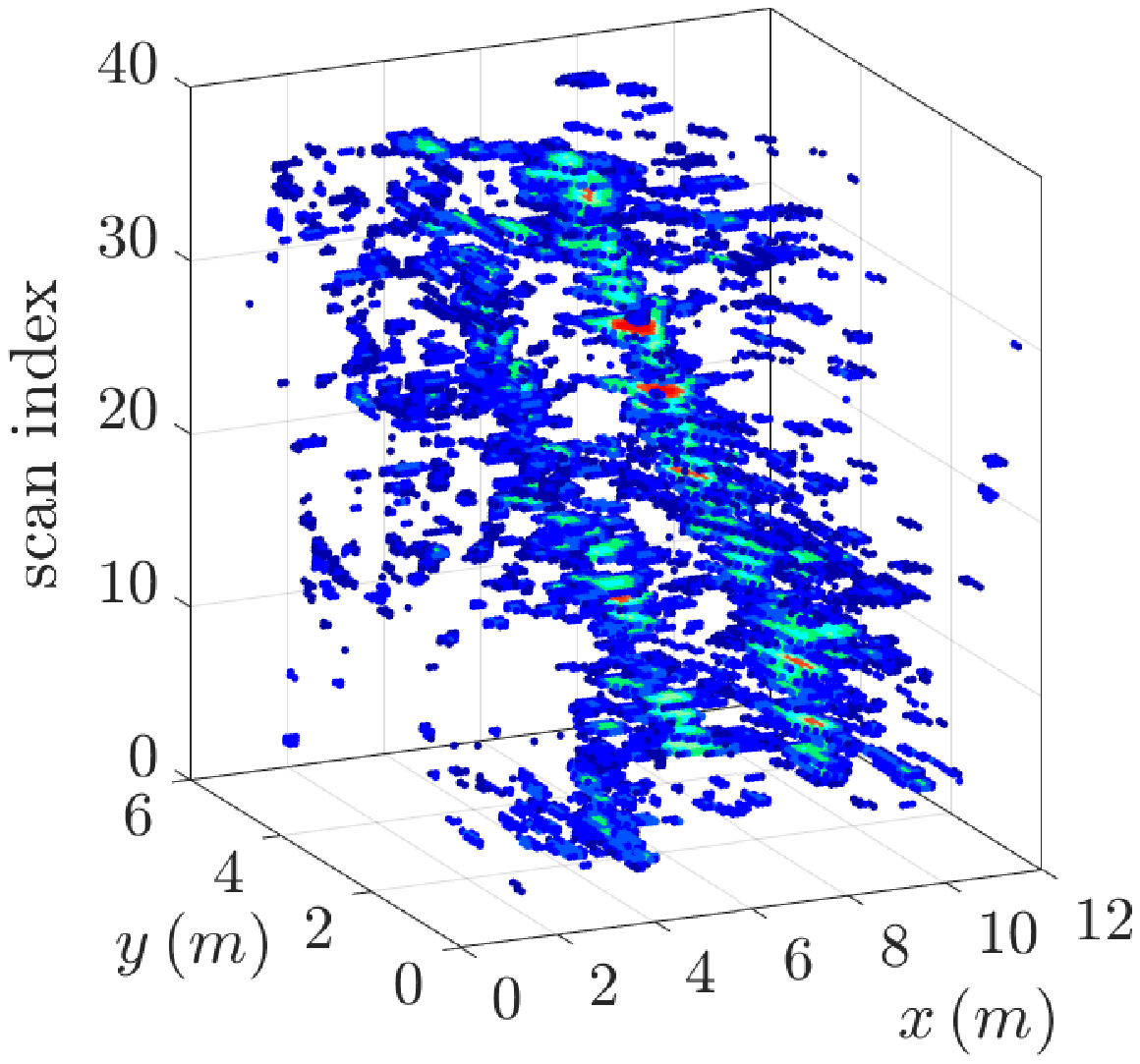}}
\quad
\subfigure[]{
\includegraphics[width=0.30\columnwidth]{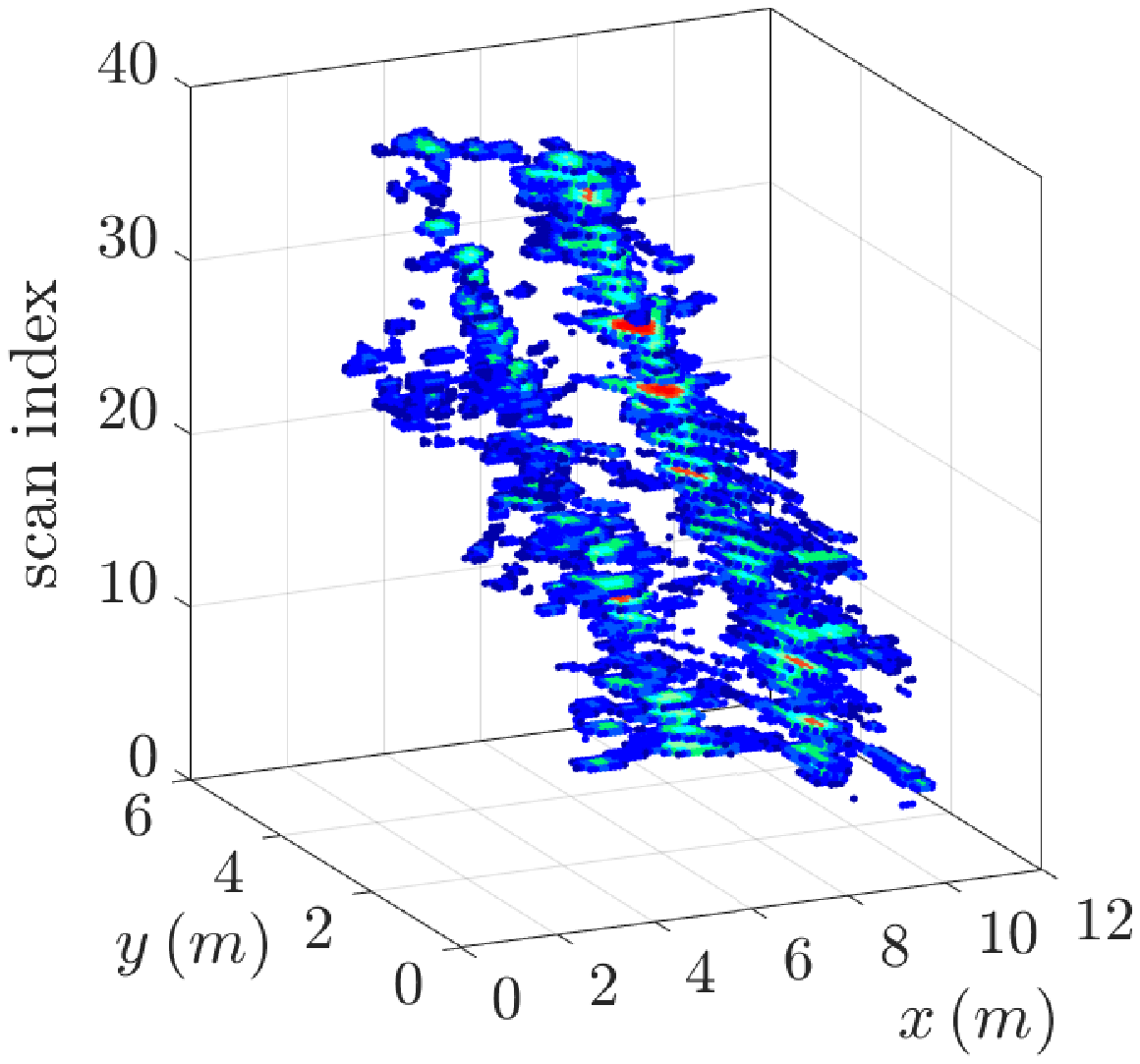}}
\quad
\subfigure[]{
\includegraphics[width=0.30\columnwidth]{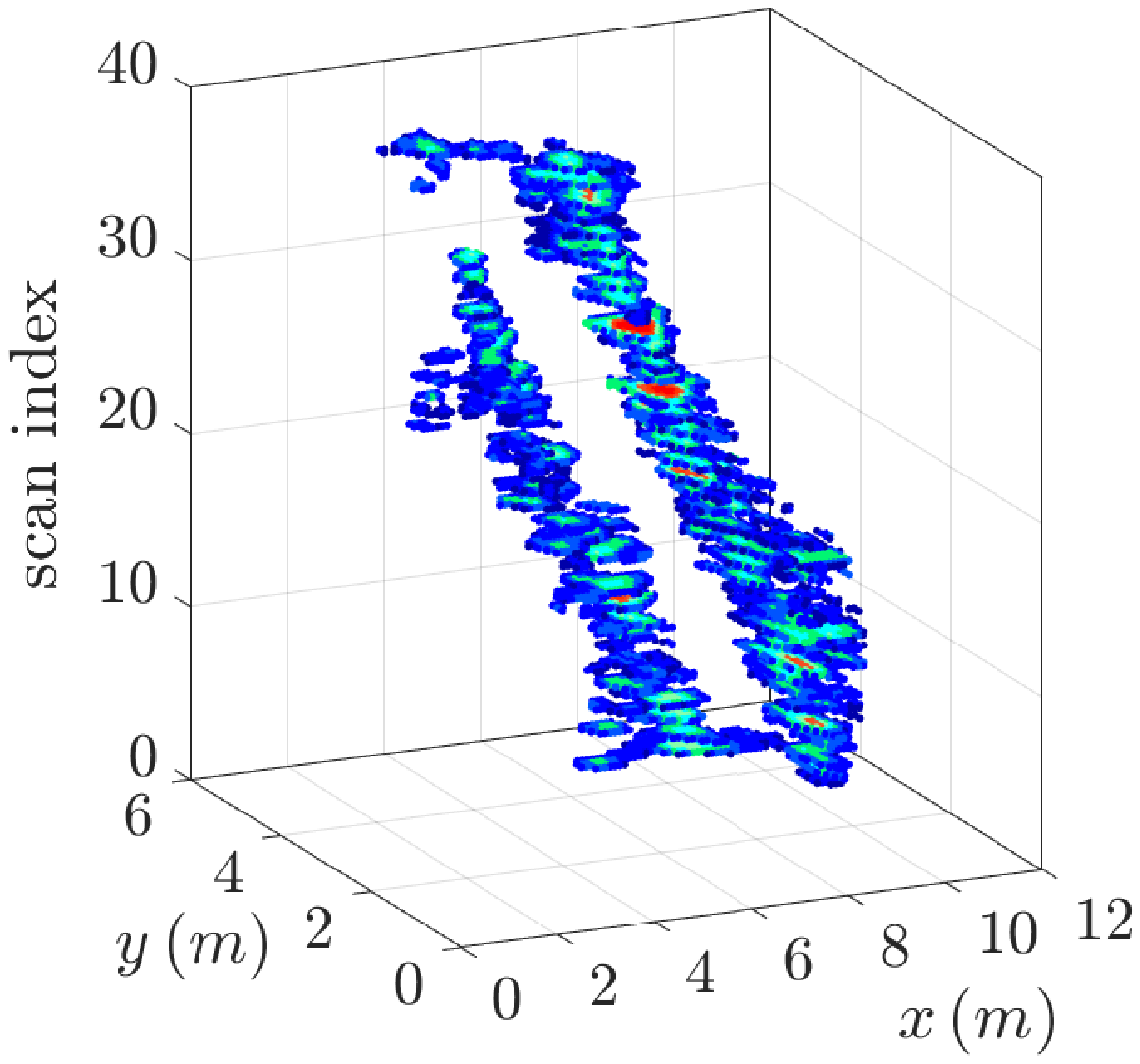}}
\quad
\subfigure[ ]{
\includegraphics[width=0.30\columnwidth]{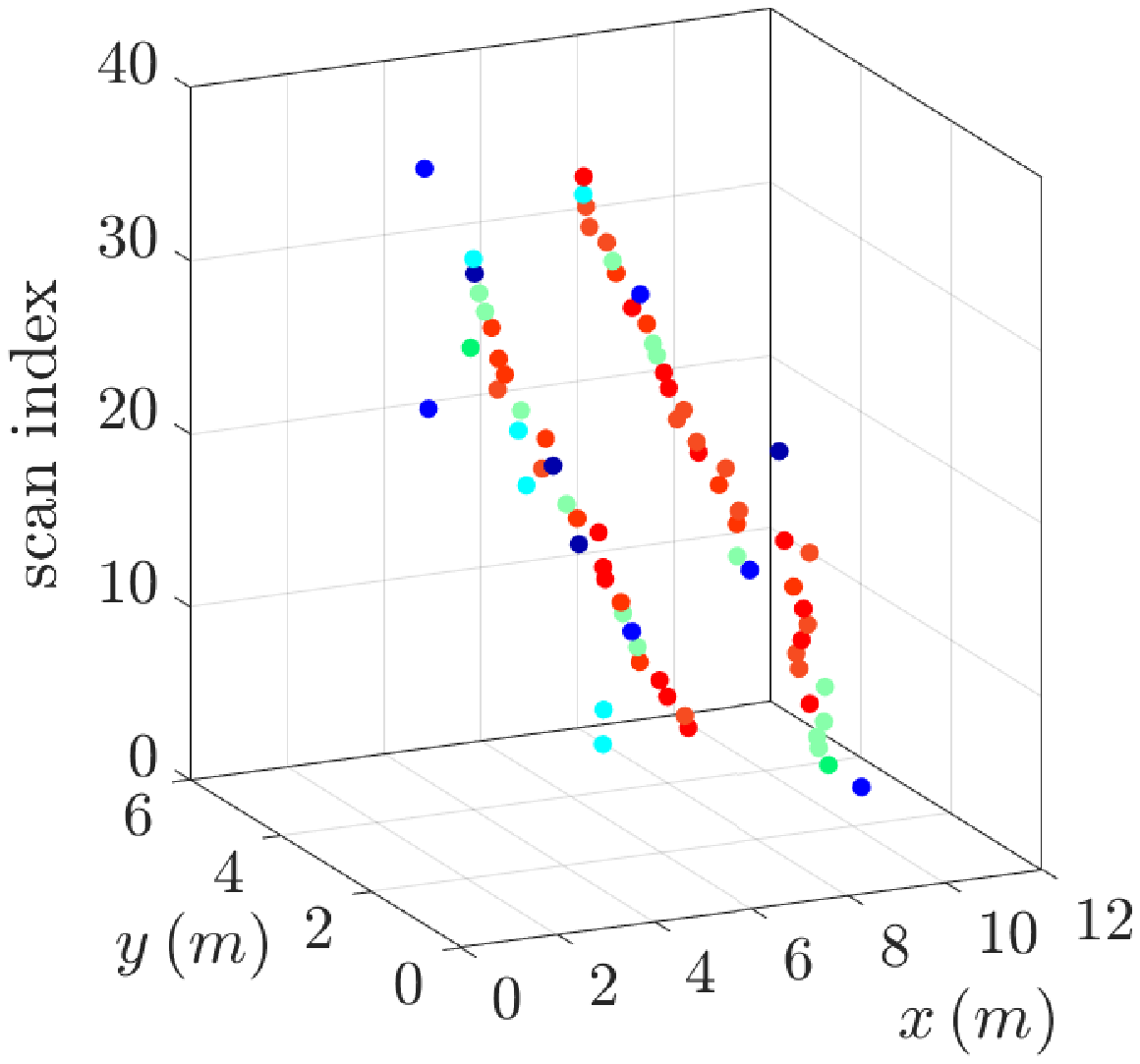}}
\quad
\subfigure[ ]{
\includegraphics[width=0.30\columnwidth]{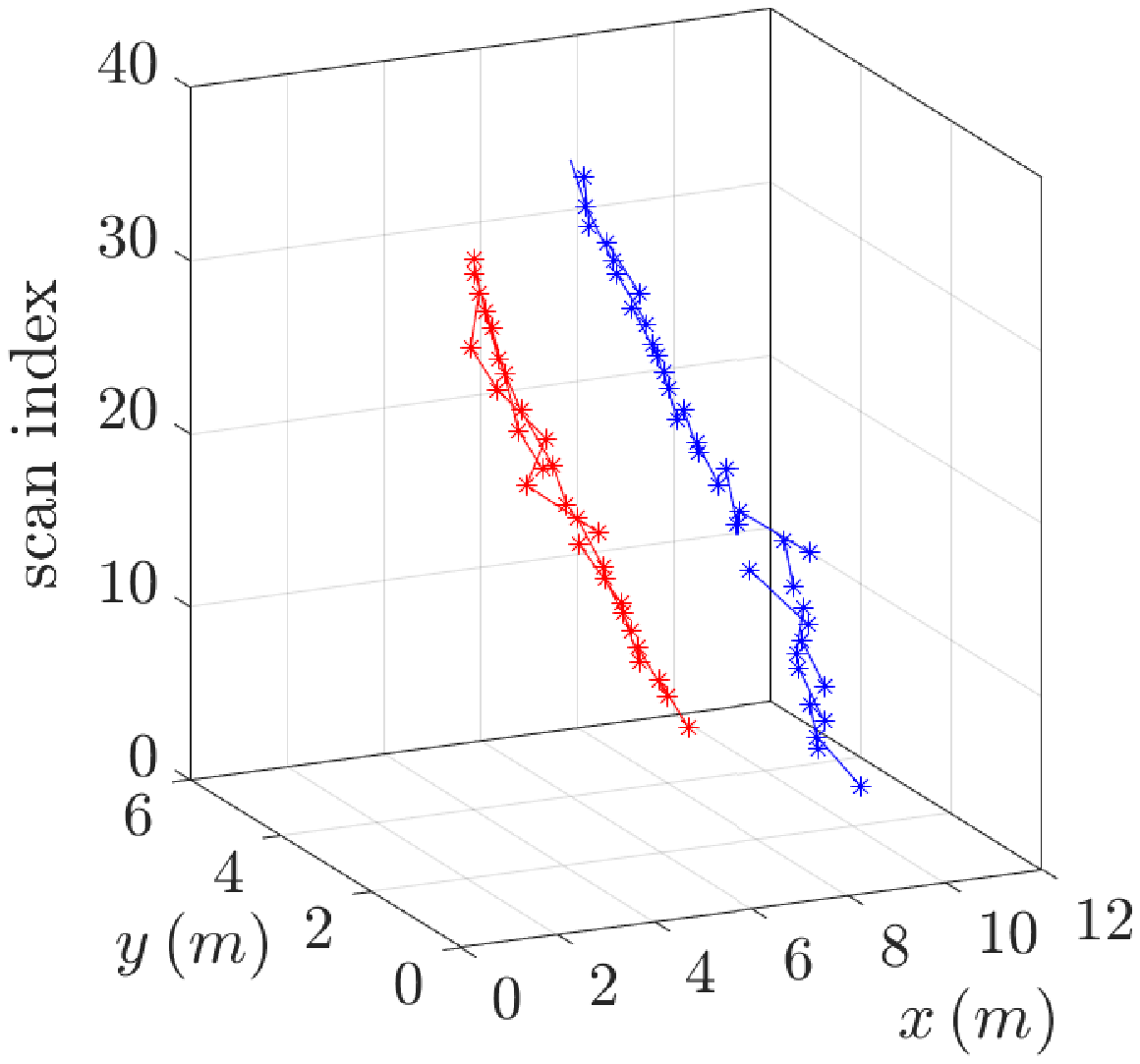}}
\quad
\subfigure[ ]{
\includegraphics[width=0.30\columnwidth]{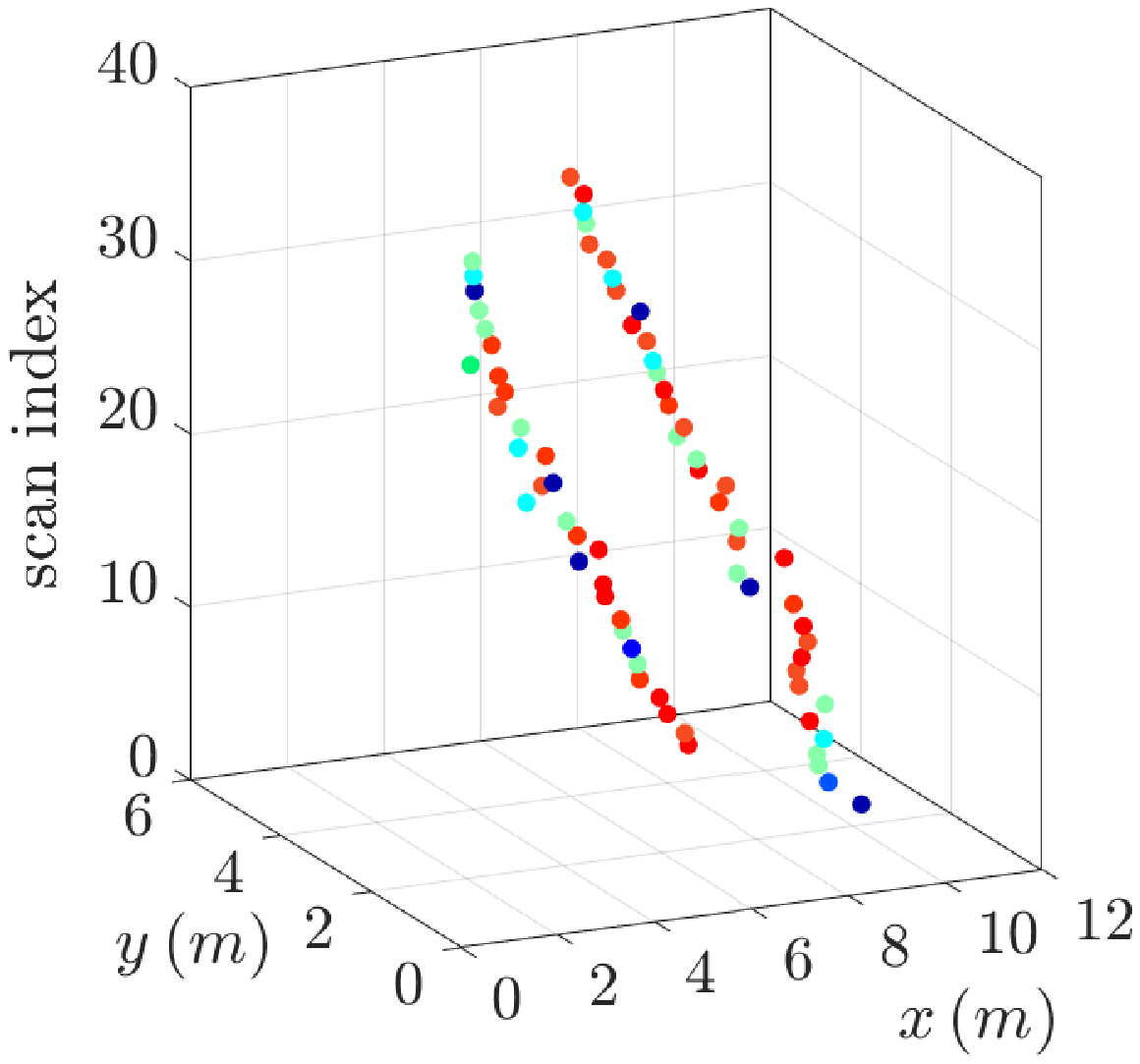}}
\quad
\caption{(a) Example of 3D data structure applying \eqref{eq:M_to_S} only. Red and blue colors denote the largest and the smallest score, respectively.  (b) The same  data  structure  obtained  by  applying  3D  region  growing.  (c)  The  same  data  structure  obtained  by  applying  also  3D  opening  operation.  (d)  The  points obtained with the proposed generation method, and (e) the corresponding tracklets. (f) Trajectory points after tracklet association (before outlier removal and trajectory smoothing). }\label{fig:Case2_cloud_points}
\end{figure}

\begin{figure}[!tb]
\centering
\includegraphics[width=0.4\columnwidth]{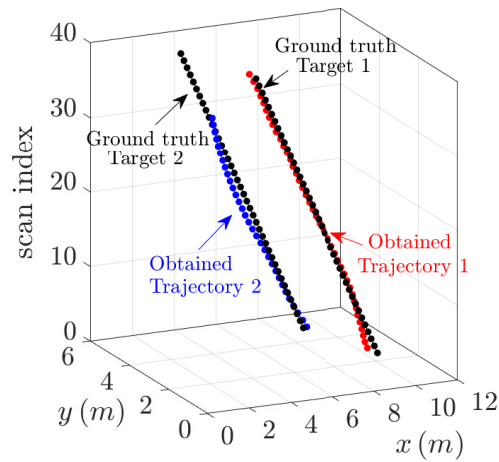}
\caption{The ground truth trajectory (black ball) and the smoothed one. Red balls indicate smoothed trajectory of target 1; blue balls indicate smoothed trajectory of target 2.}\label{fig:Case2_final_track}
\end{figure}

\begin{figure}[!tb]
\centering
\includegraphics[width=0.5\columnwidth]{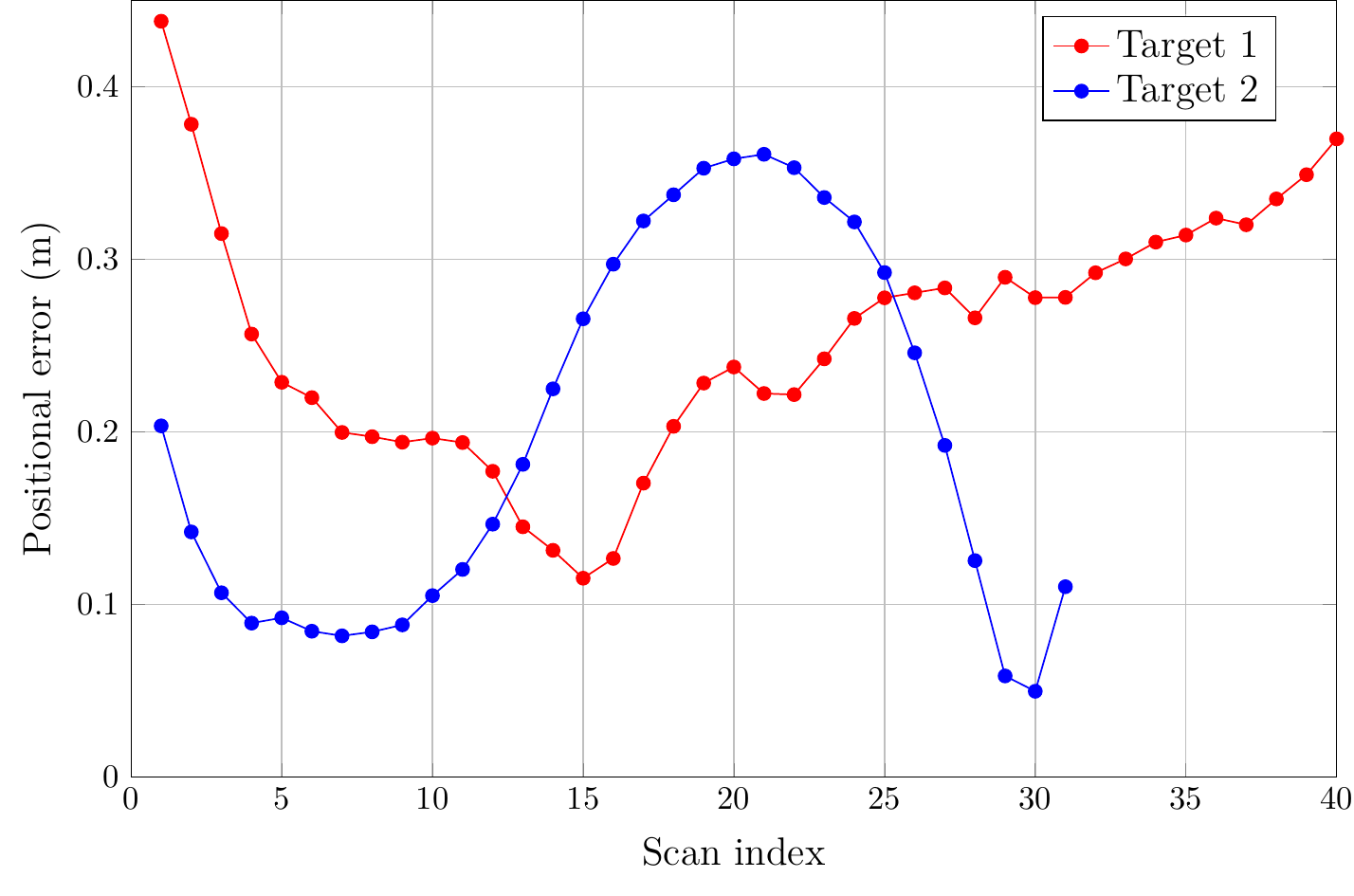}
\caption{Final positioning error (smoothed trajectories) through all scans where targets are detected. }
\label{fig:Case2_positional_error}
\end{figure}

\subsection{Case Study 2}\label{sec:case_study2}

In the second experiment, we employ a multistatic \ac{RSN}, composed of one \ac{UWB} transmitter and $\numsens=4$ \ac{UWB} sensors placed in a rectangular area of size $12\,\mathrm{m}\times 6 \,\mathrm{m}$, to detect two human targets in an indoor scenario. 
The ground truth of the two targets is provided in Fig.~\ref{fig:Case2_scenario}, where the transmitter and four sensors are indicated by the black square and the four triangles, respectively. 
The total number of scan periods in this second experiment is $K=40$.

In each bistatic pair, a single \ac{UWB} pulse transmission causes the reception of at least two pulse replicas: the direct one via the transmitter-to-receive path and the target-reﬂected one via the transmitter-to-target-to-receiver path.  
Fig.~\ref{fig:Case2_map_sensor}(a) shows an example of the actual waveform $\boldsymbol{m}_n^t$ received by the first sensor ($n=1$) in the first scan ($t=1$). 
The corresponding score map $\bm{S}_n^t$ is depicted in Fig.~\ref{fig:Case2_map_sensor}(b), in which the transmitter, sensor, and two target positions are marked with a black square, a red triangle, and two stars, respectively. 
An example of overall score map $\bm{S}^t$ (again for $t=1$) is shown in Fig~\ref{fig:Case2_map_sensor}(c). In this experiment, we applied the clutter removal procedure described in Appendix~\ref{app:clutter} to remove static clutter; the corresponding clutter-suppressed score map $\hat{\bm{S}}^t$ is shown in Fig.~\ref{fig:Case2_map_sensor}(d). As it is possible to observe, the score map after clutter suppression is much neater, and the target cells are better emphasized with respect to the surrounding ones. 

As done for the first experiment, we also show the result of the several processing steps through all scan periods for the second one. 
This is depicted in Fig.~\ref{fig:Case2_cloud_points}. 
Specifically, we illustrate how the \ac{3D} data structure $\boldsymbol{M}^t$ would appear in the last scan period without application of \ac{3D} region growing and opening operation (Fig.~\ref{fig:Case2_cloud_points}(a)); how it would appear with application of \ac{3D} region growing only (Fig.~\ref{fig:Case2_cloud_points}(b)); how it appears by applying both techniques (Fig.~\ref{fig:Case2_cloud_points}(c)). 
Scores are indicated with colors. %
Comparing Fig.~\ref{fig:Case2_cloud_points}(a) and Fig.~\ref{fig:Case2_cloud_points}(c), we see that a number of isolated cells and boundary cells are removed. %
The points generated through all scan periods by application of the technique proposed in Section~\ref{subsec:detection} are depicted in Fig.~\ref{fig:Case2_cloud_points}(d), where colors are used to indicate the point scores $p_{i,j}^k$ in \eqref{eq:measurement}; all tracklets detected by processing the generated points are shown in Fig.~\ref{fig:Case2_cloud_points}(e); 
all points obtained after tracklet association (before outlier removal and trajectory smoothing) are depicted in Fig.~\ref{fig:Case2_cloud_points}(f), where we can observe the presence of just a few residual false alarm points to be rejected by outliers removal.  

Two trajectories are obtained after tracklet association. The two smoothed trajectories are presented in Fig.~\ref{fig:Case2_final_track}, in red for target~1 and in blue for target~2. 
The ground truth of each target is also depicted in black. 
As we can observe, target~1 is detected in all $40$ scans; target~2 is also detected up to scan $31$ after which, however, it is lost. 
Actually, as from Fig.~\ref{fig:Case2_cloud_points}(d), no point was extracted from this target in the last scan periods due to extremely low echoes. This phenomenon is explained by observing, from Fig.~\ref{fig:Case2_scenario}, that target~2 is shadowed by target~1 in the last scan periods, i.e., target~1 is between the transmitter and target~2. 
The average detection rate of the two targets is $88.75\%$. Target 1 is achieved through all scan periods; the detection rate equals $100\%$ up to scan $31$. 
Remarkably, no false alarm trajectories are built despite the heavy clutter affecting the initial processing stages.

The positioning error for each of the two targets is provided in Fig.~\ref{fig:Case2_positional_error} as a function of the scan index. 
The average positioning error of target~1 (through all scans) and target~2 (up to scan $31$) are $25.4\,\mathrm{cm}$ and $19.7\,\mathrm{cm}$, respectively. The positioning error values are generally very satisfactory, always below $40\,\mathrm{cm}$ except for target~1 in the first scan period.  
To provide a benchmark for the achieved performance, we can compare the proposed approach with the method described in \cite{chiani2018sensor}; this method works very well, even in indoor scenarios, when target echoes are relatively strong but is not specifically tailored to address weak targets. In this second experiment, target~2 generates very weak echoes, and with the same \ac{UWB} waveforms, the method of \cite{chiani2018sensor} was not able to detect it in any scan. The proposed technique compares favorably with the benchmark, revealing its potential in the framework of \ac{UWB} detection and tracking. Its better performance with weak targets outcomes from the application of a \ac{TBD} approach, featuring the judicious combination of low thresholds in score map generation (in order not to cancel out echoes from weak targets) and of a sliding window based temporal processing, including generation of points, tracklet detection, and tracklet association, that allows handling a considerable number of false alarms.

The overall results of Case Study 2 are presented in Table~\ref{tab:Res_Case2}. The detection rate of two benchmark methods is lower than that achieved by the proposed technique, since two targets exist in this case study: The presence of multiple target yields a deterioration in point generation \cite{chiani2018sensor}. As from Table~\ref{tab:Res_Case2}, the proposed method can outperform the other two approaches also under a multistatic configuration.

\renewcommand{\arraystretch}{1}
\begin{table}[t]
  \centering
  \fontsize{8.5}{8}\selectfont
  \begin{threeparttable}
  \caption{Results of Case Study 2}
  \label{tab:Res_Case2}
     \begin{tabular}{p{4.4cm}<{\centering} p{2cm}<{\centering}p{2cm}<{\centering}p{2cm}<{\centering}p{1cm}<{\centering}p{1cm}<{\centering}p{1cm}<{\centering}}
    \toprule
     & $P_d^1$ \& $E_p^1$ 
     & $P_d^2$ \& $E_p^2$
     & $N_{\text{FA}}$
     & \ac{OSPA}
\cr
    \midrule% 
Proposed method ($\Delta$=0.1m)&40/40 \& 0.254&31/40 \&  0.197&0&0.282\cr
    Point generation \cite{chiani2018sensor} + \ac{TBD} \cite{YAN2021107821} &8/40 \& 0.463&0/40 \&  $\varnothing$  &0 &0.676\cr
    Point generation \cite{chiani2018sensor} + \ac{PHD} filter \cite{granstrom2012extended} &5/40 \& 0.483&1/40 \&  0.497&7/40&0.657\cr
    \bottomrule
    \end{tabular}
    \end{threeparttable}
\end{table}

\section{Conclusions}\label{sec:Conclusions}

In this paper, we proposed a complete \ac{TBD} processing chain for (monostatic or multistatic) \ac{UWB} \acp{RSN}, able to cope with weak targets. 
As opposed to track-after-detect approaches, in which target detection is first performed exploiting measurements collected by the several sensors in the same scan period and then detection points are processed by a tracking filter, the temporal dimension is here introduced in the detection process. Waveforms, pre-processed by the spatially distributed \ac{UWB} sensors, are jointly processed for score map generation under a lower threshold for weak targets. Score maps are then jointly processed over the time, in a sliding window fashion, to generate points, tracklets, and finally trajectories. The effectiveness of the proposed method has been verified with actual measurements (in outdoor and indoor environments) and human targets. Trajectory confirmation using several measurements over a time window turned out to be very beneficial to weak target detection and clutter suppression.

\appendices

\section{Clutter Suppression in Score Maps}\label{app:clutter}

The proposed clutter removal procedure relies on the score maps collected through the past $U(t) = t - \max\{1,t-V\}$ scans for some $V > 0$, namely, $\bm{S}^{t-U(t):t-1}=\{\bm{S}^{t-U(t)},\bm{S}^{t-U(t)+1},\cdots,\bm{S}^{t-1}\}$. A \emph{clutter density map} is obtained by computing the average grid cell scores through the past $U(t)$ scans. Denoting this map by $\bm{C} = \{C(i_x,i_y)_{1\leq i_x \leq N_x, 1\leq i_y \leq N_y}\}$, we have
\begin{equation} \label{eq:33}
\begin{aligned}
C(i_x,i_y)=\frac{1}{U(t)}\sum_{i=1}^{U(t)}S^{t-1}(i_x,i_y) .
\end{aligned}
\end{equation}
The score map at scan $t$ after clutter suppression, denoted by $\bm{\hat{S}}^t =\{ \hat{S}^t(i_x,i_y)_{1\leq i_x \leq N_x, 1\leq i_y \leq N_y}\}$, is now computed as 
$\hat{S}^t (i_x, i_y)=S^t(i_x, i_y)/C(i_x, i_y)$.
The score map $\bm{\hat{S}}^t$ is then employed instead of the original one $\bm{S}^t$ in the subsequent processing steps.

\section{3D Region Growing}\label{app:3DRG}

The \ac{3D} region growing algorithm described in Section~\ref{subsec:3d} is formalized in Algorithm~\ref{algo:RG}. In the pseudocode, the generic cell (element of the input \ac{3D} data structure $\boldsymbol{M}^{(t-W+1):t}$) is denoted by $Z$ and the generic seed cell by $Z_\mathrm{s}$. 
The set of seed cells is denoted by $\mathcal{S}$. 
Moreover, for any cell $Z$, $\mathcal{N}_+(Z)$ represents the set of all cells that are neighbors of $Z$ and whose score is nonzero. 
Note that there are at most $26$ such neighbors. 
The status of a seed cell $Z_s$ is denoted by $\Xi(Z_{\mathrm{s}})$ and can take two values, namely, $\Xi(Z_{\mathrm{s}})=1$ (active seed cell) and $\Xi(Z_{\mathrm{s}})=0$ (inactive seed cell). 
In the beginning, all seed cells are set to active (lines 2-4 in the algorithm).

\begin{algorithm}[!tb]
\setstretch{1.0}
\SetAlgoLined
\KwData{\ac{3D} $(N_x\times N_y\times W)$ data structure~$\boldsymbol{M}^{(t-W+1):t}$, set of seed cells $\mathcal{S}$}
\KwResult{$R^t$, $\{ \boldsymbol{\mathcal{R}}_1^t, \boldsymbol{\mathcal{R}}_2^t, \dots, \boldsymbol{\mathcal{R}}_{R^t}^t \}$}
$n \leftarrow 0$; $\mathcal{Q} \leftarrow \{\}$\; 
\ForAll{$(Z_{\mathrm{s}} \in \mathcal{S})$}{
$\Xi\left( Z_{\mathrm{s}} \right) \leftarrow 1$\;
}
\ForAll{$(Z_{\mathrm{s}} \in \mathcal{S})$}{
\If{$\left( (M (Z_{\mathrm{s}}) > 0) \wedge ( \Xi (Z_{\mathrm{s}}) \neq 0 ) \right) $}{
$n \leftarrow n + 1$\;
$\boldsymbol{\mathcal{R}}_n^t \leftarrow \{Z_{\mathrm{s}}\}$\; 
$\mathcal{P} \leftarrow \mathcal{N}_+(Z_{\mathrm{s}})$\;
\While{$(\mathcal{P}\neq\{\})$}{
$\mathcal{T} \leftarrow \{\}$\;
\ForAll{$(Z \in \mathcal{P})$}{
$\mathcal{T} \leftarrow \mathcal{T} \cup \left( \mathcal{N}_+(Z) \setminus (\mathcal{P} \cup \boldsymbol{\mathcal{R}}_n^t ) \right)$\;
$\boldsymbol{\mathcal{R}}_n^t \leftarrow \boldsymbol{\mathcal{R}}_n^t \cup \{Z\}$\;
$\mathcal{P} \leftarrow \mathcal{P} \setminus \{Z\}$\;
}
$\mathcal{P} \leftarrow \mathcal{P} \cup \mathcal{T}$\;
}
\If{\emph{($\boldsymbol{\mathcal{R}}_n^t$ satisfies \eqref{eq:region_score_condition} and \eqref{eq:region_card_condition})}}{
$\Xi(Z_{\mathrm{s}}) = 0$ $\forall\,Z_{\mathrm{s}}$ s.t. $Z_{\mathrm{s}} \in \boldsymbol{\mathcal{R}}_n^t$\;
$\mathcal{Q} \leftarrow \mathcal{Q} \cup \{\boldsymbol{\mathcal{R}}_n^t \}$\;
}
\Else{
\ForAll{$(Z \in \boldsymbol{\mathcal{R}}_n^t)$}{
$M(Z) \leftarrow 0$\;
}
$n \leftarrow n-1$\;
  }
}
}
\ForAll{$(Z \in \boldsymbol{M}^{(t-W+1):t}$ \emph{s.t.} $Z \notin \cup_{i=1}^n \boldsymbol{\mathcal{R}}_i^t)$}{$M(Z) \leftarrow 0$}
\Return $n$, $\mathcal{Q}$
\caption{3D Region Growing}\label{algo:RG}
\end{algorithm}

Seed cells can be processed in any order. 
For each seed cell $Z_{\mathrm{s}}$, a new region (containing in the beginning just the seed cell) is initialized if $Z_{\mathrm{s}}$ is active and its score is nonzero (lines~6-8). 
Moreover, the set $\mathcal{P}$ is initialized with the neighbors of $Z_{\mathrm{s}}$ having a nonzero score (line 9). 
The cycle at lines 10-18 performs the following processing. 
In every iteration, all cells in $\mathcal{P}$ are processed in any order. 
For each such cell $Z$, all neighbors of $Z$ with a nonzero score, except the ones that are already in $\mathcal{P}$ or in $\boldsymbol{\mathcal{R}}_n^t$, are accumulated in the set $\mathcal{T}$ (line 13), and then the cell $Z$ is moved from $\mathcal{P}$ to $\boldsymbol{\mathcal{R}}_n^t$ (lines~14-15). 
This way, at the end of each iteration of the while cycle, the set $\mathcal{T}$ contains all new neighbors, not discovered before, with a nonzero score.
All of these new cells are moved to $\mathcal{P}$ (line~17). 
The cycle terminates whenever $\mathcal{P}$ equals the empty set, meaning that no new neighbors with a nonzero score have been discovered.

The region $\boldsymbol{\mathcal{R}}_n^t$ grown from $Z_{\mathrm{s}}$ is now tested (lines 19-22): If it fulfills the conditions \eqref{eq:region_score_condition} and \eqref{eq:region_card_condition}, then the region is included in the set of output regions $\mathcal{Q}$ and the status of all seed cells in it is set to inactive, so that these seed cells will not be processed.
Otherwise (lines 23-28), the region is rejected and the score of all of its cells is forced to zero.
Before returning the number of obtained regions and their set, the score of all cells not belonging to any of the the regions id forced to zero.

\end{document}